\documentstyle[12pt,epsf,subeqnarray,eqsection,indent]{article}

\begin{document}

\date{November 24, 1998 \\ revised April 8, 1999 \\[1mm]
      to appear in {\em Journal of Statistical Physics} }

\def\spose#1{\hbox to 0pt{#1\hss}}
\def\ltapprox{\mathrel{\spose{\lower 3pt\hbox{$\mathchar"218$}}
 \raise 2.0pt\hbox{$\mathchar"13C$}}}
\def\gtapprox{\mathrel{\spose{\lower 3pt\hbox{$\mathchar"218$}}
 \raise 2.0pt\hbox{$\mathchar"13E$}}}
\def\inapprox{\mathrel{\spose{\lower 3pt\hbox{$\mathchar"218$}}
 \raise 2.0pt\hbox{$\mathchar"232$}}}

\font\sevenrm  = cmr7
\font\seventt  = cmtt7
\def\fancyplus{\hbox{+\kern-6.4pt\lower2.6pt\hbox{\sevenrm --}%
\kern-4pt\raise4.9pt\hbox{\sevenrm --}%
\kern-7.8pt\raise0.7pt\hbox{\seventt |}%
\kern+4pt\raise0.7pt\hbox{\seventt |}}}

\title{Antiferromagnetic Potts Models \\
       on the Square Lattice: \\
       A High-Precision Monte Carlo Study}

\author{
  \\
  {\small Sabino Jos\'e Ferreira}         \\[-0.2cm]
  {\small\it Departamento de Estat\'{\i}stica -- ICEx}  \\[-0.2cm]
  {\small\it Universidade Federal de Minas Gerais} \\[-0.2cm]
  {\small\it Avenida Ant\^onio Carlos 6627}        \\[-0.2cm]
  {\small\it Belo Horizonte, MG 30161-970 BRASIL}  \\[-0.2cm]
  {\small Internet: {\tt SABINO@EST.UFMG.BR}} \\[-0.2cm]
  \\[-2mm]  \and
  {\small Alan D. Sokal}                  \\[-0.2cm]
  {\small\it Department of Physics}       \\[-0.2cm]
  {\small\it New York University}         \\[-0.2cm]
  {\small\it 4 Washington Place}          \\[-0.2cm]
  {\small\it New York, NY 10003 USA}      \\[-0.2cm]
  {\small Internet: {\tt SOKAL@NYU.EDU} }        \\[-0.2cm]
  {\protect\makebox[5in]{\quad}}  
  \\
}
\vspace{0.5cm}

\maketitle
\thispagestyle{empty}   

\vspace{0.2cm}

\begin{abstract}
We study the antiferromagnetic $q$-state Potts model on the square lattice
for $q=3$ and $q=4$, using the Wang-Swendsen-Koteck\'y (WSK)
Monte Carlo algorithm
and a powerful finite-size-scaling extrapolation method.
For $q=3$ we obtain good control up to correlation length $\xi \sim 5000$;
the data are consistent with
$\xi(\beta) = A e^{2\beta} \beta^p (1 + a_1 e^{-\beta} + \ldots)$
as $\beta\to\infty$, with $p \approx 1$.
The staggered susceptibility behaves as $\chi_{stagg} \sim \xi^{5/3}$.
For $q=4$ the model is disordered ($\xi \ltapprox 2$)
even at zero temperature.
In appendices we prove a correlation inequality for
Potts antiferromagnets on a bipartite lattice,
and we prove ergodicity of the WSK algorithm at zero temperature
for Potts antiferromagnets on a bipartite lattice.
\end{abstract}

\vspace{0.5 cm}

\clearpage

\newcommand{\be}{\begin{equation}}
\newcommand{\ee}{\end{equation}}
\newcommand{\<}{\langle}
\renewcommand{\>}{\rangle}
\newcommand{\para}{\|}
\renewcommand{\perp}{\bot}

\def\half{ {{1 \over 2 }}}
\def\smfrac#1#2{{\textstyle\frac{#1}{#2}}}
\def\smhalf{ {\smfrac{1}{2}} }
\def\scra{{\cal A}}
\def\scrc{{\cal C}}
\def\scre{{\cal E}}
\def\scrf{{\cal F}}
\def\scrh{{\cal H}}
\def\scrm{{\cal M}}
\newcommand{\scrmvec}{\vec{\cal M}}
\def\scro{{\cal O}}
\def\scrp{{\cal P}}
\def\scrr{{\cal R}}
\def\scrs{{\cal S}}
\def\scrt{{\cal T}}
\def\ttens{{\stackrel{\leftrightarrow}{T}}}
\def\scrttens{{\stackrel{\leftrightarrow}{\cal T}}}
\def\scrv{{\cal V}}
\def\scrw{{\cal W}}
\def\scry{{\cal Y}}
\def\tauss{\tau_{int,\,\scrm^2}}
\def\taux{\tau_{int,\,{\cal M}^2}}
\newcommand{\taum}{\tau_{int,\,\vec{\cal M}}}
\def\taue{\tau_{int,\,{\cal E}}}
\newcommand{\imag}{\mathop{\rm Im}\nolimits}
\newcommand{\real}{\mathop{\rm Re}\nolimits}
\newcommand{\tr}{\mathop{\rm tr}\nolimits}
\newcommand{\sgn}{\mathop{\rm sgn}\nolimits}
\newcommand{\codim}{\mathop{\rm codim}\nolimits}
\def\textprime{{${}^\prime$}}
\newcommand{\longto}{\longrightarrow}
\def\var{ \hbox{var} }
\newcommand{\gtilde}{ {\widetilde{G}} }
\newcommand{\USp}{ \hbox{\it USp} }
\newcommand{\CP}{ \hbox{\it CP\/} }
\newcommand{\QP}{ \hbox{\it QP\/} }
\def\hboxscript#1{ {\hbox{\scriptsize\em #1}} }
\def\hboxtiny#1{ {\hbox{\scriptsize\em #1}} }   
\def\stagg{ {\hboxscript{stagg}} }
\def\smstagg{ {stagg} }

\newcommand{\plotdot}{\makebox(0,0){$\bullet$}}
\newcommand{\plotsmalldot}{\makebox(0,0){{\footnotesize $\bullet$}}}

\def\bsigma{\mbox{\protect\boldmath $\sigma$}}
\def\btau{\mbox{\protect\boldmath $\tau$}}
\def\br{{\bf r}}

\newcommand{\reff}[1]{(\ref{#1})}

\newcommand{\zed}{{\bf \rm Z}}
\newcommand{\R}{\hbox{{\rm I}\kern-.2em\hbox{\rm R}}}
\font\srm=cmr7 		
\def\szed{\hbox{\srm Z\kern-.45em\hbox{\srm Z}}}
\def\sR{\hbox{{\srm I}\kern-.2em\hbox{\srm R}}}
\def\C{{\bf C}}



\newtheorem{theorem}{Theorem}[section]
\newtheorem{corollary}[theorem]{Corollary}
\newtheorem{lemma}[theorem]{Lemma}
\def\proof{\bigskip\par\noindent{\sc Proof.\ }}
\def\qed{\hbox{\hskip 6pt\vrule width6pt height7pt depth1pt \hskip1pt}\bigskip}

%
%
\newenvironment{sarray}{
          \textfont0=\scriptfont0
          \scriptfont0=\scriptscriptfont0
          \textfont1=\scriptfont1
          \scriptfont1=\scriptscriptfont1
          \textfont2=\scriptfont2
          \scriptfont2=\scriptscriptfont2
          \textfont3=\scriptfont3
          \scriptfont3=\scriptscriptfont3
        \renewcommand{\arraystretch}{0.7}
        \begin{array}{l}}{\end{array}}

\newenvironment{scarray}{
          \textfont0=\scriptfont0
          \scriptfont0=\scriptscriptfont0
          \textfont1=\scriptfont1
          \scriptfont1=\scriptscriptfont1
          \textfont2=\scriptfont2
          \scriptfont2=\scriptscriptfont2
          \textfont3=\scriptfont3
          \scriptfont3=\scriptscriptfont3
        \renewcommand{\arraystretch}{0.7}
        \begin{array}{c}}{\end{array}}

\section{Introduction}   \label{sec1}

The Potts model \cite{Potts_52,Wu_82,Wu_84}
plays an important role in the general theory of critical phenomena,
especially in two dimensions
\cite{Baxter_82,Itzykson_collection,DiFrancesco_97},
and has applications to various condensed-matter systems \cite{Wu_82}.
Ferromagnetic Potts models have been extensively studied over the
last two decades, and much is known about their
phase diagrams \cite{Wu_82,Wu_84}
and critical exponents \cite{Itzykson_collection,DiFrancesco_97,Nienhuis_84}.
But for antiferromagnetic Potts models, many basic questions remain open:
Is there a phase transition at finite temperature, and if so, of what order?
What is the nature of the low-temperature phase(s)?
If there is a critical point, what are the critical exponents and the
universality classes?
Can these exponents be understood (for two-dimensional models)
in terms of conformal field theory?

One thing is known rigorously \cite{Kotecky_88,Salas-Sokal}:
for $q$ large enough (how large depends on the lattice in question),
the antiferromagnetic $q$-state Potts model has
a unique infinite-volume Gibbs measure and
exponential decay of correlations
at all temperatures, {\em including zero temperature}\/:
the system is disordered as a result of the large ground-state entropy.
However, for smaller values of $q$, phase transitions can and do occur.
Moreover, for these antiferromagnetic models
the nature of the phase transition is highly
lattice-dependent, in sharp contrast to the universality
typically enjoyed by ferromagnets.
Thus, one expects that for each lattice ${\cal L}$ there
will be a value $q_c({\cal L})$ such that
\begin{itemize}
   \item[(a)]  For $q > q_c({\cal L})$  the model has exponential decay
       of correlations uniformly at all temperatures,
       including zero temperature.
   \item[(b)]  For $q = q_c({\cal L})$  the model has a critical point
       at zero temperature.
   \item[(c)]  For $q < q_c({\cal L})$  any behavior is possible.
       Often (though not always) the model has a phase transition
       at nonzero temperature, which may be of either first or second order.
\end{itemize}
The problem, for each lattice, is to find $q_c({\cal L})$
and to determine the precise behavior for each $q \le q_c({\cal L})$.

In this paper we report the results of a large-scale
Monte Carlo study of the 3-state and 4-state antiferromagnetic Potts models
on the (two-dimensional) {\em square lattice}\/,
using the Wang-Swendsen-Koteck\'y (WSK) \cite{WSK_89,WSK_90,Lubin-Sokal}
cluster algorithm.\footnote{
   A preliminary version of this work has appeared previously
   \cite{swaf2d_prb}.
}
We use a powerful finite-size-scaling (FSS) extrapolation method
\cite{Luscher_91,Kim_93,Kim_LAT93,Kim_94a,Kim_94b,Kim_95,%
fss_greedy,mgsu3,fss_greedy_fullpaper}
to estimate the infinite-volume correlation length $\xi$
and staggered susceptibility $\chi_{\stagg}$.
Using lattices up to $1536 \times 1536$,
we can attain an accuracy of a few percent on $\xi$ and $\chi_{\stagg}$
at correlation lengths $\xi$ as large as 5000.
This allows us to conjecture the exact form of the critical behavior
for the 3-state model.

The $q$-state Potts model is defined by the reduced Hamiltonian
\be
   \scrh   \;=\;   -J \sum_{\< xy \>}   \delta_{\sigma_x, \sigma_y}
   \;,
 \label{ham}
\ee
where the sum runs over all nearest-neighbor pairs of lattice sites,
and each spin takes values $\sigma_x \in \{ 1, 2, \ldots, q \}$.
The antiferromagnetic case corresponds to $J = -\beta < 0$.
Henceforth we restrict attention to the model \reff{ham} on the
square lattice.

Baxter \cite{Baxter_82,Baxter_82b} has determined the exact free energy
(among other quantities) for the square-lattice Potts model
on two special curves in the $(J,q)$-plane:
\begin{eqnarray}
   e^J   & = &   1 \pm \sqrt{q}             \label{eq2}  \\[1mm]
   e^J   & = &  -1 \pm \sqrt{4-q}           \label{eq3}
\end{eqnarray}
Curve (\ref{eq2}${}_+$) is known to correspond to the ferromagnetic
 critical point,
and Baxter \cite{Baxter_82b} conjectured that
 curve (\ref{eq3}${}_+$) corresponds to the antiferromagnetic critical point.
For $q=2$ this gives the known exact value \cite{Onsager_44};
for $q=3$ it predicts a zero-temperature critical point ($J_c = -\infty$),
 in accordance with previous belief \cite{Lenard_67,Baxter_70}\footnote{
   Note also that the $q=3$ model is exactly soluble
   at zero temperature in an arbitrary magnetic field
   \cite{Baxter_70,Truong_86,Pearce_89a,Pearce_89b};
   this might increase one's suspicions that the zero-temperature zero-field
   case is critical.
};
and for $q>3$ it predicts that the putative critical point lies in the
unphysical region ($e^{J_c} < 0$), so that the entire physical region
$-\infty \le J \le 0$ lies in the disordered phase.
In other words, Baxter \cite{Baxter_82b} predicts that
$q_c(\hbox{square lattice}) = 3$,
a prediction that we will verify numerically in this paper.

Some properties of the zero-temperature critical point for $q=3$
are known (non-rigorously) as a consequence of the mapping of this model
onto a height model, whose long-wavelength behavior is that of a
massless Gaussian
\cite{Nijs_82,Kolafa_84,Park_89,Burton_Henley_97,Salas-Sokal_98}.
In particular, the critical exponents associated to the
staggered and uniform magnetizations are predicted \cite{Nijs_82,Park_89}
to be $\eta_{stagg} = 1/3$ and $\eta_u = 4/3$, respectively.\footnote{
   These predictions have recently been verified numerically
   to high precision \cite{Salas-Sokal_98}.
}
However, the {\em approach}\/ to the critical point is much less well
understood.  In what way, for example, do $\xi$ and $\chi_{\stagg}$ diverge
as $\beta\to\infty$?

Nightingale and Schick \cite{Nightingale_82},
using a phenomenological-renormalization method
based on infinite strips of width 2--8,
claimed that the correlation length diverges as
$\xi \sim \exp(c \beta^{\approx 1.3})$.
Wang, Swendsen and Koteck\'y \cite{WSK_89,WSK_90}, using Monte Carlo,
claimed to confirm this latter behavior.
But this behavior seems {\em a priori}\/ implausible to us:
the fundamental variable in the Potts model is $t = e^J$,
so an ordinary power-law critical point $\xi \sim (t-t_c)^{-\nu}$
with $t_c = 0$ would correspond to $\xi \sim e^{\nu\beta}$.
Moreover, we suspect that this model can be exactly solved
(at least in the sense of determining the exact asymptotic behavior
as $\beta\to\infty$),
in which case $\nu$ would most likely be a rational number.
We are unable to imagine any mechanism leading to
$\xi \sim \exp(c \beta^\kappa)$ with $\kappa \neq 1$.

In this paper we shall present numerical evidence that
strongly suggests the asymptotic behavior
\be
   \xi(\beta)   \;=\;
   A e^{2\beta} \beta^p
   \left[ 1 + a_1 e^{-\beta} + a_2 e^{-2\beta} + \ldots \right]
 \label{xi_asymp}
\ee
with $p \approx 1$.
The critical exponent $\nu = 2$ found here corresponds to an operator
with scaling dimension $X = 2 - 1/\nu = 3/2$, which is one of the possibilities
proposed by Saleur \cite[p.~248]{Saleur_91} ---
though not the one he considered most likely!
The multiplicative logarithmic correction $\beta^p \sim (\log t)^p$
is harder to understand theoretically.
Indeed, our data can arguably be reconciled with $p=0$
at the price of including additive corrections to scaling
based on fractional powers of $e^{-\beta}$:
\be
   \xi(\beta)   \;=\;
   A e^{2\beta}
   \left[ 1 + a_1 e^{-\lambda_1\beta} + a_2 e^{-\lambda_2\beta} + \ldots \right]
 \label{xi_asymp2}
\ee
with $\lambda_1 \approx 0.5$.
But this Ansatz too has its theoretical difficulties;
see Section \ref{sec7.1} for discussion of all these issues.
We hope, in any case,
that the numerical results presented here will serve as useful clues
toward the exact solution of this model.
The works of Saleur \cite{Saleur_90,Saleur_91}
and Henley \cite{Henley_94}
provide some tantalizing ideas in this direction.

As for the staggered susceptibility,
the predicted critical exponent $\eta = 1/3$ \cite{Nijs_82,Park_89} 
leads via the scaling law $\gamma = (2-\eta)\nu$
to the behavior $\chi_{\stagg} \sim \xi^{5/3}$.
Our numerical results are consistent with this prediction,
unmodified by any further powers of $\beta$.

\bigskip

The plan of this paper is as follows:
In Section \ref{sec2} we set the notation and recall briefly
our finite-size-scaling extrapolation method
and the Wang-Swendsen-Koteck\'y (WSK) Monte Carlo algorithm.
In Section \ref{sec3} we report our raw data.
In Section \ref{sec4} we analyze our static data for the 3-state model,
using the finite-size-scaling extrapolation method.
In Section \ref{sec5} we analyze the dynamic critical behavior
of the WSK algorithm for the 3-state model.
In Section \ref{sec6} we analyze the data for the 4-state model.
In Section \ref{sec7} we summarize our conclusions and
discuss prospects for future work.
In Appendix \ref{appA} we prove a correlation inequality for
antiferromagnetic Potts models on a bipartite lattice.
In Appendix \ref{appB} we prove the ergodicity of the WSK algorithm
at zero temperature for
antiferromagnetic Potts models on a bipartite lattice.

\section{Preliminaries}   \label{sec2}

\subsection{Definitions and Notation}   \label{sec2.1}

The $q$-state Potts model is defined by the reduced Hamiltonian
\be
   \scrh   \;=\;   -J \sum_{\< xy \>}   \delta_{\sigma_x,\sigma_y}
   \;,
\ee
where the sum runs over all nearest-neighbor pairs of lattice sites,
and each spin takes values $\sigma_x \in \{ 1, 2, \ldots, q \}$.
The antiferromagnetic case corresponds to $J = -\beta < 0$.
It is useful to represent the $q$ possible values of the spin $\sigma_x$
by unit vectors ${\bf e}^{(1)} ,\ldots, {\bf e}^{(q)} \in \R^{q-1}$
pointing from the center to the vertices of a hypertetrahedron;
these vectors satisfy
\be
   {\bf e}^{(i)} \cdot {\bf e}^{(j)}
   \;=\;
   {q \delta_{ij}  \,-\, 1   \over   q \,-\, 1}
   \;=\;
   \cases{  1                    & if $i=j$  \cr
            \noalign{\vskip 2mm}
            - \, {1 \over q-1}   & if $i \neq j$  \cr
         }
\ee
We denote this ``vectorial'' spin by $\bsigma_x \equiv {\bf e}^{(\sigma_x)}$.

The two-point correlation function $G(x,y)$ is defined by
\be
   G(x,y)   \;\equiv\;
   \< \bsigma_x \,\cdot\, \bsigma_y \>
   \;=\;
   \left\<   { {q \delta_{\sigma_x,\sigma_y} \,-\, 1}   \over  {q \,-\, 1} }
   \right\>
   \;.
 \label{def_G}
\ee
Henceforth we exploit translation invariance and write $G(x,y) = G(x-y)$.
We also define the Fourier-transformed correlation function
at wavevector (``momentum'') $p$:
\be
   \widetilde{G}(p)   \;=\;   \sum\limits_{x}  e^{ip \cdot x} \, G(x)
   \;.
\ee
On the square lattice, the relevant staggering wavevector for
antiferromagnetic Potts models is
\be
   p_{\stagg} \;\equiv\; (\pi,\pi)   \;,
\ee
in the sense that $\widetilde{G}(p)$ is maximum at $p = (\pi,\pi)$:
this follows from the correlation inequality proven in Appendix \ref{appA}.
On a finite $L \times L$ lattice with periodic boundary conditions,
we also define the four smallest nonzero wavevectors,
\begin{subeqnarray}
   p_{min,\pm 1}   \;\equiv\; (\pm 2\pi/L, 0)   \\
   p_{min,\pm 2}   \;\equiv\; (0, \pm 2\pi/L)
\end{subeqnarray}

We wish to study the following quantities:

(a)  The energy\footnote{
   Here $E$ is the mean energy per {\em link}\/
   in the antiferromagnetic model;
   we have chosen this normalization in order to have $0 \le E \le 1$,
   with $E = 0$ for an antiferromagnetic ground state
   and $E = 1$ for a ferromagnetic ground state.
   The mean energy per {\em site}\/ is of course $2E$.
}
\be
   E  \;=\;  \< \delta_{\sigma_0,\sigma_{\bf e}} \>
   \;,
\ee
where ${\bf e}$ stands for any nearest neighbor of the origin.

(b) The staggered susceptibility
\be
   \chi_{\stagg} \;=\;  \widetilde{G}(p_{\stagg})
   \;.
\ee
Note that on a finite lattice this is well-defined only if
the lattice size $L$ is {\em even}\/.

(c) The second-moment correlation length, which is defined in
finite volume by
\be
   \xi_L   \;=\;
   { \displaystyle  [(\chi_{\stagg}/F_{\stagg}) - 1] ^{1/2}
     \over
     \displaystyle  2 \sin (\pi/L)
   }    
   \;\,,
 \label{corr_len}
\ee
where
\be
   F_{\stagg} \;\equiv\;  \widetilde{G}(p_{\stagg} + p_{min,\pm 1})
              \;=\;  \widetilde{G}(p_{\stagg} + p_{min,\pm 2})
\ee
is the correlation function at the wavevectors closest to
$p_{\stagg}$.\footnote{
   See e.g.\ \cite[equations (4.11)--(4.13)]{CEPS_swwo4c2}
   for the definition of the second-moment correlation length
   in a ferromagnetic model, along with its motivation.
   Here we make the obvious transcription to an antiferromagnetic model
   that orders at momentum $p_{\stagg} = (\pi,\pi)$.
}
Again $L$ must be even.

All these quantities can be expressed as expectations involving the
following observables:
\begin{subeqnarray}
   \scrm_{\stagg}^2    & = &
      \left( \sum_x e^{i p_{\stagg} \cdot x} \, \bsigma_x \right)^2    \\[2mm]
   \scrf_{\stagg}      & = &
                      \half \left[
                      \left| \sum_x e^{i (p_{\stagg} + p_{min,+1}) \cdot x}
                             \, \bsigma_x \right| ^2
                        \,+\,
                      \left| \sum_x e^{i (p_{\stagg} + p_{min,+2}) \cdot x}
                             \, \bsigma_x \right| ^2
                   \right]  \qquad                   \\[2mm]
   \scre    &=&   \sum_{\< xy \>}  \delta_{\sigma_x, \sigma_y}
\end{subeqnarray}
Thus, we have
\begin{subeqnarray}
   \chi{\stagg}   & = &   {1 \over V}  \< \scrm_{\stagg}^2 \>   \\[2mm]
   F{\stagg}      & = &   {1 \over V}  \< \scrf_{\stagg}   \>   \\[2mm]
   E              & = &   {1 \over 2V}  \< \scre   \>
\end{subeqnarray}
where $V = L^2$ is the number of lattice sites.

In addition to studying the (static) behavior of
the antiferromagnetic Potts model,
we are also interested in studying the dynamic critical behavior
of the Wang-Swendsen-Koteck\'y (WSK) Monte Carlo algorithm.
So let $A$ be an observable
(i.e.\ a function of the spin configuration $\{ \sigma \}$).
We define the
unnormalized autocorrelation function
\be
   C_{AA}(t)  \;=\;   \< A_s A_{s+t} \>   -  \< A \> ^2  \;,
\ee
where expectations are taken {\em in equilibrium}\/,
and the corresponding normalized autocorrelation function
\be
   \rho_{AA}(t)  \;=\;  C_{AA}(t) / C_{AA}(0) \;.
\ee
We furthermore define the integrated autocorrelation time
\begin{subeqnarray}
   \tau_{int,A}  & = &
        \half \sum_{t = -\infty}^{\infty}  \rho_{AA}(t)         \\
   &=&  \half \;+\;  \sum_{t = 1}^{\infty} \rho_{AA} (t)   \;.
\end{subeqnarray}
[The factor of $\half$ is purely a matter of convention;  it is
inserted so that $\tau_{int,A} \approx \tau$ if
$\rho_{AA}(t) \approx e^{-|t|/\tau}$ with $\tau \gg 1$.]
Finally, we define the exponential autocorrelation times
\be
   \tau_{exp,A}  \;=\;
      \limsup_{t \to \infty}  {|t| \over -\log | \rho_{AA}(t)|}
\ee
and
\be
   \tau_{exp}   \;=\;  \sup_A \, \tau_{exp,A}
   \;.
\ee
Note that $\tau_{exp} = \tau_{exp,A}$
whenever the observable $A$ is not orthogonal to the
slowest mode of the system.

The integrated autocorrelation time controls the statistical error
in Monte Carlo measurements of $\< A \>$.  More precisely,
the sample mean
\be
   \bar A \;\equiv\; {1 \over n }  \sum_{t=1}^n  A_t
\ee
has variance
\begin{subeqnarray}
\var( \bar A )  &= &
{1 \over n^2} \ \sum_{r,s \,=\, 1}^n \ C_{AA} (r-s)      \\
&=& {1 \over n }\ \sum_{{t} \,=\, -(n-1)}^{n-1}
(1 -  {{|t| \over n }} ) C_{AA} (t) \label{var_observa}  \\
&\approx&  {1 \over n }\ (2 \tau_{int,A} ) \ C_{AA} (0)
\qquad {\rm for}\ n\gg \tau \label{var_observb}
\end{subeqnarray}
Thus, the variance of $\bar{A}$ is a factor $2 \tau_{int,A}$
larger than it would be if the $\{ A_t \}$ were
statistically independent.
Stated differently, the number of ``effectively independent samples''
in a run of length $n$ is roughly $n/2 \tau_{int,A}$.
The autocorrelation time $\tau_{int,A}$ (for interesting observables $A$)
is therefore a ``figure of (de)merit'' of a Monte Carlo algorithm.

The integrated autocorrelation time $\tau_{int,A}$ can be estimated
by standard procedures of statistical time-series analysis
\cite{Priestley_81,Anderson_71}.
These procedures also give statistically valid {\em error bars}\/
on $\< A \>$ and $\tau_{int,A}$.
For more details, see \cite[Appendix C]{Madras_88}
or \cite[Section 3]{Sokal_Cargese}.
In this paper we have used a self-consistent truncation window of width
$c \tau_{int,A}$, where $c=6$;
this choice is reasonable whenever the
autocorrelation function $\rho_{AA}(t)$ decays roughly exponentially,
as it does here (see Section \ref{sec5.2} below).

In setting the error bars on $\xi$ [defined in \reff{corr_len}]
we have used the triangle inequality;
such error bars are overly conservative, but we were too lazy to
measure the cross-correlations between $\scrm_{\stagg}^2$ and $\scrf_{\stagg}$.
(This was a mistake, and in future work we {\em will}\/ measure these
 cross-correlations.)

\subsection{Finite-Size-Scaling Extrapolation Method}   \label{sec2.2}

In the theory of critical phenomena we are usually interested
in infinite-volume systems,
but Monte Carlo simulations are perforce carried out on finite lattices.
One traditional approach has been to run on lattice sizes $L \gtapprox 6\xi$,
which are large enough so that the finite-size corrections are negligible.
In the past few years, however, methods have become available
for extrapolating finite-size data to $L=\infty$,
based on finite-size-scaling (FSS) theory;
these methods allow one to work, for a given lattice size $L$,
at correlation lengths $\xi$ much larger than were previously attainable.
In this subsection we review an extremely powerful and general method
of this kind,
due originally to L\"uscher, Weisz and Wolff \cite{Luscher_91}
and more recently elaborated by our group
\cite{fss_greedy,mgsu3,fss_greedy_fullpaper}.
This extrapolation method plays a crucial role in the present work,
as it allows us to reach correlation lengths $\xi$ of order 5000,
whereas in the traditional approach we would have been limited to
$\xi \ltapprox 250$.

Consider, for simplicity, a model controlled by a renormalization-group
fixed point having {\em one}\/ relevant operator.
(The parameter $\beta$ may be multidimensional,
 but only one direction in $\beta$-space should be relevant
 in the RG sense.  In other words, the continuum limit should be unique
 modulo length rescalings.)
Let us work on a periodic lattice of linear size $L$.
Let $\xi(\beta,L)$ be a suitably defined finite-volume correlation length,
such as the second-moment correlation length defined by \reff{corr_len};
and let $\scro$ be any long-distance observable
(e.g.\ the correlation length or the susceptibility).
Then finite-size-scaling theory
\cite{Barber_FSS_review,Cardy_FSS_book,Privman_FSS_book}
predicts that
\be
   {\scro(\beta,L) \over \scro(\beta,\infty)}   \;=\;
   f_{\scro} \Bigl( \xi(\beta,\infty)/L \Bigr)
   \,+\,  O \Bigl( \xi^{-\omega}, L^{-\omega} \Bigr)
   \;,
 \label{eqfss1}
\ee
where $f_{\scro}$ is a universal (though usually unknown) function
and $\omega$ is a correction-to-scaling exponent.\footnote{
   This form of finite-size scaling
   assumes hyperscaling, and thus is expected to hold only below the upper
   critical dimension of the model.
   See e.g.\ \cite[Chapter I, section 2.7]{Privman_FSS_book}.
   Note also that when we say $f_{\scro}$ is ``universal'',
   we mean only that it is the same for all models in a given
   universality class.  Of course $f_{\scro}$ varies from
   one universality class to another.
}
It follows that if
$s$ is any fixed scale factor (usually we take $s=2$), then
\be
   {\scro(\beta,sL) \over \scro(\beta,L)}   \;=\;
   F_{\scro} \Bigl( \xi(\beta,L)/L \,;\, s \Bigr)
   \,+\,  O \Bigl( \xi^{-\omega}, L^{-\omega} \Bigr)
   \;,
 \label{eqfss2}
\ee
where $F_\scro$ can easily be expressed in terms of $f_\scro$ and $f_\xi$.
(Henceforth we shall suppress the argument $s$
 if it is clear from the context.)
In other words, if we make a plot of $\scro(\beta,sL) / \scro(\beta,L)$
versus $\xi(\beta,L)/L$, then all the points should lie on a single curve,
modulo corrections of order $\xi^{-\omega}$ and $L^{-\omega}$.

Our method proceeds as follows\footnote{
   Our method \cite{fss_greedy,mgsu3,fss_greedy_fullpaper}
   is essentially identical to that of
   L\"uscher, Weisz and Wolff \cite{Luscher_91}.
   The principal difference is that L\"uscher {\em et al.}\/\
   choose carefully their runs $(\beta,L)$
   so as to produce only a few distinct values of
   $x \equiv \xi(\beta,L)/L$,
   while we attempt to cover an entire interval of $x$.
   Which approach is preferable depends on one's aims
   and on the available CPU time.   
   Also, the motivations are somewhat different:
   the primary aim of L\"uscher {\em et al.}\/\ \cite{Luscher_91}
   is to compare the asymptotic behavior of
   the finite-size-scaling functions $F_{\scro}(x)$
   at large $x$ to the perturbative predictions;
   they did not discuss the possibility of obtaining
   extrapolations to $L=\infty$ at each fixed $\beta$,
   although this is of course implicit in their method.
   The method of Kim \cite{Kim_93,Kim_LAT93,Kim_94a,Kim_94b,Kim_95}
   is also very closely related,    
   but he compares lattice size $L$ to $\infty$ rather than to $sL$;
   this is a (slight) disadvantage.
   Our method also has many features in common with that used by
   Flyvbjerg and Larsen \cite{Flyvbjerg_91a,Flyvbjerg_91b}
   to extrapolate their $1/N$-expansion finite-lattice data.
   It should be emphasized that all these methods are completely general;
   although they were historically first applied to
   asymptotically free theories \cite{Luscher_91},
   they are in no way limited to this case.
   Note also that all these methods share the property
   of working only with observable quantities ($\xi$, $\scro$ and $L$)
   and not with bare quantities ($\beta$).
   Therefore, they rely only on ``scaling'' and not on ``asymptotic scaling'';
   and they differ from other FSS-based methods such as
   phenomenological renormalization \cite{Nightingale_78}.
}:
Make Monte Carlo runs at numerous pairs $(\beta,L)$ and $(\beta,sL)$.
Plot $\scro(\beta,sL) / \scro(\beta,L)$ versus $\xi(\beta,L)/L$,
using those points satisfying both $\xi(\beta,L) \ge$ some value $\xi_{min}$
and $L \ge$ some value $L_{min}$.
If all these points fall with good accuracy on a single curve ---
thus verifying the Ansatz \reff{eqfss2} for
 $\xi \ge \xi_{min}$, $L \ge L_{min}$ ---
choose a smooth fitting function $F_{\scro}$.
Then, using the functions $F_\xi$ and $F_\scro$,
extrapolate the pair $(\xi,\scro)$ successively from
$L \to sL \to s^2 L \to \ldots \to \infty$.

We have chosen to use functions $F_{\scro}$ of the form\footnote{
   In performing this fit, one may use any basis one pleases in
   the space spanned by the functions $\{e^{-k/x}\}_{1 \le k \le n}$;
   the final result (in exact arithmetic) is of course the same.
   However, in finite-precision arithmetic the calculation may become
   numerically unstable if the condition number of the least-squares matrix
   gets too large.  In particular, this disaster occurs if we use
   as a basis the monomials $t^k$ (where $t = e^{-1/x}$).
   The trouble is that these monomials are ``almost collinear''
   in the relevant Hilbert space $L^2(\mu)$
   defined by $\mu(t) = \sum_i w_i \, \delta(t-t_i)$,
   where $t_i$ are the values of $t \equiv e^{-L/\xi(\beta,L)}$
   arising in the data pairs and
   $w_i = 1/[\hbox{error on } \scro(2L)/\scro(L)]^2$
   are the corresponding weights.
   To avoid this disaster, we should seek to use a basis
   that is closer to orthogonal
   in $L^2(\mu)$.  Of course, exactly orthogonalizing in $L^2(\mu)$
   is equivalent to diagonalizing the least-squares matrix,
   which is unfeasible;  but we can do well enough by
   using polynomials with zero constant term that are orthogonal
   with respect to the simple measure
   $w(t) = t^a (t_{max} - t)^b$ on $[0,t_{max}]$,
   where $a$ and $b$ are some chosen numbers $> -1$.
   These polynomials are Jacobi polynomials
   $f_k(t) = t P_{k-1}^{(b,a+2)}(2t/t_{max} \,-\, 1)$
   for $1 \le k \le n$ \cite[pp.~321--328]{Hassani_91}.
   The idea here is that the measure $w(t) = t^a (t_{max} - t)^b$
   should roughly approximate the measure $\mu(t)$.
   We have here used $a=0$, $b=-3/4$;
   but the performance is very insensitive to the choices of $a$ and $b$.
   This cleverness in the choice of basis
   {\em vastly}\/ improves the numerical stability of the result,
   by reducing the condition number of the matrix arising in the fit.
   Typical condition numbers using Jacobi polynomials are
   $\sim 20$ for $n=3$ and  $\sim 100$ for $n=9$.
}
\be
   F_\scro(x)   \;=\;
   1 + a_1 e^{-1/x} + a_2 e^{-2/x} + \ldots + a_n e^{-n/x}   \;.
 \label{eqfss3}
\ee
(Other forms of fitting functions can be used instead.)
This form is partially motivated by theory, which tells us
that in some cases
$F_\scro(x) \to 1$ exponentially fast as $x\to 0$.
Typically a fit of order $3 \le n \le 13$ is sufficient;
the required order depends on the range of $x$ values covered by the data
and on the shape of the curve.
Empirically, we increase $n$ until the $\chi^2$ of the fit
becomes essentially constant.
The resulting $\chi^2$ value provides
a check on the systematic errors arising from
corrections to scaling and/or from the inadequacies of the form \reff{eqfss3}.

The {\em statistical}\/ error on the extrapolated value of
$\scro_\infty(\beta) \equiv \scro(\beta,\infty)$ comes from three sources:
\begin{itemize}
   \item[(i)]  Error on $\scro(\beta,L)$, which gets multiplicatively
       propagated to $\scro_\infty$.
   \item[(ii)]  Error on $\xi(\beta,L)$, which affects the argument
       $x \equiv \xi(\beta,L)/L$ of the scaling functions
       $F_\xi$ and $F_\scro$.
   \item[(iii)]  Statistical error in our estimate of the coefficients
       $a_1, \ldots, a_n$ in $F_\xi$ and $F_\scro$.
\end{itemize}
The errors of type (i) and (ii) depend on the statistics available
at the single point $(\beta,L)$, while the error of type (iii) depends
on the statistics in the whole set of runs.
Errors (i)+(ii) [resp.\ (i)+(ii)+(iii)]
can be quantified by performing an auxiliary Monte Carlo experiment in which
the input data at $(\beta,L)$ [resp.\ the whole set of input data]
are varied randomly within their error bars
and then extrapolated
(we call this the method of ``fake data sets'').\footnote{
   In principle, $\xi$ and $\scro$ should be generated
   from a {\em joint}\/ Gaussian with the correct covariance.
   We ignored this subtlety and simply generated {\em independent}\/
   fluctuations on $\xi$ and $\scro$.
}

The discrepancies between the extrapolated values from different
lattice sizes at the same $\beta$ --- to the extent that these exceed
the estimated statistical errors --- can serve as a rough estimate
of the remaining systematic errors.
More precisely,
let $\scro_i$ ($i=1,\dots,m$) be the extrapolated values
at some given $\beta$, and let $C = (C_{ij})_{i,j=1}^m$
be the estimated covariance matrix for their statistical errors.\footnote{
   This covariance matrix is computed from the auxiliary Monte Carlo
   experiment mentioned in the preceding paragraph.
   Since this $C$ is only a statistical estimate,
   the values of $\bar{\scro}$, $\bar{\sigma}$ and $\scrr$
   will vary {\em slightly}\/ from one analysis run to the next.
}
[Errors of type (iii) induce off-diagonal terms in $C$.]
Then we form the weighted average
\be
   \bar{\scro}   \;=\;
   \left( \sum\limits_{i,j=1}^m  (C^{-1})_{ij} \scro_j \right)
   \Biggl/
   \left( \sum\limits_{i,j=1}^m  (C^{-1})_{ij} \right)
   \;,
 \label{scrobar}
\ee
the error bar on the weighted average
\be
   \bar{\sigma}   \;=\;
   \left( \sum\limits_{i,j=1}^m  (C^{-1})_{ij} \right) ^{\! -1/2}
   \;,
 \label{sigmabar}
\ee
and the residual sum-of-squares
\be
   \scrr   \;=\;
   \sum\limits_{i,j=1}^m  (\scro_i - \bar{\scro}) (C^{-1})_{ij}
                          (\scro_j - \bar{\scro})
   \;.
  \label{residual}
\ee
Under the assumptions that
\begin{itemize}
   \item[(a)] the fluctuations among the $\scro_1,\ldots,\scro_m$ are purely
      statistical [i.e.\ there are {\em no}\/ systematic errors in the
      extrapolation], and
   \item[(b)]  the statistical error bars are correct,
\end{itemize}
$\scrr$ should be distributed as a $\chi^2$ random variable
with $m-1$ degrees of freedom.
Moreover, the sum of $\scrr$ over all the values of $\beta$
should be distributed as a $\chi^2$ random variable
with $\sum (m-1)$ degrees of freedom.\footnote{
   This latter statement is not quite correct, as it ignores the
   correlations between the various $\scro_i$ at {\em different}\/ 
   $\beta$, which are induced by errors of type (iii). 
   [Correlations between different $\scro_i$ at the {\em same}\/ 
    $\beta$, which are also induced by errors of type (iii), 
    {\em are}\/ included in \reff{scrobar}--\reff{residual}.]
}
In this way, we can search for values of $\beta$ for which the extrapolations
from different lattice sizes are mutually inconsistent;
and we can test the overall self-consistency of the extrapolations.

A figure of (de)merit of the method is the relative variance on the
extrapolated value $\scro_\infty(\beta)$,
multiplied by the computer time needed to obtain it.\footnote{
   At fixed $(\beta,L)$,
   this variance-time product tends to a constant as the CPU time
   tends to infinity.  However, if the CPU time used is too small,
   then the variance-time product can be significantly larger than
   its asymptotic value, due to nonlinear cross terms between error sources
   (i) and (ii).
}
We expect this {\em relative variance-time product}\/
[for errors (i)+(ii) only] to scale as
\be
   \hbox{RVTP}(\beta,L)   \;\approx\;
   \xi_\infty(\beta)^{d+z_{int,\scro}}
      \, G_{\scro} \Bigl( \xi_\infty(\beta)/L \Bigr)
   \;,
 \label{RVTP_scaling}
\ee
where $d$ is the spatial dimension
and $z_{int,\scro}$ is the dynamic critical exponent
of the Monte Carlo algorithm being used;
here $G_\scro$ is a combination of several static and dynamic
finite-size-scaling functions,
and depends both on the observable $\scro$ and on the algorithm
but not on the scale factor $s$.
As $\xi_\infty/L$ tends to zero,
we expect $G_\scro$ to diverge as $(\xi_\infty/L)^{-d}$
(since it is wasteful to use a lattice $L \gg \xi_\infty$).
As $\xi_\infty/L$ tends to infinity,
we expect $G_\scro \sim (\xi_\infty/L)^p$ for some power $p$
(see \cite{fss_greedy_fullpaper} for details).
Note that {\em the power $p$ can be either positive or negative}\/.
If $p>0$, there is an optimum value of $\xi_\infty/L$;
this determines the best lattice size at which to perform runs
for a given $\beta$.
If $p<0$, it is most efficient to use the {\em smallest}\/ lattice size
for which the corrections to scaling are negligible compared to the
statistical errors.
[Of course, this analysis neglects errors of type (iii).
 The optimization becomes much more complicated if errors of type (iii)
 are included, as it is then necessary to optimize the set of runs
 as a whole.]

The reader is referred to \cite{mgsu3,fss_greedy_fullpaper}
for a fuller treatment of this extrapolation method,
in particular the finite-size-scaling theory
and the analysis of the propagation of statistical errors.

Let us make one final comment about the physics contained in the
scaling function $F_\xi(x)$.
At the critical point $\beta_c$, the correlation length $\xi(\beta_c,L)$
is proportional to $L$:  one thus has
\be
    \lim\limits_{L\to\infty}   {\xi(\beta_c,L)  \over  L}
    \;=\;  \hbox{some value }  x^\star
\ee
and
\be
    \lim\limits_{L\to\infty}   {\xi(\beta_c,sL)  \over  \xi(\beta_c,L)} 
    \;=\;  s
    \;.
\ee
Therefore, $x^\star$ is determined by the relation
\be
   F_\xi(x^\star;s)   \;=\;  s
   \;.
 \label{F_xi_xstar}
\ee
The constant $x^\star$ is characteristic of the massless field theory
corresponding to the given critical point,
on a continuum torus with aspect ratio 1;
for two-dimensional models it should
in principle be calculable via conformal field theory.
Likewise, for any observable $\scro$ which behaves in the critical region
like $\scro \sim \xi^{\gamma_\scro/\nu}$,
one has
\be
   F_\scro(x^\star;s)   \;=\;  s^{\gamma_\scro/\nu}
   \;.
 \label{F_O_xstar}
\ee

\subsection{Wang-Swendsen-Koteck\'y (WSK) Algorithm}
   \label{sec2.3}

About a decade ago,
Wang, Swendsen and Koteck\'y (WSK) \cite{WSK_89,WSK_90}
proposed an elegant and extraordinarily efficient Monte Carlo algorithm
for simulating the antiferromagnetic $q$-state Potts model
on an arbitrary finite graph $G$.
The elementary update of their algorithm goes as follows:
Choose at random two distinct ``colors'' $i,j \in \{1,\ldots,q\}$;
freeze all the spins $\sigma_x$ currently taking values $k \neq i,j$,
and allow the remaining spins to take value either $i$ or $j$.
The induced model is thus an antiferromagnetic {\em Ising}\/ model
(in zero magnetic field\footnote{
   The key point here is that the interaction energies
   $i$--$k$ and $j$--$k$ are equal.
   This guarantees that the induced Ising model
   has zero magnetic field.
})
on a subgraph of $G$;
this model can be updated by any legitimate Monte Carlo algorithm,
such as the Swendsen-Wang algorithm \cite{Swendsen_87}
or Wolff's single-cluster variant \cite{Wolff_89a}
(we use the former).
It is easy to see that this algorithm leaves invariant the
Gibbs measure of the underlying Potts model.

At zero temperature ($\beta=\infty$) the antiferromagnetic Potts model
reduces to the equal-weight distribution on $q$-colorings of $G$,
and the WSK algorithm becomes:  independently for each connected cluster
of $i$--$j$ spins, either leave that cluster as is
or else flip it (interchanging $i$ and $j$).

The WSK algorithm is trivially seen to be ergodic at any
nonzero temperature.  However, the ergodicity at zero temperature
is a very subtle problem, which has thus far been only partially resolved.
Lubin and Sokal \cite{Lubin-Sokal} showed that for $q=3$
the WSK algorithm is {\em non-ergodic}\/ at zero temperature
on periodic square lattices of size $3m \times 3n$
where $m$ and $n$ are relatively prime.
On the other hand, in Appendix \ref{appB} we shall prove
that the WSK algorithm is {\em ergodic}\/ for all $q$
whenever the graph $G$ is bipartite;
in particular, this happens on periodic square lattices of size
$m \times n$ whenever $m$ and $n$ are both {\em even}\/.
For other cases the ergodicity is an open problem.

It should be noted that the problem of ergodicity at zero temperature
is not merely a theoretical one, even if all our simulations are
performed at nonzero temperature.
If the algorithm is non-ergodic at $\beta=\infty$,
then the autocorrelation time must diverge as $\beta\to\infty$,
{\em even on a fixed finite lattice}\/.\footnote{
   More precisely, it is the {\em exponential}\/ autocorrelation time
   $\tau_{exp}$ (corresponding to the slowest mode in the system)
   which must diverge.  The integrated autocorrelation time
   $\tau_{int,A}$ for any given observable $A$ need not diverge;
   that depends on the choice of $A$.
   Moreover, even if it does diverge, the divergence could be very weak,
   if $A$ has ``weak overlap'' with the slowest mode.
   On the other hand, the divergence of $\tau_{exp}$ already
   calls into question the convergence to equilibrium,
   by raising the specter of ``metastability''.
}
This behavior, if it occurs, could be a severe impediment to
high-precision Monte Carlo study of the model.
So it is fortunate that our model here lies precisely in the
situation for which ergodicity has been proven:
the periodic square lattice with $L$ even.

\section{Summary of our Runs}  \label{sec3}

We simulated the square-lattice Potts antiferromagnets for $q=3$ and $q=4$,
using the WSK algorithm with standard (multi-cluster) Swendsen-Wang updates
of the induced Ising model.

For $q=3$ we ran on $L \times L$ periodic lattices with
$L = 32,64,128,256,512,1024,1536$
at 149 different pairs $(\beta,L)$
in the range $2.0 \le \beta \le 6.0$
(corresponding to $5 \ltapprox \xi_\infty \ltapprox 20000$).
Our data cover rather densely the range
$0.09 \ltapprox x \equiv \xi(\beta,L)/L \ltapprox 0.63$.
Each data point comprises
between $2 \times 10^5$ and $2.2 \times 10^7$ iterations
of the WSK algorithm,
which corresponds to anywhere from $40000\tau$ to $5 \times 10^6 \tau$.
We discarded the first 10000 iterations of each run,
which ought to be more than enough for equilibration ($> 2000\tau$!);
we also made spot checks for evidence of initialization bias
after the discard interval, and found none.
Most of the runs used a random initial configuration (``hot start'').
In some cases with $L=32,64$ we made multiple independent runs,
some of which used an antiferromagnetically ordered initial configuration
(``cold start'');
statistical tests showed complete agreement between the runs.
The data from independent runs were merged statistically in the usual way.
The raw data for $q=3$ can be found in Table~\ref{table:q=3}.

The CPU time for our program is $4.4 L^2$ $\mu$s/sweep
on an IBM RS-6000/370,
and the total CPU time was about 2.5 years on this same machine.
(This is an ``equivalent'' figure:  in fact our runs were performed
on a variety of mostly slower machines in both New York and Belo Horizonte,
so the actual total CPU time was more than this.)

For $q=4$ it sufficed to make a small number of runs for
$L=32,64$;  the total CPU time was less than 3 days
on an IBM RS-6000/370.
The raw data can be found in Table~\ref{table:q=4}.

\section{Data Analysis:  $q=3$, Static Quantities}   \label{sec4}

In this section we analyze the static data for $q=3$.
Concerning the
correlation length $\xi$ and the staggered susceptibility $\chi_{\stagg}$,
we first extrapolate these quantities to infinite volume
(Section \ref{sec4.1}) and then analyze the behavior as
$\beta\to\infty$ of the extrapolated data (Section \ref{sec4.2}).
We conclude by taking a brief look at the energy $E$ (Section \ref{sec4.3}).

\subsection{Extrapolation to Infinite Volume}   \label{sec4.1}

We shall extrapolate $\xi$ and $\chi_{\stagg}$ to infinite volume
using the method of Section \ref{sec2.2} with scale factor $s=2$.
This method is specified by three parameters:
the cut points $\xi_{min}$ and $L_{min}$,
and the interpolation order $n$.
We shall therefore vary these parameters systematically and study the
systematic errors attributable to them.

\subsubsection{Correlation Length}   \label{sec4.1.1}

In Table~\ref{table_chisq_xi} we report the quality of the fit ---
chi-squared ($\chi^2$), number of degrees of freedom (DF),
$\chi^2$/DF, and the corresponding confidence level\footnote{
   ``Confidence level'' is the probability that
   $\chi^2$ would exceed the observed value, assuming that the
   underlying statistical model is correct.  An unusually low confidence level
   (e.g.\ less than 5\%) thus suggests that the underlying statistical model
   is {\em incorrect}\/.  Here this may be due to an inadequate
   interpolation Ansatz (too low $n$) or to
   corrections to scaling (too low $\xi_{min}$ or $L_{min}$).
   Another possible cause of unusually low confidence levels
   will be discussed in Section \ref{sec4.1.2}.
}
--- for the function $F_{\xi} \equiv \xi(\beta,2L)/ \xi(\beta,L)$
as a function of the interpolation order $n$ and the cut point $L_{min}$;
here we have used $\xi_{min} = 10$.
(We tried also $\xi_{min} = 20$ and the results are virtually unchanged.)
A reasonable $\chi^2$ is obtained when $n \ge 5$ and $L_{min} \ge 64$;
a slightly better $\chi^2$/DF is obtained by taking $L_{min} = 128$.
Further increases in $n$ and/or $L_{min}$ do {\em not}\/ improve
the $\chi^2$/DF.\footnote{
   Indeed, for $L_{min} = 256$ the $\chi^2$/DF is {\em worse}\/
   than for $L_{min} = 64,128$; it is, in fact, as bad as for $L_{min} = 32$!
   We do not understand the reason for this behavior,
   which may be simply a statistical fluctuation.
   \protect\label{note_Lmin_256}
}
Our preferred fit is therefore
$n=5$, $\xi_{min} = 10$ and $L_{min} = 128$:
we get
\begin{eqnarray}
   F_\xi(x)   & = &   1  \,+\, 0.896737 e^{-1/x}   \,+\, 29.243141 e^{-2/x}
      \,-\, 253.811947 e^{-3/x}   \nonumber \\
   &  & \quad   +\,  2092.996892 e^{-4/x}  \,-\, 5334.794958 e^{-5/x}
   \;,
 \label{fit_FSS_xi}
\end{eqnarray}
which is plotted in Figure~\ref{fig:xi_FSSplot}.
This fit is reliable only in the interval where there are data points
contributing to it, namely $x \le x_{max} \approx 0.629781$.

We remind the reader that our raw-data error bars on $\xi$
are {\em overestimates}\/, as a result of our use of the triangle
inequality;  therefore, the $\chi^2$ values reported in
Table~\ref{table_chisq_xi} are {\em underestimates}\/,
and only their {\em relative}\/ magnitudes can be considered to be reliable.
The absolute quality of the fit is, therefore, not as good as it looks.
This will be further discussed below in connection with $\chi_{\stagg}$
(Section \ref{sec4.1.2}).

We can make a crude estimate of the universal value $x^\star$
defined by $F_\xi(x^\star) = 2$ [cf.\ \reff{F_xi_xstar}].
On the one hand, $F_\xi(x_{max}) = 1.987350$,
so $x^\star > x_{max} \approx 0.629781$.
On the other hand, $F'_\xi(x_{max}) = 3.080022$,
so if we extrapolate linearly for $x \ge x_{max}$,
we get $x^\star = 0.633888$.
This is not far from the value $x^\star = 0.633983$
obtained by taking \reff{fit_FSS_xi} seriously even for $x > x_{max}$.
So it is a fair guess that $x^\star \approx 0.633888$.\footnote{
   After completion of this work, Salas and Sokal \cite{Salas-Sokal_98}
   obtained $x^\star = 0.63457 \pm 0.00033$ by high-precision
   simulation of this model {\em at}\/ $\beta = \infty$.
}

In Figure~\ref{fig:xi_DeviazoniCurva}(a) we plot
the {\em deviations}\/ from our preferred fit together with their error bars.
The points with $L=32$ show weak ($< 0.01$)
but apparently statistically significant deviations,
of positive sign,
in the interval $0.52 \ltapprox x \ltapprox 0.625$:
see the blow-up of this region in
Figure~\ref{fig:xi_DeviazoniCurva}(b).\footnote{
   The $L=32$ point at $x \approx 0.633$ lies outside the range of $x$
   covered by the fit ($x_{max} \approx 0.630$ when $L_{min} = 128$),
   so the negative deviation exhibited by this point may not be meaningful.
}
It is a reasonable guess that these deviations arise from
systematic corrections to scaling.
However, a careful test of this hypothesis would require higher statistics
and more densely spaced points;
in addition, it would be useful to obtain data with
extremely high statistics ($\sim 10^8$ sweeps)
on smaller lattices than we have bothered to use here ($L=16$ and even $L=8$),
in order to observe a stronger correction-to-scaling ``signal''.
The points with $L \ge 64$ do not seem to show
any systematic deviations from the fit.
Because we observe statistically significant corrections to scaling
on only one lattice size, we are unable to make any firm statement
about the $L$-dependence of the correction-to-scaling term
(which we expect to be of the form $L^{-\Delta}$,
 where $\Delta > 0$ is a correction-to-scaling exponent).
All we can say is that $\Delta$ is not {\em too}\/ small,
since otherwise the correction to scaling would be observable
also on the $L=64$ lattice.
But we are unable to say whether, for example,
$\Delta \approx 1$ or $\Delta \approx 2$.
Let us remark that Salas and Sokal \cite{Salas-Sokal_98}
have recently predicted, on the basis of the height representation
of the zero-temperature model
\cite{Nijs_82,Kolafa_84,Park_89,Burton_Henley_97,Salas-Sokal_98},
that $\Delta = 2$ (at least when $\beta=\infty$).

We now compute the extrapolated values $\xi_\infty$,
using the function $F_\xi$ given by our preferred fit
as well as by several alternative fits that use
more or less stringent choices of $L_{min}$.
(In all cases we take $n=5$;
 the results for $n=6$ or $n=7$ differ in almost all cases by
 $< 0.4$ standard deviations.)
The statistical error bars on $\xi_\infty$ are computed by an auxiliary
Monte Carlo process using ``fake data sets'',
as described in Section \ref{sec2.2},
and include errors of types (i)+(ii)+(iii).
In particular, the correlations between the
extrapolated values $\xi_\infty$ from different lattice sizes
at the same $\beta$ (though not those at different $\beta$ values)
are taken account of in this computation.
For each $\beta$ we compute the weighted average \reff{scrobar}
of the various estimates $\xi_\infty$,
along with the statistical error bar \reff{sigmabar}
and the $\scrr$ value \reff{residual}.
These are reported in Table~\ref{table:DatiEstrapolatiMediatiConChiQuadro}.
The statistical errors on $\xi_\infty$ are of order
1\% (resp.\ 2\%, 3\%, 5\%)
at $\xi_\infty \approx 1000$ (resp.\ 2000, 5000, 10000).
These extrapolated values from different lattice sizes
at the same $\beta$ are found to be consistent within statistical errors:
only two of the 42 $\beta$ values has an $\scrr$ value
too large at the 5\% level;
and summing all $\beta$ values we have
$\sum\scrr = 43.45$ (75 DF, level = 99.9\%).
This unusually low chi-squared is probably due,
at least in part, to our overestimation of the raw-data error bars on $\xi_L$,
which leads to an overestimation of the error bars on $\xi_\infty$.

The discrepancies between the extrapolations with $L_{min} = 64, 128, 256$
are in almost all cases less than half the quoted statistical error.\footnote{
   On the other hand, the extrapolations with $L_{min} = 32$ frequently
   deviate from our preferred $L_{min} = 128$ extrapolation by as much as
   $\approx 1 \sigma$.
   These deviations are probably a correction-to-scaling effect
   on the borderline of statistical significance.
}
We are thus reasonably confident that we have obtained quantitative control
over the systematic errors due to corrections to scaling,
and that their effect can be at most to double the quoted statistical errors.

\subsubsection{Staggered Susceptibility}   \label{sec4.1.2}

Next we carry out an analogous analysis for the staggered
susceptibility $\chi_{\stagg}$.
Note that for this observable our raw-data error bars are reliable,
so the absolute $\chi^2$ values can be taken seriously.

In Table~\ref{table_chisq_chi}
we report the quality of the fit for the function
$F_{\chi_{\smstagg}} \equiv \chi_{\stagg}(\beta,2L)/ \chi_{\stagg}(\beta,L)$
as a function of the interpolation order $n$ and the cut point $L_{min}$;
here we have used $\xi_{min} = 10$.
(We tried also $\xi_{min} = 20$ and the results are virtually unchanged.)
The $\chi^2$/DF is smallest when $n \ge 6$ and $L_{min} \ge 128$.\footnote{
   The $\chi^2$/DF gets {\em worse}\/ for $L_{min} = 256$,
   just as it does for the correlation length
   (see footnote \protect\ref{note_Lmin_256} above).
}
However, this $\chi^2$/DF is $\approx 2$,
rather than the $\approx 1$ that it ought to be;
and as a result, the confidence levels are extremely low (of order $10^{-6}$).
We do not understand this behavior, but we can make a few observations:

(a) Clearly, the explanation cannot be either an inadequate fitting function
or corrections to scaling, because increasing $n$ and/or $L_{min}$
and/or $\xi_{min}$ does {\em not}\/ improve the fit.

(b) One possible explanation might be that our raw-data error bars are
underestimated;  but we tried various alternative statistical methods,
such as breaking up long runs into sub-runs, and all gave compatible
error bars.  So we do {\em not}\/ think that this is the problem.

(c) It is worth noting that nearly all the points with poor $\chi^2$
come from the region of large $x$, particularly $x \gtapprox 0.59$.
Indeed, if we look separately at the contributions to $\chi^2$
(using the fit with $n=6$, $\xi_{min} = 10$, $L_{min} = 128$)
coming from the intervals $x \le 0.53$, $0.53 < x < 0.59$ and $x \ge 0.59$,
we find that the first interval contains 29 points which together contribute
24.31 to the $\chi^2$ ($\chi^2$/DF = 0.84, level = 71.4\%),
the second interval contains 10 points which together contribute
16.65 to the $\chi^2$ ($\chi^2$/DF = 1.67, level = 8.2\%),
while the third interval contains 32 points which together contribute
90.49 to the $\chi^2$ ($\chi^2$/DF = 2.83, level = $1.7 \times 10^{-7}$).

(d) This behavior can also be seen in Figure~\ref{fig:chi_DeviazoniCurva},
where we show the deviations from
the fit $n=6$, $\xi_{min} = 10$, $L_{min} = 128$
together with their error bars.
The points with $L=32$ show weak ($< 0.02$)
but statistically significant deviations, of positive sign,
in the interval $0.53 \ltapprox x \ltapprox 0.625$:
see the blow-up of this region in
Figure~\ref{fig:chi_DeviazoniCurva}(b).\footnote{
   The $L=32$ point at $x \approx 0.633$ lies outside the range of $x$
   covered by the fit ($x_{max} \approx 0.630$ when $L_{min} = 128$),
   so the negative deviation exhibited by this point may not be meaningful.
}
However, for $L \ge 64$ the corrections to scaling have become
completely invisible.
On the other hand, for $x > 0.58$
[see the enlarged view in Figure~\ref{fig:chi_DeviazoniCurva}(c)]
we see that several points
deviate from the fitting curve by 2--4 standard deviations;
but there does {\em not}\/ seem to be any systematic trend
to these deviations, nor do the {\em absolute}\/ deviations
seem to be larger for smaller $L$.
So, once again, it is unlikely that these deviations are caused by
corrections to scaling, or indeed by any process that causes
a {\em systematic}\/ bias.

(e) One {\em possible}\/ cause of the unusually high $\chi^2$
is the following:
We treated the points in the fit
--- which of course correspond to {\em pairs}\/ $(\beta,L)/(\beta,2L)$ ---
as statistically independent;
but this is not quite right, as the same raw-data point $(\beta,2L)$
can contribute to {\em two}\/ pairs,
namely $(\beta,L)/(\beta,2L)$ and $(\beta,2L)/(\beta,4L)$,
being in the numerator of the first and in the denominator of the second.
As a result, these pairs of pairs are significantly {\em anti}\/correlated
--- one would expect a correlation coefficient of $\approx -1/2$,
if all three raw-data points have roughly the same relative error ---
and they will thus tend to deviate from each other by {\em more}\/ than
would have been predicted from independent fluctuations with the
given error bars.  Furthermore, when $\xi(\beta,L)/L$ is close to $x^\star$,
$\xi(\beta,2L)/2L$ is in turn not much smaller than $\xi(\beta,L)/L$,
so this anticorrelation acts on pairs of points having
relatively {\em nearby}\/ values of $x$.
This could explain why the large contributions to $\chi^2$
come almost exclusively from the region $x \gtapprox 0.59$,
and why they are apparently completely random.
Unfortunately, it seemed unfeasible for us to invert the large matrices
(of order $\approx 70$) that would be needed to take proper account
of these correlations.
Suffice it to say that {\em if}\/ this is the correct explanation,
then the observed large $\chi^2$ is simply spurious,
and the fit is in reality good after all!
Moreover, in this case the error bars on the extrapolated values
$\xi_\infty$ and $\chi_{\stagg,\infty}$ will be correct,
as the method of ``fake data sets'' {\em does}\/ take proper account
of the aforementioned correlations.

Modulo these caveats, therefore,
we take as our preferred fit the one with
$n=6$, $\xi_{min} = 10$ and $L_{min} = 128$:
we get
\begin{eqnarray}
   F_{\chi_{\smstagg}}(x)   & = &
      1  \,+\,  2.234450 e^{-1/x}  \,+\, 80.120833 e^{-2/x}
         \,-\,  966.050470 e^{-3/x}     \nonumber \\
   & & \quad     +\, 10728.802555 e^{-4/x}  \,-\,  49077.175871 e^{-5/x}
        \,+\, 73776.084137 e^{-6/x} \;,     \nonumber \\
 \label{fit_FSS_chi}
\end{eqnarray}
which is plotted in Figure~\ref{fig:chi_FSSplot}.
This fit is reliable in the interval $x \le x_{max} \approx 0.629781$.
Using the value $x^\star \approx 0.633888$ derived from $F_\xi$,
we have
\be
   F_{\chi_{\smstagg}}(x^\star)   \;=\;
   3.172742   \;=\;   2^{1.66573}
\ee
if we extrapolate \reff{fit_FSS_chi} linearly for $x \ge x_{max}$,
or
\be
   F_{\chi_{\smstagg}}(x^\star)   \;=\;
   3.172345   \;=\;   2^{1.66555}
\ee
if we take \reff{fit_FSS_chi} seriously also for $x > x_{max}$.
Either way, this is in excellent agreement with the prediction \cite{Park_89}
$\gamma/\nu = 5/3$  [cf.\ \reff{F_O_xstar}].

We can now compute the extrapolated values $\chi_{\smstagg,\infty}$,
using the functions $F_\xi$ and $F_{\chi_{\smstagg}}$
given by our preferred fit
as well as by several alternative fits that use
more or less stringent choices of $L_{min}$.
(In all cases we take $n=5$ for $F_\xi$ and $n=6$ for $F_{\chi_{\smstagg}}$;
 the results for larger $n$ differ in almost all cases by
 $< 0.6$ standard deviations.)
The statistical error bars on $\chi_{\smstagg,\infty}$
are computed as before by an auxiliary Monte Carlo process,
and include errors of types (i)+(ii)+(iii).
For each $\beta$ we compute the weighted average \reff{scrobar}
of the various estimates $\chi_{\smstagg,\infty}$,
along with the statistical error bar \reff{sigmabar}
and the $\scrr$ value \reff{residual}.
These are reported in Table~\ref{table:DatiEstrapolatiMediatiConChiQuadro}.
The statistical errors on $\chi_{\smstagg,\infty}$ are of order
2\% (resp.\ 3\%, 5\%, 8\%)
at $\xi_\infty \approx 1000$ (resp.\ 2000, 5000, 10000),
that is, about twice as big as those on $\xi_\infty$.
These extrapolated values from different lattice sizes
at the same $\beta$ are found to be consistent within statistical errors:
only two of the 42 $\beta$ values has an $\scrr$ value
too large at the 5\% level;
and summing all $\beta$ values we have
$\sum\scrr = 50.41$ (75 DF, level = 98.7\%).
We don't know why this chi-squared is so small;
it may be due in part
to our overestimation of the raw-data error bars on $\xi_L$,
which leads to an overestimation of
the errors of type (ii) on $\chi_{\smstagg,\infty}$.

The discrepancies between the extrapolations with $L_{min} = 64, 128, 256$
are again less than half the quoted statistical error in
nearly all cases.\footnote{
   The extrapolations with $L_{min} = 32$ frequently
   deviate from our preferred $L_{min} = 128$ extrapolation by as much as
   $\approx 1.2 \sigma$.
   These deviations are, once again, probably a correction-to-scaling effect
   on the borderline of statistical significance.
}
We are thus reasonably confident that we have obtained quantitative control
over the systematic errors due to corrections to scaling,
and that their effect can be at most to double the quoted statistical errors.

\subsection{Analysis of Extrapolated Data}   \label{sec4.2}

In this section we analyze the behavior as
$\beta\to\infty$ of the extrapolated data for $\xi$ and $\chi_{\stagg}$.
In all cases we use the preferred extrapolations, namely the ones with
$\xi_{min} = 10$, $L_{min} = 128$, and $n=5$ (resp.\ 6) for
$\xi$ (resp.\ $\chi_{\stagg}$).

\subsubsection{Correlation Length}   \label{sec4.2.1}

Our data are in clear agreement with the prediction of a critical point
at $\beta=\infty$.
The correlation length $\xi_\infty$ rises roughly like $e^{2\beta}$,
and we initially thought that this was the exact asymptotic behavior.
However, at $\beta \gtapprox 3.4$ ($\xi_\infty \gtapprox 75$),
$\xi_\infty$ begins to rise {\em faster}\/ than this
(Figure~\ref{fig1}a),
and this rise shows no sign of abating at least up to
$\beta \approx 5.9$ ($\xi_\infty \approx 15000$).
We therefore guessed a multiplicative logarithmic correction,
i.e.\ $\xi_\infty \sim e^{2\beta} \beta^p$ for some power $p>0$:
see Figures~\ref{fig1}b,c for $p=1/2$ and $p=1$, respectively.

In order to distinguish between these scenarios,
we need to make some assumption on the form of
the additive corrections to the leading asymptotic behavior.
Unfortunately we do not know how to carry out a low-temperature expansion
around the (critical) zero-temperature state;
but the simplest hypothesis is that there exists an expansion
in powers of $e^{-\beta}$, which corresponds to a minimum energy cost
of one unit for an ``overturned'' spin.
That is, we expect
\be
   \xi_\infty(\beta)   \;=\;
   A e^{2\beta} \beta^p
   \left[ 1 + a_1 e^{-\beta} + a_2 e^{-2\beta} + \ldots \right]
   \;.
 \label{eq5}
\ee
If we accept this Ansatz, a value $p \approx 1$ is clearly favored
(Figure~\ref{fig2}).
A fit to the first two terms of \reff{eq5} with $p=1$,
using the data points with
$\beta \ge 2.95$ ($e^{-\beta} \ltapprox 0.052$),
yields $A = 0.01814 \pm 0.00006$ and
$A a_1 = 0.20051 \pm 0.00225$ (hence $a_1 \approx 15$)
with $\chi^2 = 13.10$ (35 DF, level = 99.97\%).

On the other hand, Chris Henley (private communication)
has suggested to us that the corrections to scaling
might contain {\em fractional}\/ powers of $e^{-\beta}$:
\be
   \xi_\infty(\beta)   \;=\;
   A e^{2\beta} \beta^p
   \left[ 1 + a_1 e^{-\lambda_1\beta} + a_2 e^{-\lambda_2\beta} + \ldots \right]
 \label{eq5_Delta}
\ee
with $\lambda_1 < 1$.
Since $e^{-\lambda_1\beta} \sim \xi^{-\lambda_1/2}$,
such behavior would ordinarily arise from
a correction-to-scaling exponent $\Delta = \lambda_1/2$;
and our data for the finite-size-scaling functions show
no evidence of a correction-to-scaling exponent
anywhere near this small (see Section \ref{sec4.1.1}).
Let us nevertheless consider the Ansatz \reff{eq5_Delta} open-mindedly
and see whether it can accommodate $p=0$.
In Figure~\ref{fig_xi_p=0_Delta}a,b we plot $\xi_\infty(\beta)/e^{2\beta}$
versus $e^{-\lambda\beta}$ for $\lambda = 1$ and 0.5, respectively.
With $\lambda = 1$, the plot shows both strong curvature and
a rather high slope near the origin;
for $\beta \ge 4.50$ ($e^{-\beta} \ltapprox 0.011$)
the data can be fit well by a straight line with
$A \approx 0.110$ and $a_1 \approx -15$.
With $\lambda = 0.5$, the curvature and slope are less radical;
the data for $\beta \ge 4.50$ ($e^{-0.5\beta} \ltapprox 0.105$)
can be fit well by a straight line with
$A \approx 0.122$ and $a_1 \approx -2.4$.
However, even this latter plot is nowhere near as convincing
as Figure~\ref{fig2}.

Finally, an anonymous referee
has suggested to us that the corrections to scaling
might be in {\em powers of $1/\beta$}\/:
\be
   \xi_\infty(\beta)   \;=\;
   A e^{2\beta} \beta^p
   \left[ 1 + {a_1 \over \beta} + {a_2 \over \beta^2} + \ldots \right]
 \label{eq5_referee}
\ee
(see Section~\ref{sec7.1} for theoretical discussion).
In Figure~\ref{fig_referee}a,b,c we plot
$\xi_\infty(\beta)/(e^{2\beta} \beta^p)$ versus $1/\beta$
for $p=0$, 1/2, 1.
Clearly, our data ---
which lie in the range $0.15 \ltapprox 1/\beta \ltapprox 0.4$ ---
are so far from asymptotic that no conclusion can be drawn.
In particular, the plot for $p=0$ (resp.\ $p=1$)
shows such a large negative (resp.\ positive) slope
at the smallest available value of $1/\beta$
that the extrapolated intercept at $1/\beta = 0$
differs by nearly a factor of 2 from the last data point;
and the plot for $p=1/2$ shows a large third derivative
(i.e.\ abrupt change from parabolic to flat) at $1/\beta \approx 0.2$.

Let us note in passing that our data fit much less well the Ansatz
$\xi_\infty \sim \exp(c \beta^\kappa)$
used by some previous workers \cite{Nightingale_82,WSK_89,WSK_90}.
Indeed, a log-log plot of $\log \xi_\infty$ versus $\beta$
(Figure~\ref{fig3})
shows significant curvature:
the apparent exponent $\kappa$ varies from $\approx 1.69$ at small $\beta$
to $\approx 1.38$ at larger $\beta$.
(This decrease is consistent with our conjecture
 that the true asymptotic value of the exponent $\kappa$ is 1.)
Moreover, as noted in the Introduction,
we have been unable to imagine any theoretical mechanism leading to
$\xi \sim \exp(c \beta^\kappa)$ with $\kappa \neq 1$.

\subsubsection{Staggered Susceptibility}   \label{sec4.2.2}

The staggered susceptibility is consistent with the believed
exact behavior \cite{Park_89} $\chi_{\stagg,\infty} \sim \xi_\infty^{5/3}$,
unmodified by any further powers of $\beta$.
To test this behavior quantitatively,
we need to set error bars on the ratio
$\chi_{\stagg,\infty} / \xi_\infty^{5/3}$.
Since unfortunately we do not know the covariance between
our estimates of $\xi_\infty$ and $\chi_{\stagg,\infty}$,
the best we can do is to use the triangle inequality
to set an {\em upper bound}\/ on the error bar for the ratio;
this upper bound is of course a gross overestimate of the true error,
since the estimates of $\xi_\infty$ and $\chi_{\stagg,\infty}$
are presumably strongly positively correlated.
As a result, the error bars in all fits will be grossly overestimated,
and the $\chi^2$ will be grossly underestimated;
only the {\em relative}\/ values of $\chi^2/\hbox{DF}$
have any significance.

In Figure~\ref{fig_chioverxi53}
we plot $\chi_{\stagg,\infty}/\xi_\infty^{5/3}$ versus $\beta$
(note the very narrow vertical scale);
the error bars are those given by the triangle inequality,
{\em reduced by a factor of 10 for visual clarity}\/.
We see that $\chi_{\stagg,\infty}/\xi_\infty^{5/3}$
varies slightly with $\beta$,
but appears to be tending to a constant $\approx 2.67$ as $\beta\to\infty$.

If we fit $\chi_{\stagg,\infty}/\xi_\infty^{5/3}$ to the Ansatz $A \beta^p$,
the estimates of the power $p$ are extremely small, of order 0.02,
and statistically consistent with zero.
This confirms our belief that there are no additional powers of $\beta$
in the ratio $\chi_{\stagg,\infty}/\xi_\infty^{5/3}$.
If we impose $p=0$ and fit $\chi_{\stagg,\infty}/\xi_\infty^{5/3}$
to a constant $A$,
we obtain $A = 2.657 \pm 0.051$
($\chi^2 = 0.007$, 13 DF, level $> 99.9\%$)
using the data from $\beta \ge 4.70$.
Of course, the error bar on $A$ here is a gross overestimate,
and the $\chi^2$ value is a gross underestimate.

We can, of course, investigate directly the behavior of
$\chi_{\stagg,\infty}$ as a function of $\beta$, without reference to $\xi$.
This approach has the advantage that the error bars on $\chi_{\stagg,\infty}$
are reliable.
In Figure~\ref{fig1_chi} we plot
$\chi_{\stagg,\infty} / (e^{(10/3)\beta} \beta^q)$
versus $\beta$ for $q=0,{5\over6},{5\over3}$.
The behavior is qualitatively similar to that observed in
Figure~\ref{fig1} for $\xi$,
although the variation is somewhat sharper.
If we try again the Ansatz
\be
   \chi_{\stagg,\infty}(\beta)   \;=\;
   B e^{(10/3)\beta} \beta^q
   \left[ 1 + b_1 e^{-\beta} + b_2 e^{-2\beta} + \ldots \right]
   \;,
 \label{eq5_chi}
\ee
a value $q \approx 5/3$ is favored
(Figure~\ref{fig2_chi}).
The curvature is greater than in the corresponding plot for $\xi$,
but the linearity is still reasonable for $e^{-\beta} \ltapprox 0.03$.
A fit to \reff{eq5_chi} with $q=5/3$,
using the data points with $\beta \ge 3.60$ ($e^{-\beta} \le 0.0273$),
yields $B = 0.00329 \pm 0.00003$ and $Bb_1 = 0.06661 \pm 0.00124$
(hence $b_1 \approx 20$)
with $\chi^2 = 17.55$ (27 DF, level = 92\%). 
We omit the plots based on a fractional-power additive correction to scaling,
which are similar to those shown for the correlation length
(Figure~\ref{fig_xi_p=0_Delta}).

\subsection{Behavior of the Energy}   \label{sec4.3}

On theoretical grounds we expect that the infinite-volume energy per site
has a low-temperature expansion of the form
\be
   E_\infty(\beta)   \;=\;   c_1 e^{-\beta} + c_2 e^{-2\beta} + \ldots
   \;.
 \label{eq6}
\ee
Unfortunately we do not have access to $E_\infty$,
as we have made no attempt to extrapolate the energies
(which are {\em short}\/-distance quantities)
to infinite volume.
But examination of the finite-volume energies $E_L$
indicates that for $L \ge 512$ (resp.\ 1024)
the remaining $L$-dependence
is less than $\approx 2 \times 10^{-4}$ (resp.\ $5 \times 10^{-5}$);
so the energies appear to have {\em almost}\/ reached
their infinite-volume limits.
In Figure~\ref{fig4} we plot $E/e^{-\beta}$ versus $e^{-\beta}$,
using different symbols to represent different lattice sizes.
A fit to the $L=1024$ points with
$\beta \ge 5.00$ ($e^{-\beta} \le 0.0067$)
yields $c_1 = 0.21777 \pm 0.00003$ and $c_2 = 1.65303 \pm 0.00664$
($\chi^2 = 5.09$, 8 DF, level = 75\%).
The $L=1536$ points are also compatible with this fit,
and the $L=512$ points differ only slightly from it.
This provides good support for the Ansatz \reff{eq6}.

We would like to make a warning concerning the use of
reweighting methods\footnote{
   These methods are sometimes called ``histogram''
   (or ``multiple-histogram'') methods,
   but in fact the reweighting is most conveniently carried out
   {\em without}\/ forming histograms!
}
\cite{Ferrenberg_88,Ferrenberg_89,Swendsen_93,Ferrenberg_95}
in which Monte Carlo runs at one temperature $\beta$
are employed to generate data at another temperature $\beta'$
by reweighting with the factor $\exp[-(\Delta\beta)H]$,
where $\Delta\beta \equiv \beta' - \beta$.
This reweighting is of course always valid in principle;
but one must be aware that the statistical error bars
on the reweighted data grow rapidly as $|\Delta\beta|$ grows,
and the maximum $|\Delta\beta|$ for which one can obtain a
not-too-large error bar gets smaller for larger $L$:
\be
   |\Delta\beta|   \;\ltapprox\;
   \cases{L^{-d/2}     & near a non-phase-transition point  \cr
          L^{-1/\nu}   & near a critical point \cr
          L^{-d}       & near a first-order phase-transition point  \cr
         }
\ee
Our Monte Carlo data illustrate this point in a very striking way.
In all of our runs with $L \ge 128$ ---
totalling more than 60 million measurements ---
we did not observe even a single configuration
(after the discard interval)
with energy $E=0$.
In other words, none of our histograms for $L \ge 128$
--- even those at our largest $\beta$ value, namely $\beta=6.0$ ---
have {\em any}\/ overlap with the
zero-temperature probability distribution.
It follows that reweighting to zero temperature in these cases
is {\em nonsense}\/.
For $L = 64$, our runs at $\beta=4.50$ and 5.00
(but not smaller $\beta$) do show {\em some}\/ configurations with $E=0$,
but the sample size of such configurations is very small:
about 0.2\% at $\beta = 5.0$.
Reweighting to zero temperature in these cases
would thus produce {\em enormous error bars}\/
(at least if the error bars are computed correctly!).
Only for $L \le 32$ do we have a significant number of zero-energy
configurations:  for example, for $L=32$ and $\beta = 5.0$,
we found that 21.6\% of the configurations have $E=0$.
Furthermore, even reweighting to nonzero temperatures is fraught
with severe dangers.  For example, although our run lengths are anywhere
from $2 \times 10^5$ to $10^6$ measurements,
the energy histograms of a pair of runs typically
show {\em no}\/ overlap if
$|\Delta\beta| \gtapprox 0.3$ (resp.\ 0.4, 0.5--0.6, 0.5--0.9, 1.2--1.4)
for $L=1536$ (resp.\ 1024, 512, 256, 128).\footnote{
   The allowable $|\Delta\beta|$ depends slightly on $\beta$,
   getting smaller at smaller $\beta$.
   For example, at $L=512$ the $\beta=5.7$ run shows no overlap with
   the $\beta=5.1$ run, which in turn shows no overlap with 
   the $\beta=4.65$ run, which in turn shows no overlap with 
   the $\beta=4.2$ run.
}
It follows that reweighting beyond these limits is nonsense,
and reweighting near these limits leads to huge statistical errors.

\section{Data Analysis:  $q=3$, Dynamic Quantities}   \label{sec5}

In this section we analyze the dynamic critical behavior of the
WSK algorithm for the 3-state antiferromagnetic Potts model
on the square lattice.

\subsection{Integrated Autocorrelation Times}  \label{sec5.1}

Examination of Table~\ref{table:q=3} indicates that the
autocorrelation times $\tau_{int,\scrm_{\smstagg}^2}$ and
$\tau_{int,\scre}$ are
bounded uniformly in $\beta$ and $L$.
(Indeed, their values are very small:
$\tau_{int,\scrm_{\smstagg}^2} < 5$ and $\tau_{int,\scre} < 4$.)
We conclude that
{\em critical slowing-down is completely eliminated}\/.

We can study the dynamic critical behavior in more detail
by applying the standard dynamic finite-size-scaling Ansatz
\be
  \tau_{int,A}(\beta,L)  \;\approx\;
     \xi(\beta,L)^{z_{int,A}} \, g_A \Bigl( \xi(\beta,L)/L  \Bigr)
 \label{dyn_FSS_Ansatz}
\ee
to the observables $A = \scrm_{\stagg}^2$ and $\scre$.
Here $z_{int,A}$ is a dynamic critical exponent,
$g_A$ is an unknown scaling function, and
$g_A(0) = \lim_{x \downarrow 0} g_A(x)$
is supposed to be finite and nonzero.\footnote{
   We emphasize that the dynamic critical exponent $z_{int,A}$
   is in general {\em different}\/ from the exponent $z_{exp}$
   associated with the exponential autocorrelation time $\tau_{exp}$
   \cite{Sokal_Cargese,CPS_90,Sokal_LAT90}.
}
Usually we would
determine $z_{int,A}$ by plotting $\tau_{int,A}/\xi(L)^{z_{int,A}}$
versus $\xi(L)/L$ and adjusting $z_{int,A}$
until the points fall as closely
as possible onto a single curve (with priority to the larger $L$ values).
But in our case the situation is much simpler:
the dynamic critical exponents $z_{int,\scrm_{\smstagg}^2}$
and $z_{int,\scre}$ are {\em zero}\/.

In Figure~\ref{fig:magsq_dyn_FSSplot} we show the
dynamic finite-size-scaling plot for $\tau_{int,\scrm_{\smstagg}^2}$,
assuming $z_{int,\scrm_{\smstagg}^2} = 0$.
The data collapse is amazingly good, especially for a plot with
no free parameters.  Indeed, in all our Monte Carlo work on
dynamic critical phenomena
we have never observed a data collapse this good,
even when we had the freedom to adjust $z_{int,A}$.

For $\tau_{int,\scre}$, by contrast, the data collapse is {\em not}\/
so good:  see Figure~\ref{fig:ener_dyn_FSSplot},
where we again assume $z_{int,\scre} = 0$.
Clearly there are {\em huge}\/ corrections to dynamic finite-size-scaling
for this observable.  Even so, the points do appear to be converging
as $L\to\infty$ to a limiting curve (which can be roughly traced using
the $L=512$ and $L=1024$ points).

\subsection{Autocorrelation Functions}
   \label{sec5.2}

Now we want to test the more detailed dynamic finite-size-scaling Ansatz
\be
   \rho_{AA}(t;\beta,L)  \;\approx\;
     |t|^{-p_A}
     h_A \Bigl( t / \tau_{exp,A}(\beta,L) \,;\,  \xi(\beta,L)/L  \Bigr)
   \;,
 \label{dyn_FSS_2}
\ee
where $p_A$ is an unknown exponent and $h_A$ is an unknown scaling function.
If $p_A = 0$, then \reff{dyn_FSS_2} can equivalently be written as
\be
   \rho_{AA}(t;\beta,L)  \;\approx\;
     \widehat{h}_A \Bigl( t / \tau_{int,A}(\beta,L) \,;\,
                          \xi(\beta,L)/L  \Bigr)
   \;.
 \label{dyn_FSS_3}
\ee
In this latter situation\footnote{
  Contrary to much belief, $z_{int,A}$ {\em need not} equal $z_{exp}$.
  Indeed, if $p_A>0$, we have $z_{int,A} = (1-p_A) z_{exp} < z_{exp}$.
  See \cite{Sokal_Cargese,Sokal_LAT90} for further discussion.
},
$\tau_{int,A}$ and $\tau_{exp,A}$ have the {\em same}\/ dynamic critical
exponent $z_{int,A}=z_{exp}$, and we furthermore have
\be
 \frac{\tau_{int,A}}{\tau_{exp,A}} \;\approx\; F_A\left(\xi(\beta,L)/L\right)
\ee
where
\be
  F_A(x) \;\equiv\; \lim_{t\to +\infty}\, \frac{1}{t}\,
     \log {\widehat h}_A(t;x) \;.
\ee

Let us now test the Ansatz \reff{dyn_FSS_3}
for the observable $A = \scrm_{\stagg}^2$.
(We restrict attention to this observable,
 since we already know that for $A = \scre$ the
 dynamic finite-size-scaling behavior is poor.)
In Figure~\ref{scaled_dynamic_plot_allpoints}
we plot $\rho_{\scrm_{\stagg}^2 \scrm_{\stagg}^2}(t)$
versus $t / \tau_{int,\scrm_{\smstagg}^2}$,
using {\em all}\/ the data points.
The points fall {\em roughly}\/ on a single curve before falling into
the statistical noise (which we expect to be of order
$(n/\tau_{int,\scrm_{\smstagg}^2})^{-1/2}$ where $n$ is the run length,
hence of order $\pm 0.005$).
However, even at small $t / \tau_{int,\scrm_{\smstagg}^2}$
there are clear deviations from a single curve,
indicating that the scaling function $\widehat{h}_{\scrm_{\stagg}^2}$
depends in a nontrivial way on its second argument $\xi(L)/L$.
Therefore, in Figure~\ref{scaled_dynamic_plot_subdivided}a--f
we show the same plot with the data subdivided into ``slices'' of $\xi(L)/L$;
the slices are chosen empirically so that the data points
within a slice fall reasonably well onto a single curve
modulo statistical noise.
On each plot we also draw, for reference,
a line corresponding to a pure exponential decay
$\tau_{int,\scrm_{\smstagg}^2} = \tau_{exp,\scrm_{\smstagg}^2}$.
The data support the Ansatz \reff{dyn_FSS_3} reasonably well,
with each range of $\xi(L)/L$ defining roughly a single curve
(until that curve falls into the statistical noise).
The curves for small $\xi(L)/L$ are close to straight
(i.e.\ close to a pure exponential),
while the curves for larger $\xi(L)/L$ are increasingly convex.\footnote{
   It is amusing to note that a similar behavior was observed
   in our study of the multi-grid Monte Carlo algorithm for
   the two-dimensional $O(3)$ $\sigma$-model \cite{mgon}.
   We wonder whether it is a general phenomenon.
}
(Note that the rescaled horizontal axis ensures that the
  total area under each curve is 1. Therefore, the more convex curves must
  be below the straight curve for small $t/\tau_{int,\scrm_{\smstagg}^2}$,
  but above it for large $t/\tau_{int,\scrm_{\smstagg}^2}$.)
This means that the ratio
$\tau_{int,\scrm_{\smstagg}^2}/\tau_{exp,\scrm_{\smstagg}^2}$
is close to 1 for small $\xi(L)/L$,
and less than 1 for larger $\xi(L)/L$.
It is {\em conceivable}\/ that
$\tau_{int,\scrm_{\smstagg}^2}/\tau_{exp,\scrm_{\smstagg}^2}$
tends to 1 as $\xi(L)/L \to 0$;
if true, this would mean that $\scrm_{\stagg}^2$
truly becomes the ``slowest mode'' in the limit $L\to\infty$, $\xi/L \to 0$.

\subsection{Relative Variance-Time Product}  \label{sec5.3}

Finally, let us look at the scaling \reff{RVTP_scaling}
of the relative variance-time product.
For each observable $\scro = \xi, \chi_{\stagg}$
we proceed as follows:
For each run $(\beta,L)$ we form the relative variance
$(\Delta \scro_\infty/ \scro_\infty)^2$
on the extrapolated infinite-volume value coming from that run;
we then multiply by $L^2 \times (\#\hbox{iterations} - \#\hbox{discard})$,
which is a normalized measure of the CPU time invested in that run
(after the discard interval);
the result is, by definition, $\hbox{RVTP}(\beta,L)$.
We then divide $\hbox{RVTP}(\beta,L)$ by $\xi_\infty(\beta)^2$
[since $d=2$ and $z_{int,\scro} = 0$]
and plot it versus $\xi_\infty(\beta)/L$.\footnote{
   In this latter computation, $\xi_\infty(\beta)$ is taken to be
   our best estimate
   (reported in Table~\ref{table:DatiEstrapolatiMediatiConChiQuadro}),
   based on averaging {\em all}\/ the runs at the given $\beta$.
}
The results are reported in Figure~\ref{fig_RVTP}a for $\xi$
and Figure~\ref{fig_RVTP}b for $\chi_{\stagg}$.
The scaling is reasonably good, though not perfect;
this is not surprising, since our computed RVTP includes errors of types
(i) + (ii) + (iii)
while the scaling formula \reff{RVTP_scaling} refers only to errors of
types (i) + (ii).\footnote{
   The much cleaner graph shown in \cite[Figure 2]{fss_greedy}
   was computed by using the theoretical formulae for the
   propagation of errors under extrapolation \cite{mgsu3,fss_greedy_fullpaper},
   and includes {\em only}\/ errors of types (i) + (ii).
}${}^,$\footnote{
   The anomalous $L=1536$ point at $\xi_\infty/L \approx 16.6$
   corresponds to $\beta=6.0$, where our statistics are poor
   and the extrapolation has almost broken down.
   Its relative error (33\% for $\xi$ and 71\% for $\chi_{\stagg}$)
   is so large that both the extrapolated value and its error estimate
   are unreliable.
}
Indeed, the fact that we see even modestly good scaling
indicates that the errors of type (iii) are not dominant.

We see that for $\xi$ the optimal value of $\xi_\infty/L$ is $\approx 1$,
but the minimum is very flat:  any value in the range
$0.5 \ltapprox \xi_\infty/L \ltapprox 10$
is almost equally good.
It is only for $\xi_\infty/L \ltapprox 0.3$ that the RVTP rises sharply,
by a factor of 10 or more.
In other words, the only region in which one should {\em not}\/ run
is the region in which one traditionally {\em always}\/ ran,
namely $\xi_\infty/L \ltapprox 1/6$.
For $\chi_{\stagg}$ the story is similar, but the minimum is somewhat sharper:
the optimum is at $\xi_\infty/L \approx 0.4$,
and the RVTP rises by about a factor of 3 (resp.\ 10)
as $\xi_\infty/L$
increases to $\approx 10$ (resp.\ decreases to $\approx 0.1$).

\section{Data Analysis: $q=4$}   \label{sec6}

For $q=4$ the story is very brief:
simulations on $L=32,64$ agree within statistical error
and show that $\xi \ltapprox 2$
uniformly as $\beta \to \infty$ (Table~\ref{table:q=4}).
Clearly there is no critical point in the physical region.
Physically, there is so much entropy that the correlations decay
exponentially even at zero temperature.
This can be proven rigorously to occur on the square lattice for $q \ge 7$
\cite{Salas-Sokal},
and our simulations confirm Baxter's \cite{Baxter_82b} prediction
that it occurs in fact for $q > 3$.

The autocorrelation times $\tau_{int,\scrm_{\smstagg}^2}$ and
$\tau_{int,\scre}$ of the WSK algorithm
are bounded uniformly in $\beta$ and $L$,
and indeed are almost constant:
$\tau_{int,\scrm_{\smstagg}^2} \approx 2.6$
and $\tau_{int,\scre} \approx 3.5$.

\section{Discussion and Conclusions}   \label{sec7}

In this section we summarize our conclusions and discuss their
theoretical implications.  We conclude by mentioning a few possible
directions for future work.

\subsection{Behavior of the Correlation Length}   \label{sec7.1}

The numerical data presented in this paper (Section \ref{sec4.2.1})
show clearly that the correlation length diverges as $\beta\to\infty$
approximately like $\xi(\beta) \sim e^{2\beta}$.
Since the fundamental variable in the Potts model is $t = e^J = e^{-\beta}$,
an ordinary power-law critical point $\xi \sim (t-t_c)^{-\nu}$
with $t_c = 0$ would correspond to $\xi(\beta) \sim e^{\nu\beta}$.
Therefore, our result can be interpreted as indicating a
power-law critical point at zero temperature with critical exponent $\nu=2$.
The fact that $\nu$ is here a rational number
reinforces our suspicion that this two-dimensional model
can be solved exactly,
at least in the sense of determining the exact asymptotic behavior
as $\beta\to\infty$.
The exponent $\nu = 2$ corresponds to an operator
with scaling dimension $X = 2 - 1/\nu = 3/2$,
which is one of the possibilities
proposed by Saleur \cite[p.~248]{Saleur_91} ---
albeit not the one he considered most likely!

On closer examination, however, the ratio
$\xi(\beta)/e^{2\beta}$ appears {\em not}\/ to be
asymptotically constant as $\beta \to\infty$ (see Figure~\ref{fig1}a);
rather, it begins to rise when $\beta \approx 3.4$ ($\xi \approx 75$),
and this rise shows no sign of abating at least up to
$\beta \approx 5.9$ ($\xi \approx 15000$).
Indeed, our data are compatible with an asymptotic behavior
\be
   \xi(\beta)   \;=\;
   A e^{2\beta} \beta^p
   \left[ 1 + a_1 e^{-\beta} + a_2 e^{-2\beta} + \ldots \right]
 \label{eq7.1}
\ee
with $p \approx 1$ (see Figure~\ref{fig2}).
This corresponds to a power-law critical point with
multiplicative logarithmic correction $\beta^p \sim |\log (t-t_c)|^p$.
The problem is to make theoretical sense of such a behavior.

In the preliminary report of this work \cite{swaf2d_prb},
we asserted that a multiplicative logarithmic correction
$\beta^p \sim (\log t)^p$ with $p$ integer
(particularly $p=1$) can occur
in the renormalization-group framework
as a result of ``resonance'' between operators whose
scaling dimensions are rationally related.
This assertion is (we now realize) only half-true.
For the susceptibility, specific heat and similar observables,
it is indeed true that multiplicative logarithmic corrections
with positive integer powers $p$ can occur as a result of resonance
\cite{Wegner_72,Wegner-Riedel,Wegner_D+G}.
However, we have been unable to devise any renormalization-group scenario
in which the {\em correlation length}\/ acquires such a
multiplicative logarithmic correction in the absence of marginal operators.
For example:

1) Suppose that we hypothesize a scenario with
one relevant variable $t$ (eigenvalue $\lambda > 0$)
and one irrelevant variable $u$ (eigenvalue $-\lambda$),
satisfying the RG flow equations
\begin{subeqnarray}
   {dt \over dl}   & = &   \lambda t  \,+\, t^2 u   \\[2mm]
   {du \over dl}   & = &   -\lambda u
\end{subeqnarray}
where $t(0)$ and $u(0)$ are the couplings in the Hamiltonian,
and $t(l)$ and $u(l)$ are the renormalized couplings after
modes of momentum $\gtapprox e^{-l}$ have been integrated out.
The solution is
\begin{subeqnarray}
   t(l)  & = &  e^{\lambda l} \left( {1 \over t(0)} - Al \right) ^{\! -1}
      \\[2mm]
   u(l)  & = &  A e^{-\lambda l}
\end{subeqnarray}
where $A = u(0)$.
Let us now choose $l$ so that $t(l) = 1$;
this implies that $e^l = \xi/\xi_1$,
where $\xi_1$ is the correlation length at
$t=1$ and $u= A e^{-\lambda l} \approx 0$.
Hence
\be
   {\xi \over \xi_1}   \;=\;
   \left[  {1 \over t(0)}  \,-\, A \log(\xi/\xi_1) \right] ^{\! 1/\lambda}
   \;=\; t^{-1/\lambda} [1 + O(t \log t)]  
   \;,
\ee
so that the presence of the irrelevant operator $u$ (i.e.\ $A \neq 0$)
induces only an {\em additive}\/ correction to scaling $O(t \log t)$.

2) Suppose, alternatively, that we hypothesize two relevant operators $t,u$
with eigenvalues $n\lambda$ and $\lambda$, respectively,
where $n$ is an integer $\ge 1$ and $\lambda > 0$,
satisfying the RG flow equations
\begin{subeqnarray}
   {dt \over dl}   & = &   n\lambda t  \,+\, u^n   \\[2mm]
   {du \over dl}   & = &   \lambda u
\end{subeqnarray}
The solution is
\begin{subeqnarray}
   t(l)  & = &  [t(0) + u(0)^n l] e^{n \lambda l}     \\[2mm]
   u(l)  & = &  u(0) e^{\lambda l}
\end{subeqnarray}
Since there are two relevant operators,
generically {\em two}\/ couplings have to be adjusted in order to
place the system at a critical point.
But if some symmetry were to cause $t(0)$ to be exactly zero
[or at least $\ltapprox u(0)^n$], then a critical point can be reached
by adjusting $u(0)$ alone.
So assume this, and let us analyze the resulting critical behavior.
Let us choose $l$ so that $t(l) = 1$;
this implies that $e^l = \xi/\xi_1$,
where $\xi_1$ is the correlation length at
$t=1$ and $u= l^{-1/n} \approx 0$.
Hence
\be
   \left( {\xi \over \xi_1} \right) ^{\! n\lambda}
   \log\! \left( {\xi \over \xi_1} \right)
   \;=\;  u(0)^{-n}
   \;,
\ee
so that
\be
   {\xi \over \xi_1}   \;=\;  u^{-1/\lambda} \, |\log u|^{-1/n\lambda}
      \left[ 1 \,+\, O\biggl( {\log \log u  \over  \log u} \biggr) \right]
   \;.
\ee
(One expects a further correction-to-scaling term $O(|\log u|^{-1/n})$
 arising from the fact that $u(l) = l^{-1/n} \neq 0$.)
Hence there is a multiplicative logarithmic correction,
but its power is {\em negative}\/;
and there are very-slowly-decaying (logarithmic)
additive corrections to scaling.

3) In the presence of a {\em marginally irrelevant}\/ operator $u$,
multiplicative logarithmic corrections of either sign can
be obtained.\footnote{
   For a pedagogical discussion, see \cite[section 3.6]{LeBellac_91}.
   Concrete manifestations of this phenomenon can be found in
   the four-dimensional $N$-vector model \cite{Brezin_73,Kogon_81},
   the three-dimensional tricritical $N$-component model \cite{Wegner-Riedel},
   the three-dimensional $N$-component ferromagnet with
      strong dipolar interactions \cite{Larkin_69,Brezin_76},
   and the two-dimensional 4-state Potts ferromagnet
      \cite{Nauenberg_80,Cardy_80,Salas-Sokal_potts4}.
}
Suppose, for example, that the flow equations are
\begin{subeqnarray}
   {dt \over dl}   & = &   \gamma(u) t    \\[2mm]
   {du \over dl}   & = &   -\beta(u)
\end{subeqnarray}
Then the solution is given by the implicit equation
\be
   l  \;=\;  \int_{u(l)}^{u(0)} {du' \over \beta(u')}
\ee
together with
\be
   t(l)   \;=\;  t(0) \exp\!\left( \int_{u(l)}^{u(0)}
                                   {\gamma(u') \over \beta(u')} \, du'
                            \right)
   \;.
\ee
If we now assume that
\begin{subeqnarray}
   \gamma(u)  & = &  \lambda \,+\, \gamma_1 u \,+\, \gamma_2 u^2 \,+\,
                                      \gamma_3 u^3 \,+\,  \ldots   \\[2mm]
   \beta(u)   & = & \hphantom{\lambda \,+\, \gamma_1 u \,+\,} \;
                       \beta_2 u^2 \,+\, \beta_3 u^3 \,+\,  \ldots
\end{subeqnarray}
with $\lambda > 0$ and $\beta_2 > 0$,
and assume further that $u(0) > 0$ is small enough so that
$\beta(u) > 0$ for $0 < u < u(0)$,
we then find that
\begin{subeqnarray}
   t(l)  & = &  {\rm const} \times t(0) \,
                e^{\lambda l} \, l^{\gamma_1/\beta_2}
             \left[ 1 \,+\, {\beta_3 \gamma_1 \over \beta_2^2} {\log l \over l}
                      \,+\, O\biggl( {1 \over l} \biggr)  \right]
      \\[3mm]
   u(l)  & = &  {1 \over \beta_2 l}
                \left[ 1 \,-\, {\beta_3 \over \beta_2^2} {\log l \over l}
                         \,+\, O\biggl( {1 \over l} \biggr)  \right]
\end{subeqnarray}
Setting now $t(l) = 1$ and $e^l = \xi/\xi_1$ as before, we obtain
\be
   {\xi \over \xi_1}   \;=\;  t^{-1/\lambda} \,
       |\log t|^{-\gamma_1/\beta_2\lambda}
       \left[ 1 \,+\, {\beta_3 \gamma_1 \over \beta_2^3 \lambda}
                      {\log |\log t| \over |\log t|}
                \,+\, O\biggl( {1 \over |\log t|} \biggr)  \right]
   \;.
\ee
The exponent of the multiplicative logarithmic correction,
$-\gamma_1/\beta_2\lambda$, can thus be of either sign.
Note, however, the presence of very-slowly-decaying
additive corrections to scaling of the form
$O(\log\log t/\log t)$ [with a universal coefficient]
and $O(1/\log t)$ [with a nonuniversal coefficient].

Thus, we have been unable to devise any renormalization-group scenario
in which the correlation length acquires a
multiplicative logarithmic correction with exponent $p > 0$
in the absence of marginal operators;
and we suspect that no such scenario exists.
On the other hand, if marginal operators were present in our model
(as in scenario \#3), then one of their effects would be to induce
$1/\log L$ corrections to scaling in the
finite-size-scaling functions\footnote{
   See \cite[Section 3]{Salas-Sokal_potts4} for a detailed theoretical study
   of the $1/\log L$ corrections in the finite-size-scaling functions
   when a marginally irrelevant operator is present;
   and see \cite[Figure 6]{Salas-Sokal_potts4} for an illustration of their
   practical effect in the two-dimensional 4-state Potts ferromagnet.
},
and we see no evidence of corrections decaying anywhere near so slowly.
On the contrary, the corrections to scaling are here almost undetectable,
and they appear to decay like $L^{-\Delta}$ with $\Delta \gtapprox 1$
(see Section~\ref{sec4.1.1}).

Finally, an anonymous referee has pointed out to us
the possibility that the important microscopic variable is not
$t = e^{-\beta}$ but rather $t = 1/\beta$,
and that this $t$ is a {\em marginally relevant}\/ operator:
\be
   {dt \over dl}   \;=\;
   B(t)  \;=\;  b_2 t^2 \,+\, b_3 t^3 \,+\, b_4 t^4 \,+\, \ldots
 \label{RG_referee}
\ee
with $b_2 > 0$, i.e.\ a situation of asymptotic freedom.
(This sounds implausible at first sight for a discrete-spin model,
 but is not impossible.)
The solution to \reff{RG_referee} is given by
\be
   l  \;=\;  \int_{t(0)}^{t(l)} {dt' \over B(t')}
   \;.
\ee
If we set $t(l) = 1$ and $t(0) = 1/\beta$
and use $e^l = \xi/\xi_1$ as before, we obtain
\be
   \xi   \;=\; A \, e^{\beta/b_2} \, \beta^{-b_3/b_2^2}
               \left[ 1 \,+\, O\biggl( {1 \over \beta} \biggr)  \right]
 \label{xi_referee}
\ee
where $A$ is a nonperturbative constant,
and the corrections in inverse powers of $\beta$
can be computed from the coefficients $b_4, b_5, \ldots\;$.
On the other hand, the corrections to finite-size scaling
are determined (as usual) by irrelevant operators,
and so decay as inverse powers of $L$
(provided there are no marginally irrelevant operators).
So this scenario, unlike scenario \#3, is not ruled out by our
finite-size-scaling data (Section~\ref{sec4.1.1}).
However, if \reff{xi_referee} is indeed the true behavior,
then our data are so far from asymptotic that they
give no useful information about $A$ and $p \equiv -b_3/b_2^2$
(see Section~\ref{sec4.2.1} and in particular Figure~\ref{fig_referee}).
So we cannot rule out this scenario,
but neither can we obtain any evidence in its favor.

In summary, we really do not understand the theoretical basis
for an asymptotic behavior of the form \reff{eq7.1} with $p > 0$.

An alternative possibility is that the true asymptotic behavior is
$\xi(\beta) \sim e^{2\beta}$ {\em without}\/ multiplicative logarithmic
corrections.  This could happen in either of two ways:

a)  {\em The rise seen in Figure~\ref{fig1}a at $\beta \gtapprox 3.4$
might be spurious}\/, i.e.\ an artifact of some undetected systematic error
in our extrapolation method.  Indeed, it is suspicious that this rise
begins at roughly the same correlation length ($\xi \approx 75$)
where our extrapolation method begins to play a central role.\footnote{
   From Figure~\ref{fig:xi_FSSplot} we see that the finite-size corrections
   are negligible when $\xi/L < 1/6$.
   Inspection of Table~\ref{table:q=3} shows that
   we have raw data satisfying $\xi/L < 1/6$ up to $\beta = 3.50$
   ($\xi \approx 94$), but not beyond that.
}
To test whether this rise is real,
we produced a traditional finite-size-scaling plot in which
$\xi/e^{2 \beta}$ is plotted versus $\xi/L$ (Figure~\ref{fig_trad_FSS}).
One sees clearly that the points do {\em not}\/ fall on a single curve,
and that it is the {\em largest}\/ lattices that deviate the most
from the others
(see particularly the range $0.25 \ltapprox \xi/L \ltapprox 0.4$);
the data are most definitely {\em not}\/ compatible with convergence
to a limiting FSS curve.
Rather, the deviations reflect precisely the rise of $\xi/e^{2 \beta}$
when $\xi \gtapprox 100$.
We therefore think that the rise observed in Figure~\ref{fig1}a is real.

b)  {\em The rise seen in Figure~\ref{fig1}a might level off at some
 $\beta > 6$.}\/
This is perfectly possible, but it would mean that the corrections
to the leading asymptotic behavior either have unusually strong amplitude
or else decay more slowly than the hypothesized $e^{-\beta}$.
If the corrections behave as in \reff{eq7.1},
the coefficient $a_1$ would have to be approximately $-15$
in order to obtain a decent fit between
$\beta=4.5$ ($e^{-\beta} \approx 0.011$)
and $\beta = 6.0$ ($e^{-\beta} \approx 0.002$),
and the coefficients $a_2$ and $a_3$ would have to be large as well
(see Figure~\ref{fig_xi_p=0_Delta}a).
On the other hand, if we allow additive corrections to scaling that
are {\em fractional}\/ powers of $e^{-\beta}$,
\be
   \xi(\beta)   \;=\;
   A e^{2\beta}
   \left[ 1 + a_1 e^{-\lambda_1\beta} + a_2 e^{-\lambda_2\beta} + \ldots \right]
   \;,
 \label{eq7.delta}
\ee
then the data for $\beta \ge 4.5$ can be fit reasonably well
with $\lambda_1 \approx 0.5$ and $a_1 \approx -2.4$
(see Figure~\ref{fig_xi_p=0_Delta}b).
Now such nonanalytic corrections to scaling can arise routinely
from irrelevant operators:  in the case at hand,
an additive correction $e^{-\lambda_1\beta} \sim \xi^{-\lambda_1/2}$
corresponds to a correction-to-scaling exponent
$\Delta = \lambda_1/2 \approx 1/4$.
Unfortunately, our study of the finite-size-scaling function
(Section \ref{sec4.1.1})
gives no indication of any correction to scaling with $\Delta \ltapprox 1$.
So it is highly unlikely that a correction $e^{-\lambda_1\beta}$
with $\lambda_1 \approx 0.5$ could arise from this mechanism.
We do not know whether some other mechanism might lead to such behavior.

In conclusion, two distinct Ans\"atze on the large-$\beta$ asymptotic behavior
of the correlation length ---
\reff{eq7.1} with $p \approx 1$,
and \reff{eq7.delta} with $\lambda_1 \approx 0.5$ ---
are compatible with our data,
but both present difficulties of theoretical interpretation.
We hope that someone will be able to sort this out,
and that the numerical results presented here will serve as useful clues
toward the exact solution of this model,
possibly with the help of the methods of conformal field theory
\cite{Itzykson_collection,DiFrancesco_97}.
Our estimate of the universal quantity $x^\star \approx 0.633888$
(see also \cite{Salas-Sokal_98})
provides another constraint
in determining the universality class of this model.

\subsection{Prospects for Future Work}   \label{sec7.2}

Here are some possible directions for
future work on the 3-state square-lattice Potts antiferromagnet:

1)  Study small lattices (e.g.\ $L=4,8,16,32,64$) with very high statistics
in order to obtain quantitative information on the corrections to scaling.
Salas and Sokal \cite{Salas-Sokal_98} have recently done this
at $\beta=\infty$ (corresponding to $\xi(L)/L = x^\star \approx 0.634$),
but the corrections to scaling at this value of $\xi(L)/L$ are quite weak.
As can be seen in
Figures~\ref{fig:xi_DeviazoniCurva}(b) and \ref{fig:chi_DeviazoniCurva}(b),
the corrections are much stronger in the interval
$0.54 \ltapprox \xi(L)/L \ltapprox 0.62$,
so it would be useful to get higher statistics in this region.

2)  Study large lattices (e.g.\ $L=1024,1536,2048$) at $\xi/L < 1/6$
(where the finite-size effects are negligible,
 see Figure~\ref{fig:xi_FSSplot})
in order to verify that the observed rise of $\xi(\beta)/e^{2\beta}$
at $\beta \gtapprox 3.4$ ($\xi \gtapprox 75$)
is real and not merely an artifact of our extrapolation method.
Using $L=2048$, we could hope to reach
$\beta \approx 4.1$ ($\xi \approx 320$)
with ``essentially infinite-volume'' simulations.
By this time the rise of $\xi(\beta)/e^{2\beta}$ is about 4\%
(see Figure~\ref{fig1}a) and so should be clearly detectable.

Future studies of this model should also correct two defects
in the present work:

3)  We measured here the susceptibility and correlation length associated
to the {\em staggered}\/ magnetization ${\cal M}_{\stagg}$,
which is the most relevant operator in this model ($\eta_{\stagg} = 1/3$)
and hence has the most strongly divergent susceptibility
($\gamma_{\stagg}/\nu = 2 - \eta_{\stagg} = 5/3$).
What we failed to notice (until the runs had already been made
and it was therefore too late!)\ is that the {\em uniform}\/ magnetization
${\cal M}_{u}$ is also a relevant operator
($\eta_{u} = 4/3$ \cite{Nijs_82,Burton_Henley_97,Salas-Sokal_98})
with a divergent susceptibility ($\gamma_u/\nu = 2/3$).
Future studies should measure it as well.
Finally, the staggered polarization ${\cal P}_{\stagg}$
(see \cite{Burton_Henley_97,Salas-Sokal_98} for the definition)
is also a relevant operator ($\eta_{{\bf P}_{stagg}} = 3$)
albeit with a non-divergent susceptibility
($\gamma_{{\bf P}_{stagg}}/\nu = -1$).
By measuring its correlation function at several different low momenta,
it ought to be possible to check the prediction $\eta_{{\bf P}_{stagg}} = 3$
despite the non-divergence of the corresponding susceptibility.

4) In the present paper we didn't bother to measure the cross-correlations
between $\scrm_{\stagg}^2$ and $\scrf_{\stagg}$;
as a result, we were unable to assign statistically valid error bars to $\xi$
(instead we used the triangle inequality to obtain an overly conservative
error bar).  This was a serious mistake, as it prevented us from
distinguishing clearly between statistical fluctuations and
systematic errors (arising from corrections to scaling or other sources):
see Section~\ref{sec4.1.1}.  Measuring the cross-correlations would
allow one to determine the correct error bars not only on $\xi$
but also on combinations such as $\chi/\xi^{5/3}$;
this could lead to a very sensitive test of the conjecture that
$\chi/\xi^{5/3} \to {\rm const}$ as $\beta \to \infty$
(Section~\ref{sec4.2.2}).


\appendix
\section{Proof of a Correlation Inequality for Antiferromagnetic Potts
   Models on a Bipartite Graph}   \label{appA}

Let $V$ be a finite set of sites;
we shall consider a $q$-state Potts model on $V$
consisting of spins $\sigma_i \in \{1,\ldots,q\}$ for $i \in V$,
interacting via a Hamiltonian
\be
   \scrh   \;=\;   - \sum_{\< ij \>}  J_{ij}  \delta_{\sigma_i, \sigma_j}
   \;.
 \label{ham_potts}
\ee
Here the sum runs over all pairs $i,j \in V$ (each pair counted once),
and $\{ J_{ij} \} _{i,j \in V}$ is some specified set of couplings.

In this appendix we shall prove some correlation inequalities
for the Potts model \reff{ham_potts}.
Our technique will be to embed a field of Ising spins
into the given Potts model --- using, in fact, precisely the
Wang-Swendsen-Koteck\'y (WSK) embedding discussed in Section \ref{sec2.3} ---
and to exploit the well-known Griffiths inequalities
\cite{Sylvester_76,Szasz_78} for the induced Ising model.
It is amusing that the WSK embedding can be used
both as a Monte Carlo algorithm and as an analytical technique.
Similar proofs of correlation inequalities based on the embedding
of Ising or $XY$ variables can be found in
\cite{Mack_79,Frohlich_79,Patrascioiu_92a,Aizenman_94,CEPS_RPN_static}.

We begin by introducing a redundant (but perfectly legitimate)
parametrization of the $q$-state Potts spin $\sigma_i$
in terms of a $q$-state Potts spin $\omega_i$
and an Ising spin $\tau_i$.
Let $\omega_i$ be uniformly distributed in $\{ 1,\ldots,q \}$,
and let $\tau_i$ be uniformly distributed in $\{ -1, +1 \}$;
we then define
\be
   \sigma_i   \;=\;
   \cases{   {\displaystyle 3+\tau_i \over \displaystyle 2}
                                   &  if $\omega_i = 1$ or 2  \cr
             \noalign{\vskip 2mm}
             \omega_i              &  if $\omega_i \ge 3$ \cr
         }
 \label{def_sigma}
\ee
It is easy to see that $\sigma_i$ is uniformly distributed
in $\{ 1,\ldots,q \}$:
indeed, each of the $q$ possible values of $\sigma_i$
arises from exactly two of the $2q$ possible values of $(\omega_i,\tau_i)$.
Performing this construction independently at each site $i \in V$,
we construct the {\em a priori}\/ measure for the spins
$\{ \omega_i, \tau_i \} _{i \in V}$,
and thus also for the spins $\sigma_i$ defined by \reff{def_sigma}
as functions of $(\omega_i,\tau_i)$.
The desired probability distribution is then obtained by multiplying this
{\em a priori}\/ measure by $Z^{-1} \exp[-\scrh(\{\sigma\})]$,
where $\scrh$ is given by \reff{ham_potts}
and $\{\sigma\}$ is defined by \reff{def_sigma}.

Let us first compute the probability distribution
of the set of spins $\{\tau\}$
{\em conditioned on the set of spins $\{\sigma\}$}\/.
Here the Boltzmann weight factor $Z^{-1} \exp[-\scrh(\{\sigma\})]$
plays no role, since the $\{\sigma\}$ are being held fixed;
the conditional distribution is the same as in the {\em a priori}\/ measure.
We thus have, independently for each site $i$,
\be
   \tau_i   \;=\;
   \cases{ 2\sigma_i - 3                      & if $\sigma_i = 1$ or 2   \cr
           \pm 1 \hbox{ with equal probability}   & if $\sigma_i \ge 3$  \cr
         }
\ee
It follows that the conditional expectation of $\tau_i \tau_j$
given $\{\sigma\}$ is
\begin{subeqnarray}
   E(\tau_i \tau_j | \{\sigma\})   & = &
      I(\sigma_i \le 2) \, I(\sigma_j \le 2) \,
         (2 \delta_{\sigma_i, \sigma_j} - 1)        \\[2mm]
   & = &
   \cases{ +1  & if $\sigma_i=\sigma_j=1$ or $\sigma_i=\sigma_j=2$ \cr
           -1  & if $\sigma_i=1,\,\sigma_j=2$ or $\sigma_i=2,\,\sigma_j=1$ \cr
           0   & in all other cases
         }
  \label{tau_cond}
\end{subeqnarray}
where $I(\cdots)$ denotes the indicator function of the specified event.
Next we want to take the unconditional expectation of \reff{tau_cond}.
Since the probability distribution of the $\{ \sigma \}$ is invariant
under global permutations of $\{ 1,\ldots,q \}$,
we have in particular that the joint probability distribution
of $\sigma_i$ and $\sigma_j$ is given by
\be
   \hbox{Prob}(\sigma_i, \sigma_j)   \;=\;
   {p \over q} \, \delta_{\sigma_i, \sigma_j}   \;+\;
   {1-p \over q(q-1)} \, (1 - \delta_{\sigma_i, \sigma_j})
 \label{dist_sigma_ij}
\ee
where $p = \< \delta_{\sigma_i, \sigma_j} \>$.
Averaging \reff{tau_cond} over the distribution \reff{dist_sigma_ij},
we find that
\be
   \< \tau_i \tau_j \>    \;=\;
   {2 (pq-1) \over q(q-1)}   \;=\;
   {2 \over q}
   \left\<    { {q \delta_{\sigma_i,\sigma_j} \,-\, 1}   \over  {q \,-\, 1} }
   \right\>
   \;.
 \label{potts_identity}
\ee
Note that on the right-hand side of this identity we have
precisely the two-point correlation function of
our Potts model [cf.\ \reff{def_G}], multiplied by $2/q$.

On the other hand let us compute the probability distribution
of the set of spins $\{\tau\}$
{\em conditioned on the set of spins $\{\omega\}$}\/.
Note first that
\be
   \delta_{\sigma_i, \sigma_j}   \;=\;
   I(\omega_i \ge 3) \, I(\omega_j \ge 3) \, \delta_{\omega_i, \omega_j}
   \;+\;
   I(\omega_i \le 2) \, I(\omega_j \le 2) \, {1 \,+\, \tau_i \tau_j \over 2}
   \;.
\ee
Inserting this into \reff{ham_potts},
we see that the model of spins $\{\tau\}$
conditioned on $\{\omega\}$
is an Ising model with interactions
\be
   J_{ij}^{\hboxscript{eff}}   \;\equiv\;
   \smhalf J_{ij} \, I(\omega_i \le 2) \, I(\omega_j \le 2)
   \;.
  \label{J_eff}
\ee
(The factor $\half$ comes simply from the difference between the
conventional Ising and Potts normalizations of couplings.)
Below, we shall apply various correlation inequalities
to this induced Ising model.

Let us consider some special cases:

\medskip

{\bf Example 1} (ferromagnetic Potts model).
Assume that $J_{ij} \ge 0$ for all $i,j \in V$.
Then the couplings $J_{ij}^{\hboxscript{eff}}$ are also $\ge 0$.
So Griffiths' first inequality \cite{Sylvester_76}
applied to the induced Ising model \reff{J_eff}
implies that
\be
   E(\tau_i \tau_j | \{\omega\})   \;\ge\;   0
   \qquad\hbox{for all }   \{\omega\}
   \;.
\ee
In particular,
averaging this over $\{\omega\}$ we deduce that $\< \tau_i \tau_j \> \ge 0$;
and hence by the identity \reff{potts_identity} we have
\be
   \left\<  { {q \delta_{\sigma_i,\sigma_j} \,-\, 1}   \over  {q \,-\, 1} }
   \right\>
   \;\ge\;
   0
   \;.
 \label{G_I_ferro}
\ee
This is a ``Griffiths' first inequality for ferromagnetic Potts models''.

Of course, this inequality can be proven somewhat more simply
using the Fortuin-Kasteleyn representation:
the left-hand side of \reff{G_I_ferro} is equal to a
connection probability in the Fortuin-Kasteleyn bond variables,
and thus is manifestly nonnegative.
Indeed, this proof is valid for all real $q \ge 0$.
So nothing much is gained by WSK embedding in this case;
the real value of the method arises in the next case:

\medskip
 
{\bf Example 2} (antiferromagnetic Potts model on a bipartite graph).
Assume that the set of sites $V$ can be partitioned as $V = A \cup B$
in such a way that
\be
   J_{ij}   \;\,\cases{  \ge 0       & if $i,j \in A$  \cr
                         \noalign{\vskip 1mm}
                         \ge 0       & if $i,j \in B$  \cr
                         \noalign{\vskip 1mm}
                         \le 0       & if $i \in A$, $j \in B$  \cr
                      }
\ee
(In particular, the pure antiferromagnet would have
 $J_{ij} = 0$ for $i,j \in A$ and for $i,j \in B$.)
Now define
\be
   \tau'_i   \;=\;  \cases{ \tau_i     & if $i \in A$  \cr
                           -\tau_i     & if $i \in B$  \cr
                          }
\ee
It follows that the model of spins $\{\tau'\}$
conditioned on $\{\omega\}$
is an Ising model with interactions
\be
   J_{ij}^{\prime\hboxscript{eff}}   \;\equiv\;
   \smhalf J_{ij} \, I(\omega_i \le 2) \, I(\omega_j \le 2)
   \,\times\,
   \left\{\! \begin{array}{ll}
               +1    & \hbox{if } i,j \in A  \\
               +1    & \hbox{if } i,j \in B  \\
               -1    & \hbox{if } i \in A, \, j \in B
           \end{array}
   \!\right\}
   \;\ge\;  0
   \;.
\ee
So Griffiths' first inequality \cite{Sylvester_76}
applied to this ferromagnetic Ising model implies that
\be
   E(\tau'_i \tau'_j | \{\omega\})   \;\ge\;   0
   \qquad\hbox{for all }   \{\omega\}
   \;.
\ee
Averaging over $\{\omega\}$ we deduce that $\< \tau'_i \tau'_j \> \ge 0$,
and hence by the identity \reff{potts_identity} that
\be
   \left\<  { {q \delta_{\sigma_i,\sigma_j} \,-\, 1}   \over  {q \,-\, 1} }
   \right\>
   \;=\;
   {q \over 2} \, \< \tau_i \tau_j \>
   \;\,\cases{ \ge 0       & if $i,j \in A$  \cr
               \noalign{\vskip 1mm}
               \ge 0       & if $i,j \in B$  \cr
               \noalign{\vskip 1mm}
               \le 0       & if $i \in A$, $j \in B$  \cr
             }
 \label{G_I_antiferro}
\ee
This is a kind of
``Griffiths' first inequality for antiferromagnetic Potts models''.

In particular, for antiferromagnetic Potts models on the square lattice,
\reff{G_I_antiferro} implies that the 
two-point correlation function $G(x,y)$ has sign $(-1)^{|x-y|}$,
where $|x-y|$ denotes the $\ell^1$ norm.
One consequence of this is that the Fourier-transformed
two-point function $\widetilde{G}(p)$ satisfies
\be
   |\widetilde{G}(p)|   \;\le\;   \widetilde{G}\bigl( (\pi,\pi) \bigr)
   \;.
\ee

\medskip 

{\bf Example 3} (comparison-to-Ising inequality).
Let us return to the general case of arbitrary $\{J_{ij}\}$.
From \reff{J_eff} it follows immediately that
\be
   |J_{ij}^{\hboxscript{eff}}|   \;\le\;
   \smhalf |J_{ij}|
   \;,
\ee
so by Griffiths' comparison inequality \cite{Szasz_78}
we have
\be
   \bigl| E(\tau_i \tau_j | \{\omega\}) \bigr|   \;\le\;
   \< \epsilon_i \epsilon_j \> _{Ising, \{ |J/2| \}}
   \qquad\hbox{for all }   \{\omega\}
   \;,
\ee
where $\< \,\cdots\, \> _{Ising, \{ |J/2| \}}$ denotes the
expectation in an Ising model with couplings $\{ \half |J_{ij}| \}$.
Averaging over $\{\omega\}$ and using \reff{potts_identity},
we conclude that
\be
   \left\<  { {q \delta_{\sigma_i,\sigma_j} \,-\, 1}   \over  {q \,-\, 1} }
   \right\>
   \;\le\;
   {q \over 2} \, \< \epsilon_i \epsilon_j \> _{Ising, \{ |J/2| \}}
   \;.
 \label{upper_bound}
\ee
Thus, the two-point function in an arbitrary Potts model
can be bounded {\em above}\/ by the corresponding two-point function
in a {\em ferromagnetic Ising}\/ model.\footnote{
   Of course, for a {\em ferromagnetic}\/ Potts model
   a much stronger result is true,
   namely that each correlation function $G(x,y;q)$
   is a {\em decreasing}\/ function of $q$ (for $q \ge 1$)
   at fixed $\{J_{ij}\}$.
   This follows from the Fortuin-Kasteleyn representation
   combined with the FKG inequality for the random-cluster model
   \cite{Aizenman-Chayes-Chayes-Newman,Fortuin_72}.
   So the real interest of \reff{upper_bound}
   is for the {\em antiferromagnetic}\/ or
   {\em mixed ferromagnetic/antiferromagnetic}\/ Potts models.
}

For example, one consequence of \reff{upper_bound} is that
the transition temperature of the $q=3$ Potts antiferromagnet
on the triangular lattice must satisfy
\be
   \beta_{trans}(\hbox{TRI $q=3$ Potts AF})
   \;\ge\;
   2 \beta_{crit}(\hbox{TRI Ising ferro})
   \;=\;
   \half \log 3 \;=\; 0.549306\ldots \;.
\ee
This bound is of course satisfied by the numerical estimate
$\beta_{trans}(\hbox{TRI $q=3$ Potts AF}) \approx 1.594$ \cite{Adler_95}.
Likewise, for the $q=3,4$ Potts antiferromagnets on the
simple-cubic lattice, one must have
\be
   \beta_{trans}(\hbox{SC $q=3,4$ Potts AF})
   \;\ge\;
   2 \beta_{crit}(\hbox{SC Ising ferro})
   \;,
\ee
which is satisfied by the numerical estimates
$\beta_{trans}(\hbox{SC $q=3$ Potts AF}) \approx 0.816$
 \cite{WSK_90,Gottlob_94a,Gottlob_94b,Kolesik_95,Heilmann_96},
$\beta_{trans}(\hbox{SC $q=4$ Potts AF}) \approx 1.43$ \cite{Ueno_89},
and
$\beta_{crit}(\hbox{SC Ising ferro}) \approx 0.222$ \cite{Talapov_96}.

\section{Proof of Ergodicity of the WSK Algorithm at $T=0$
   on a Bipartite Graph}    \label{appB}

Let $G=(V,E)$ be a finite undirected graph
with vertex set $V$ and edge set $E$.
Then $G$ is said to be {\em bipartite}\/ if the vertex set $V$
can be partitioned as $V = A \cup B$ in such a way that
every edge $e \in E$ has one endpoint in $A$ and
the other endpoint in $B$ (i.e.\ there are no $A$--$A$ or $B$--$B$ edges).

{\bf Example 1.}   Any finite subset of the simple (hyper-)cubic lattice
$\zed^d$, with {\em free}\/ boundary conditions, defines a bipartite graph.

{\bf Example 2.}   A box of size $L_1 \times L_2 \times \ldots \times L_d$
in $\zed^d$, with {\em periodic}\/ boundary conditions,
defines a bipartite graph if and only if all the side lengths
$L_1, L_2, \ldots, L_d$ are {\em even}\/.

The goal of this Appendix is to prove the following theorem:

\begin{theorem}
Let $G$ be a bipartite finite undirected graph,
and let $q$ be an integer $\ge 2$.
Then the Wang-Swendsen-Koteck\'y algorithm
for the $q$-state Potts antiferromagnet on $G$ at zero temperature
(i.e.\ for $q$-colorings of $G$) is ergodic.
\end{theorem}

\proof
We will prove, by induction on $q$, that the WSK algorithm for
$q$-colorings (hereafter called WSK--$q$)
is ergodic on $G$ {\em and on all its subgraphs}\/.

\medskip

{\em Case $q=2$.}\/
Since $G$ is bipartite, so are all its subgraphs $H$.
The WSK algorithm acts independently on each connected component of $H$,
so it suffices to prove the ergodicity for each connected component.
But a connected bipartite graph has precisely two 2-colorings,
which are related by a global interchange of the two colors;
and this global interchange can trivially be realized by a WSK move.

\medskip

{\em Inductive step.}\/
Let $q \ge 3$,
and suppose that WSK--$(q-1)$ is ergodic on $G$ and all its subgraphs;
we shall prove the same for WSK--$q$.
So let $H = (V',E')$ be a subgraph of $G$,
and define $A' = A \cap V'$, $B' = B \cap V'$.
Define the {\em reference configuration}\/ to be the
$q$-coloring of $H$ in which all sites in $A'$ are colored 1
and all sites in $B'$ are colored 2.
Let $\{ \sigma_x \}_{x \in V'}$ be an arbitrary $q$-coloring of $H$,
which we call the {\em target configuration}\/.
We shall show that the reference configuration can be
transformed into the target configuration by a finite sequence
of WSK--$q$ moves.
(This is sufficient to prove ergodicity,
 since the inverse of a WSK move is also a WSK move.)

(a) {\em Step 1.}\/
Choose $1,q$ as the pair of colors to be used in the WSK move,
and focus attention on all sites $x \in A'$ such that the target
configuration has $\sigma_x = q$.  All these sites
are of course currently colored 1, and all their neighbors (which are in $B'$)
are currently colored 2.  So each of these sites
is a singleton (i.e.\ a one-site connected component)
in the subgraph of $H$ formed by those sites currently colored 1 or $q$.
Therefore, each of these sites can be recolored $q$ by a WSK move,
while leaving all other sites unchanged.

(b) {\em Step 2.}\/ 
Choose $2,q$ as the pair of colors to be used in the WSK move, 
and focus attention on all sites $x \in B'$ such that the target
configuration has $\sigma_x = q$.  All these sites 
are of course currently colored 2;
and all their neighbors
(which are in $A'$ {\em and have target color $\neq q$}\/)
are currently colored 1.  So each of these sites
is a singleton
in the subgraph of $H$ formed by those sites currently colored 2 or $q$.
Therefore, each of these sites can be recolored $q$ by a WSK move,
while leaving all other sites unchanged.

(c) {\em Step 3.}\/
All sites with target color $q$ are now colored $q$,
while the remaining sites are now colored 1 or 2.
The latter sites (along with their corresponding edges)
define a subgraph $K \subset H$;
and by the inductive hypothesis they can be given their target colors
(which lie in $\{1,2,\ldots,q-1\}$)
by a sequence of WSK--$(q-1)$ moves
(which are of course also WSK--$q$ moves).
\qed

{\bf Note.}  After completion of this work, we learned that the same theorem
(as well as some generalizations of it) was obtained independently by
Burton and Henley \cite{Burton_Henley_97}.  Their proof is essentially
the same as ours.

\section*{Acknowledgments}

We wish to thank John Cardy, Chris Henley, Gustavo Mana, Paul Pearce,
Andrea Pelissetto, Jes\'us Salas and Hubert Saleur
for helpful conversations and/or correspondence.
The authors' research was supported in part by
the Conselho Nacional de Desenvolvimento Cient\'{\i}fico e Tecnol\'ogico--CNPq,
the Financiadora de Estudos e Projetos (FINEP--MCT),
the Funda\c{c}\~ao de Amparo \`a Pesquisa do Estado de Minas Gerais (FAPEMIG),
and U.S.\ National Science Foundation grants DMS-9200719 and PHY-9520978.

%
%
%

\clearpage


%
%
%
%
\begin{table}[h]
\hspace*{-2.5cm}
\footnotesize
\begin{tabular}{|r|r||r|r||r@{ (}r@{) }|r@{ (}r@{) }|r@{ (}r@{) }||r@{ (}r@{) }|r@{ (}r@{) }|}
\hline\hline  \\[-4mm]
\multicolumn{1}{c|}{$L$} &
\multicolumn{1}{c||}{$\beta$} &
   \multicolumn{1}{c|}{Total} &
   \multicolumn{1}{c||}{Discard} &
      \multicolumn{2}{c|}{$\xi$}  &
      \multicolumn{2}{c|}{$\chi_{\stagg}$}  &
         \multicolumn{2}{c||}{$E$}  &
         \multicolumn{2}{c|}{$\tau_{int,\scrm_{\smstagg}^2}$}  &
         \multicolumn{2}{c|}{$\tau_{int,\scre}$}    \\[1mm]
\hline
32  &  2.00  &  200000  &  10000  & 
    5.572  &  0.032  &  49.02  &  0.22  & 
    0.07027100  &  0.00005919  &  2.53  &  0.05  &  3.16  &  0.06   \\
32  &  2.25  &  200000  &  10000  & 
    8.222  &  0.036  &  89.75  &  0.38  & 
    0.04771217  &  0.00005094  &  3.39  &  0.07  &  3.35  &  0.07   \\
32  &  2.50  &  6000000  &  60000  & 
    11.594  &  0.008  &  145.77  &  0.09  & 
    0.03207236  &  0.00000761  &  4.17  &  0.02  &  3.49  &  0.01   \\
32  &  2.70  &  6000000  &  60000  & 
    14.116  &  0.009  &  186.45  &  0.09  & 
    0.02354641  &  0.00000625  &  4.31  &  0.02  &  3.36  &  0.01   \\
32  &  3.00  &  6000000  &  60000  & 
    16.834  &  0.009  &  227.51  &  0.08  & 
    0.01529180  &  0.00000460  &  4.19  &  0.02  &  3.06  &  0.01   \\
32  &  3.20  &  6000000  &  60000  & 
    17.956  &  0.009  &  243.94  &  0.07  & 
    0.01171553  &  0.00000381  &  4.14  &  0.02  &  2.91  &  0.01   \\
32  &  3.50  &  22000000  &  220000  & 
    19.015  &  0.005  &  259.29  &  0.03  & 
    0.00804395  &  0.00000155  &  4.12  &  0.01  &  2.76  &  0.01   \\
32  &  4.00  &  6000000  &  60000  & 
    19.812  &  0.009  &  271.55  &  0.06  & 
    0.00449722  &  0.00000209  &  4.22  &  0.02  &  2.63  &  0.00   \\
32  &  4.50  &  6000000  &  60000  & 
    20.130  &  0.009  &  276.87  &  0.06  & 
    0.00259774  &  0.00000153  &  4.34  &  0.02  &  2.56  &  0.00   \\
32  &  5.00  &  6000000  &  60000  & 
    20.263  &  0.009  &  279.45  &  0.06  & 
    0.00153121  &  0.00000116  &  4.41  &  0.02  &  2.55  &  0.00   \\
\hline
64  &  2.00  &  200000  &  10000  & 
    5.579  &  0.074  &  49.43  &  0.20  & 
    0.07048701  &  0.00002977  &  1.75  &  0.03  &  3.21  &  0.07   \\
64  &  2.25  &  200000  &  10000  & 
    8.534  &  0.064  &  97.32  &  0.42  & 
    0.04814820  &  0.00002498  &  2.07  &  0.03  &  3.27  &  0.07   \\
64  &  2.50  &  2000000  &  20000  & 
    13.181  &  0.021  &  196.58  &  0.27  & 
    0.03288755  &  0.00000638  &  2.96  &  0.02  &  3.31  &  0.02   \\
64  &  2.60  &  200000  &  10000  & 
    15.628  &  0.073  &  256.47  &  1.15  & 
    0.02829046  &  0.00001911  &  3.58  &  0.08  &  3.32  &  0.07   \\
64  &  2.70  &  2000000  &  20000  & 
    18.349  &  0.024  &  327.45  &  0.42  & 
    0.02432931  &  0.00000549  &  3.86  &  0.03  &  3.37  &  0.02   \\
64  &  2.80  &  200000  &  10000  & 
    21.167  &  0.083  &  403.52  &  1.53  & 
    0.02098848  &  0.00001660  &  4.15  &  0.10  &  3.50  &  0.07   \\
64  &  2.90  &  200000  &  10000  & 
    23.956  &  0.090  &  477.13  &  1.65  & 
    0.01814113  &  0.00001506  &  4.49  &  0.11  &  3.35  &  0.07   \\
64  &  3.00  &  2000000  &  20000  & 
    26.715  &  0.029  &  549.54  &  0.51  & 
    0.01572752  &  0.00000419  &  4.46  &  0.04  &  3.23  &  0.02   \\
64  &  3.10  &  200000  &  10000  & 
    29.067  &  0.098  &  609.52  &  1.61  & 
    0.01371684  &  0.00001249  &  4.49  &  0.11  &  3.21  &  0.07   \\
64  &  3.20  &  2000000  &  20000  & 
    31.188  &  0.031  &  661.13  &  0.48  & 
    0.01197822  &  0.00000345  &  4.44  &  0.04  &  3.04  &  0.02   \\
64  &  3.30  &  200000  &  10000  & 
    32.883  &  0.098  &  702.93  &  1.46  & 
    0.01050532  &  0.00001010  &  4.31  &  0.10  &  2.94  &  0.06   \\
64  &  3.40  &  200000  &  10000  & 
    34.326  &  0.101  &  736.49  &  1.40  & 
    0.00924902  &  0.00000937  &  4.26  &  0.10  &  2.92  &  0.06   \\
64  &  3.50  &  2000000  &  20000  & 
    35.517  &  0.031  &  764.61  &  0.42  & 
    0.00816692  &  0.00000265  &  4.36  &  0.03  &  2.83  &  0.01   \\
64  &  3.60  &  200000  &  10000  & 
    36.605  &  0.102  &  789.12  &  1.29  & 
    0.00722325  &  0.00000786  &  4.21  &  0.10  &  2.77  &  0.05   \\
64  &  3.70  &  200000  &  10000  & 
    37.183  &  0.102  &  804.16  &  1.26  & 
    0.00641653  &  0.00000739  &  4.38  &  0.11  &  2.79  &  0.05   \\
64  &  3.80  &  200000  &  10000  & 
    38.082  &  0.102  &  822.54  &  1.21  & 
    0.00571353  &  0.00000668  &  4.24  &  0.10  &  2.63  &  0.05   \\
64  &  3.90  &  200000  &  10000  & 
    38.398  &  0.101  &  832.77  &  1.19  & 
    0.00508284  &  0.00000638  &  4.31  &  0.10  &  2.70  &  0.05   \\
64  &  4.00  &  2000000  &  20000  & 
    38.786  &  0.031  &  842.08  &  0.36  & 
    0.00453673  &  0.00000183  &  4.21  &  0.03  &  2.65  &  0.01   \\
64  &  4.50  &  2000000  &  20000  & 
    39.891  &  0.032  &  871.56  &  0.34  & 
    0.00261704  &  0.00000134  &  4.32  &  0.03  &  2.57  &  0.01   \\
64  &  5.00  &  2000000  &  20000  & 
    40.325  &  0.032  &  884.30  &  0.34  & 
    0.00153744  &  0.00000100  &  4.43  &  0.04  &  2.54  &  0.01   \\
\hline
128  &  2.00  &  200000  &  10000  & 
    5.442  &  0.244  &  49.37  &  0.20  & 
    0.07048002  &  0.00001449  &  1.52  &  0.02  &  3.05  &  0.06   \\
128  &  2.25  &  200000  &  10000  & 
    8.817  &  0.174  &  98.61  &  0.41  & 
    0.04817305  &  0.00001255  &  1.70  &  0.03  &  3.30  &  0.07   \\
128  &  2.50  &  200000  &  10000  & 
    13.335  &  0.136  &  201.66  &  0.84  & 
    0.03295390  &  0.00001007  &  1.81  &  0.03  &  3.20  &  0.07   \\
128  &  2.60  &  200000  &  10000  & 
    15.967  &  0.131  &  271.66  &  1.18  & 
    0.02837582  &  0.00000939  &  2.10  &  0.04  &  3.25  &  0.07   \\
128  &  2.70  &  200000  &  10000  & 
    19.267  &  0.127  &  368.40  &  1.59  & 
    0.02450218  &  0.00000868  &  2.27  &  0.04  &  3.26  &  0.07   \\
128  &  2.80  &  200000  &  10000  & 
    23.237  &  0.129  &  500.07  &  2.20  & 
    0.02118313  &  0.00000790  &  2.63  &  0.05  &  3.20  &  0.07   \\
128  &  2.90  &  200000  &  10000  & 
    27.867  &  0.138  &  672.92  &  3.01  & 
    0.01836736  &  0.00000724  &  3.11  &  0.06  &  3.14  &  0.06   \\
128  &  2.95  &  200000  &  10000  & 
    30.271  &  0.143  &  767.77  &  3.43  & 
    0.01711320  &  0.00000701  &  3.41  &  0.07  &  3.20  &  0.07   \\
128  &  3.00  &  200000  &  10000  & 
    33.203  &  0.151  &  886.35  &  3.92  & 
    0.01595897  &  0.00000679  &  3.79  &  0.08  &  3.25  &  0.07   \\
128  &  3.10  &  200000  &  10000  & 
    38.836  &  0.162  &  1121.29  &  4.57  & 
    0.01389865  &  0.00000626  &  4.04  &  0.09  &  3.21  &  0.07   \\
128  &  3.20  &  200000  &  10000  & 
    44.917  &  0.176  &  1382.17  &  5.10  & 
    0.01213496  &  0.00000569  &  4.40  &  0.11  &  3.13  &  0.06   \\
128  &  3.30  &  200000  &  10000  & 
    50.858  &  0.187  &  1630.23  &  5.32  & 
    0.01064182  &  0.00000528  &  4.58  &  0.11  &  3.12  &  0.06   \\
128  &  3.40  &  200000  &  10000  & 
    55.937  &  0.195  &  1845.96  &  5.34  & 
    0.00935202  &  0.00000477  &  4.68  &  0.12  &  2.95  &  0.06   \\
128  &  3.45  &  200000  &  10000  & 
    58.310  &  0.196  &  1935.27  &  5.23  & 
    0.00877467  &  0.00000460  &  4.66  &  0.11  &  2.95  &  0.06   \\
128  &  3.50  &  1000000  &  10000  & 
    60.607  &  0.087  &  2029.01  &  2.21  & 
    0.00824378  &  0.00000193  &  4.48  &  0.05  &  2.93  &  0.03   \\
128  &  3.60  &  200000  &  10000  & 
    64.455  &  0.202  &  2178.61  &  4.87  & 
    0.00728348  &  0.00000400  &  4.54  &  0.11  &  2.82  &  0.05   \\
128  &  3.70  &  200000  &  10000  & 
    67.788  &  0.203  &  2304.67  &  4.61  & 
    0.00646723  &  0.00000375  &  4.41  &  0.11  &  2.80  &  0.05   \\
128  &  3.80  &  200000  &  10000  & 
    70.305  &  0.207  &  2403.28  &  4.43  & 
    0.00574268  &  0.00000342  &  4.47  &  0.11  &  2.73  &  0.05   \\
128  &  3.90  &  200000  &  10000  & 
    72.160  &  0.207  &  2474.84  &  4.24  & 
    0.00511211  &  0.00000320  &  4.40  &  0.11  &  2.72  &  0.05   \\
128  &  3.95  &  200000  &  10000  & 
    73.286  &  0.201  &  2513.98  &  4.10  & 
    0.00482635  &  0.00000313  &  4.29  &  0.10  &  2.72  &  0.05   \\
128  &  4.00  &  200000  &  10000  & 
    73.922  &  0.207  &  2542.02  &  4.14  & 
    0.00455744  &  0.00000300  &  4.48  &  0.11  &  2.71  &  0.05   \\
\hline
\end{tabular}
\vspace{3mm}
\caption{
   Our Monte Carlo data for the 3-state Potts antiferromagnet.
   ``Total'' is the total number of WSK iterations performed;
   ``Discard'' is the number of iterations discarded for equilibration.
   Error bars (one standard deviation) are shown in parentheses.
   {\bf [First page of 3-page table]}
}
\label{table:q=3}
\end{table}
\clearpage \addtocounter{table}{-1}
\begin{table}[h]
\hspace*{-2.5cm}
\footnotesize
\begin{tabular}{|r|r||r|r||r@{ (}r@{) }|r@{ (}r@{) }|r@{ (}r@{) }||r@{ (}r@{) }|r@{ (}r@{) }|}
\hline\hline  \\[-4mm]
\multicolumn{1}{c|}{$L$} &
\multicolumn{1}{c||}{$\beta$} &
   \multicolumn{1}{c|}{Total} &
   \multicolumn{1}{c||}{Discard} &
      \multicolumn{2}{c|}{$\xi$}  &
      \multicolumn{2}{c|}{$\chi_{\stagg}$}  &
         \multicolumn{2}{c||}{$E$}  &
         \multicolumn{2}{c|}{$\tau_{int,\scrm_{\smstagg}^2}$}  &
         \multicolumn{2}{c|}{$\tau_{int,\scre}$}    \\[1mm]
\hline
128  &  4.10  &  200000  &  10000  & 
    75.419  &  0.204  &  2594.69  &  3.94  & 
    0.00406992  &  0.00000280  &  4.32  &  0.10  &  2.69  &  0.05   \\
128  &  4.20  &  200000  &  10000  & 
    76.508  &  0.205  &  2636.64  &  3.92  & 
    0.00363639  &  0.00000262  &  4.52  &  0.11  &  2.64  &  0.05   \\
128  &  4.30  &  1000000  &  10000  & 
    77.427  &  0.090  &  2670.89  &  1.64  & 
    0.00325960  &  0.00000107  &  4.29  &  0.04  &  2.62  &  0.02   \\
128  &  4.40  &  1000000  &  10000  & 
    78.050  &  0.089  &  2697.29  &  1.61  & 
    0.00291971  &  0.00000101  &  4.27  &  0.04  &  2.61  &  0.02   \\
128  &  4.50  &  1000000  &  10000  & 
    78.772  &  0.090  &  2723.43  &  1.59  & 
    0.00262197  &  0.00000095  &  4.32  &  0.04  &  2.59  &  0.02   \\
128  &  4.60  &  1000000  &  10000  & 
    79.136  &  0.090  &  2742.07  &  1.59  & 
    0.00235324  &  0.00000090  &  4.37  &  0.05  &  2.59  &  0.02   \\
128  &  4.70  &  1000000  &  10000  & 
    79.505  &  0.091  &  2756.45  &  1.58  & 
    0.00211402  &  0.00000084  &  4.41  &  0.05  &  2.56  &  0.02   \\
128  &  4.80  &  1000000  &  10000  & 
    79.952  &  0.090  &  2771.60  &  1.56  & 
    0.00190149  &  0.00000080  &  4.40  &  0.05  &  2.56  &  0.02   \\
128  &  4.90  &  1000000  &  10000  & 
    80.091  &  0.091  &  2781.79  &  1.56  & 
    0.00171139  &  0.00000075  &  4.42  &  0.05  &  2.56  &  0.02   \\
128  &  5.00  &  1000000  &  10000  & 
    80.174  &  0.091  &  2788.86  &  1.56  & 
    0.00154123  &  0.00000071  &  4.42  &  0.05  &  2.53  &  0.02   \\
128  &  5.10  &  1000000  &  10000  & 
    80.443  &  0.090  &  2797.74  &  1.54  & 
    0.00138704  &  0.00000067  &  4.37  &  0.05  &  2.52  &  0.02   \\
128  &  5.20  &  1000000  &  10000  & 
    80.612  &  0.091  &  2805.76  &  1.54  & 
    0.00125012  &  0.00000064  &  4.40  &  0.05  &  2.55  &  0.02   \\
\hline
256  &  2.50  &  200000  &  10000  & 
    13.104  &  0.422  &  201.34  &  0.81  & 
    0.03295828  &  0.00000517  &  1.57  &  0.02  &  3.34  &  0.07   \\
256  &  2.60  &  200000  &  10000  & 
    16.185  &  0.356  &  273.13  &  1.08  & 
    0.02839657  &  0.00000461  &  1.56  &  0.02  &  3.18  &  0.07   \\
256  &  2.70  &  200000  &  10000  & 
    19.126  &  0.325  &  369.05  &  1.51  & 
    0.02450313  &  0.00000423  &  1.69  &  0.03  &  3.15  &  0.06   \\
256  &  2.80  &  200000  &  10000  & 
    23.045  &  0.293  &  502.20  &  2.08  & 
    0.02119791  &  0.00000393  &  1.76  &  0.03  &  3.16  &  0.06   \\
256  &  2.90  &  200000  &  10000  & 
    28.119  &  0.269  &  693.32  &  2.93  & 
    0.01837763  &  0.00000362  &  1.90  &  0.03  &  3.16  &  0.06   \\
256  &  3.00  &  200000  &  10000  & 
    34.305  &  0.258  &  954.62  &  4.09  & 
    0.01599052  &  0.00000333  &  2.09  &  0.04  &  3.13  &  0.06   \\
256  &  3.10  &  200000  &  10000  & 
    41.701  &  0.260  &  1313.21  &  5.88  & 
    0.01394816  &  0.00000305  &  2.48  &  0.04  &  3.10  &  0.06   \\
256  &  3.20  &  200000  &  10000  & 
    50.453  &  0.269  &  1803.02  &  8.21  & 
    0.01219458  &  0.00000281  &  2.96  &  0.06  &  3.05  &  0.06   \\
256  &  3.30  &  200000  &  10000  & 
    60.839  &  0.286  &  2439.41  &  10.78  & 
    0.01069957  &  0.00000263  &  3.35  &  0.07  &  3.08  &  0.06   \\
256  &  3.40  &  200000  &  10000  & 
    72.254  &  0.316  &  3198.35  &  13.71  & 
    0.00940398  &  0.00000238  &  3.99  &  0.09  &  2.94  &  0.06   \\
256  &  3.50  &  1000000  &  10000  & 
    84.685  &  0.150  &  4036.07  &  6.91  & 
    0.00829150  &  0.00000096  &  4.38  &  0.05  &  2.92  &  0.03   \\
256  &  3.60  &  200000  &  10000  & 
    96.720  &  0.365  &  4854.04  &  16.67  & 
    0.00732740  &  0.00000203  &  4.48  &  0.11  &  2.85  &  0.06   \\
256  &  3.70  &  200000  &  10000  & 
    107.817  &  0.380  &  5595.86  &  16.87  & 
    0.00648727  &  0.00000188  &  4.59  &  0.11  &  2.82  &  0.05   \\
256  &  3.80  &  200000  &  10000  & 
    117.698  &  0.392  &  6228.98  &  16.33  & 
    0.00576244  &  0.00000175  &  4.50  &  0.11  &  2.81  &  0.05   \\
256  &  3.90  &  200000  &  10000  & 
    125.908  &  0.414  &  6756.45  &  16.21  & 
    0.00512698  &  0.00000163  &  4.77  &  0.12  &  2.78  &  0.05   \\
256  &  4.00  &  200000  &  10000  & 
    132.919  &  0.413  &  7173.99  &  15.25  & 
    0.00456912  &  0.00000151  &  4.62  &  0.11  &  2.72  &  0.05   \\
256  &  4.10  &  200000  &  10000  & 
    138.468  &  0.409  &  7524.03  &  14.19  & 
    0.00407866  &  0.00000140  &  4.40  &  0.11  &  2.64  &  0.05   \\
256  &  4.20  &  200000  &  10000  & 
    143.448  &  0.409  &  7812.70  &  13.50  & 
    0.00364854  &  0.00000133  &  4.30  &  0.10  &  2.70  &  0.05   \\
256  &  4.30  &  200000  &  10000  & 
    146.776  &  0.404  &  8024.93  &  12.99  & 
    0.00326371  &  0.00000124  &  4.29  &  0.10  &  2.66  &  0.05   \\
256  &  4.40  &  200000  &  10000  & 
    150.305  &  0.412  &  8207.19  &  12.71  & 
    0.00292610  &  0.00000116  &  4.36  &  0.10  &  2.58  &  0.05   \\
256  &  4.50  &  1000000  &  10000  & 
    152.769  &  0.180  &  8361.37  &  5.39  & 
    0.00262364  &  0.00000048  &  4.33  &  0.04  &  2.62  &  0.02   \\
256  &  4.60  &  1000000  &  10000  & 
    154.560  &  0.181  &  8472.30  &  5.32  & 
    0.00235575  &  0.00000045  &  4.39  &  0.05  &  2.56  &  0.02   \\
256  &  4.65  &  1000000  &  10000  & 
    154.914  &  0.181  &  8512.61  &  5.26  & 
    0.00223281  &  0.00000043  &  4.36  &  0.05  &  2.57  &  0.02   \\
256  &  4.70  &  1000000  &  10000  & 
    155.773  &  0.179  &  8560.34  &  5.21  & 
    0.00211659  &  0.00000042  &  4.35  &  0.05  &  2.57  &  0.02   \\
256  &  4.80  &  1000000  &  10000  & 
    157.217  &  0.179  &  8645.33  &  5.05  & 
    0.00190282  &  0.00000040  &  4.25  &  0.04  &  2.54  &  0.02   \\
256  &  4.90  &  1000000  &  10000  & 
    157.807  &  0.179  &  8696.27  &  5.02  & 
    0.00171308  &  0.00000038  &  4.27  &  0.04  &  2.53  &  0.02   \\
256  &  5.00  &  1000000  &  10000  & 
    158.676  &  0.180  &  8750.44  &  5.01  & 
    0.00154114  &  0.00000036  &  4.35  &  0.05  &  2.54  &  0.02   \\
256  &  5.10  &  1000000  &  10000  & 
    159.453  &  0.182  &  8796.18  &  5.02  & 
    0.00138804  &  0.00000034  &  4.43  &  0.05  &  2.57  &  0.02   \\
256  &  5.20  &  1000000  &  10000  & 
    160.061  &  0.182  &  8833.00  &  4.98  & 
    0.00125104  &  0.00000032  &  4.42  &  0.05  &  2.56  &  0.02   \\
256  &  5.30  &  1000000  &  10000  & 
    160.395  &  0.182  &  8864.52  &  4.96  & 
    0.00112784  &  0.00000030  &  4.43  &  0.05  &  2.53  &  0.02   \\
256  &  5.40  &  1000000  &  10000  & 
    160.835  &  0.182  &  8892.90  &  4.93  & 
    0.00101687  &  0.00000029  &  4.40  &  0.05  &  2.52  &  0.02   \\
\hline
512  &  2.80  &  500000  &  10000  & 
    23.428  &  0.577  &  505.57  &  1.25  & 
    0.02119445  &  0.00000123  &  1.54  &  0.01  &  3.22  &  0.04   \\
512  &  2.90  &  500000  &  10000  & 
    27.679  &  0.508  &  690.13  &  1.74  & 
    0.01838339  &  0.00000113  &  1.61  &  0.01  &  3.17  &  0.04   \\
512  &  3.00  &  500000  &  10000  & 
    33.500  &  0.442  &  949.08  &  2.41  & 
    0.01598973  &  0.00000104  &  1.64  &  0.02  &  3.14  &  0.04   \\
512  &  3.10  &  200000  &  10000  & 
    41.719  &  0.619  &  1325.44  &  5.48  & 
    0.01394838  &  0.00000152  &  1.73  &  0.03  &  3.07  &  0.06   \\
512  &  3.20  &  200000  &  10000  & 
    50.521  &  0.559  &  1830.24  &  7.59  & 
    0.01220209  &  0.00000139  &  1.81  &  0.03  &  2.99  &  0.06   \\
512  &  3.30  &  500000  &  10000  & 
    61.816  &  0.330  &  2552.67  &  6.83  & 
    0.01070391  &  0.00000080  &  2.03  &  0.02  &  3.00  &  0.04   \\
512  &  3.40  &  500000  &  10000  & 
    75.572  &  0.320  &  3563.37  &  9.72  & 
    0.00941670  &  0.00000075  &  2.29  &  0.02  &  2.99  &  0.04   \\
512  &  3.50  &  500000  &  10000  & 
    92.478  &  0.330  &  4976.76  &  14.11  & 
    0.00830439  &  0.00000068  &  2.76  &  0.03  &  2.90  &  0.04   \\
\hline 
\end{tabular}
\vspace{3mm}
\caption{ {\bf [Second page of 3-page table]}}
\end{table}
\clearpage \addtocounter{table}{-1}
\begin{table}[h]
\hspace*{-2.5cm}
\footnotesize
\begin{tabular}{|r|r||r|r||r@{ (}r@{) }|r@{ (}r@{) }|r@{ (}r@{) }||r@{ (}r@{) }|r@{ (}r@{) }|}
\hline\hline  \\[-4mm]
\multicolumn{1}{c|}{$L$} &
\multicolumn{1}{c||}{$\beta$} &
   \multicolumn{1}{c|}{Total} &
   \multicolumn{1}{c||}{Discard} &
      \multicolumn{2}{c|}{$\xi$}  &
      \multicolumn{2}{c|}{$\chi_{\stagg}$}  &
         \multicolumn{2}{c||}{$E$}  &
         \multicolumn{2}{c|}{$\tau_{int,\scrm_{\smstagg}^2}$}  &
         \multicolumn{2}{c|}{$\tau_{int,\scre}$}    \\[1mm]
\hline
512  &  3.60  &  500000  &  10000  & 
    112.345  &  0.350  &  6836.58  &  19.37  & 
    0.00734161  &  0.00000063  &  3.26  &  0.04  &  2.82  &  0.03   \\
512  &  3.65  &  500000  &  10000  & 
    123.233  &  0.366  &  7933.08  &  22.46  & 
    0.00690976  &  0.00000061  &  3.60  &  0.05  &  2.85  &  0.03   \\
512  &  3.70  &  500000  &  10000  & 
    135.191  &  0.382  &  9165.68  &  25.34  & 
    0.00650506  &  0.00000058  &  3.87  &  0.05  &  2.81  &  0.03   \\
512  &  3.80  &  500000  &  10000  & 
    159.602  &  0.412  &  11762.98  &  30.00  & 
    0.00577529  &  0.00000054  &  4.29  &  0.06  &  2.74  &  0.03   \\
512  &  3.90  &  500000  &  10000  & 
    184.606  &  0.444  &  14484.50  &  32.72  & 
    0.00513675  &  0.00000050  &  4.52  &  0.07  &  2.73  &  0.03   \\
512  &  4.00  &  1000000  &  10000  & 
    208.293  &  0.333  &  16999.84  &  23.79  & 
    0.00457741  &  0.00000033  &  4.66  &  0.05  &  2.69  &  0.02   \\
512  &  4.10  &  500000  &  10000  & 
    228.543  &  0.493  &  19137.93  &  33.65  & 
    0.00408612  &  0.00000044  &  4.76  &  0.07  &  2.70  &  0.03   \\
512  &  4.15  &  200000  &  10000  & 
    238.737  &  0.802  &  20119.40  &  53.26  & 
    0.00386233  &  0.00000068  &  4.76  &  0.12  &  2.62  &  0.05   \\
512  &  4.20  &  500000  &  10000  & 
    247.620  &  0.501  &  21047.86  &  32.20  & 
    0.00365104  &  0.00000041  &  4.70  &  0.07  &  2.70  &  0.03   \\
512  &  4.30  &  1000000  &  10000  & 
    262.509  &  0.359  &  22500.19  &  21.64  & 
    0.00326851  &  0.00000027  &  4.62  &  0.05  &  2.63  &  0.02   \\
512  &  4.40  &  1000000  &  10000  & 
    274.947  &  0.359  &  23690.60  &  20.38  & 
    0.00292813  &  0.00000025  &  4.51  &  0.05  &  2.62  &  0.02   \\
512  &  4.50  &  1000000  &  10000  & 
    284.440  &  0.363  &  24612.46  &  19.59  & 
    0.00262622  &  0.00000024  &  4.52  &  0.05  &  2.58  &  0.02   \\
512  &  4.60  &  1000000  &  10000  & 
    292.529  &  0.362  &  25377.99  &  18.57  & 
    0.00235787  &  0.00000022  &  4.40  &  0.05  &  2.56  &  0.02   \\
512  &  4.65  &  1000000  &  10000  & 
    296.361  &  0.362  &  25722.12  &  18.29  & 
    0.00223424  &  0.00000022  &  4.42  &  0.05  &  2.59  &  0.02   \\
512  &  4.70  &  1000000  &  10000  & 
    298.694  &  0.360  &  25963.60  &  17.91  & 
    0.00211828  &  0.00000021  &  4.36  &  0.05  &  2.56  &  0.02   \\
512  &  4.80  &  1000000  &  10000  & 
    303.862  &  0.362  &  26460.25  &  17.54  & 
    0.00190436  &  0.00000020  &  4.40  &  0.05  &  2.58  &  0.02   \\
512  &  4.90  &  1000000  &  10000  & 
    308.228  &  0.363  &  26871.14  &  17.09  & 
    0.00171327  &  0.00000019  &  4.36  &  0.05  &  2.56  &  0.02   \\
512  &  5.00  &  1000000  &  10000  & 
    311.295  &  0.359  &  27167.08  &  16.42  & 
    0.00154191  &  0.00000018  &  4.24  &  0.04  &  2.55  &  0.02   \\
512  &  5.10  &  1000000  &  10000  & 
    313.277  &  0.362  &  27407.74  &  16.52  & 
    0.00138888  &  0.00000017  &  4.39  &  0.05  &  2.58  &  0.02   \\
512  &  5.20  &  1000000  &  10000  & 
    316.141  &  0.361  &  27635.16  &  16.17  & 
    0.00125163  &  0.00000016  &  4.37  &  0.05  &  2.53  &  0.02   \\
512  &  5.30  &  1000000  &  10000  & 
    317.469  &  0.362  &  27803.43  &  16.16  & 
    0.00112815  &  0.00000015  &  4.44  &  0.05  &  2.56  &  0.02   \\
512  &  5.40  &  1000000  &  10000  & 
    318.518  &  0.364  &  27951.61  &  16.08  & 
    0.00101697  &  0.00000014  &  4.45  &  0.05  &  2.53  &  0.02   \\
512  &  5.50  &  1000000  &  10000  & 
    320.114  &  0.364  &  28082.10  &  15.88  & 
    0.00091721  &  0.00000014  &  4.44  &  0.05  &  2.53  &  0.02   \\
512  &  5.60  &  600000  &  10000  & 
    320.704  &  0.470  &  28166.84  &  20.43  & 
    0.00082793  &  0.00000017  &  4.42  &  0.06  &  2.49  &  0.03   \\
512  &  5.70  &  1000000  &  10000  & 
    321.343  &  0.363  &  28247.64  &  15.64  & 
    0.00074740  &  0.00000012  &  4.39  &  0.05  &  2.51  &  0.02   \\
\hline
1024  &  3.50  &  200000  &  10000  & 
    93.782  &  1.169  &  5025.28  &  21.00  & 
    0.00830634  &  0.00000055  &  1.79  &  0.03  &  2.92  &  0.06   \\
1024  &  4.00  &  200000  &  10000  & 
    253.855  &  1.161  &  26336.19  &  115.46  & 
    0.00458170  &  0.00000038  &  3.53  &  0.08  &  2.76  &  0.05   \\
1024  &  4.20  &  200000  &  10000  & 
    351.988  &  1.424  &  43182.70  &  167.96  & 
    0.00365419  &  0.00000033  &  4.66  &  0.11  &  2.64  &  0.05   \\
1024  &  4.50  &  200000  &  10000  & 
    481.901  &  1.582  &  64990.58  &  167.30  & 
    0.00262718  &  0.00000027  &  4.72  &  0.12  &  2.58  &  0.05   \\
1024  &  4.70  &  200000  &  10000  & 
    543.182  &  1.645  &  74421.27  &  152.23  & 
    0.00211861  &  0.00000024  &  4.66  &  0.11  &  2.57  &  0.05   \\
1024  &  4.80  &  435000  &  10000  & 
    564.506  &  1.102  &  77670.79  &  95.63  & 
    0.00190502  &  0.00000015  &  4.45  &  0.07  &  2.56  &  0.03   \\
1024  &  4.90  &  350000  &  10000  & 
    581.321  &  1.233  &  80085.31  &  103.34  & 
    0.00171359  &  0.00000016  &  4.49  &  0.08  &  2.57  &  0.04   \\
1024  &  5.00  &  200000  &  10000  & 
    597.074  &  1.646  &  82401.80  &  128.63  & 
    0.00154236  &  0.00000020  &  4.25  &  0.10  &  2.47  &  0.04   \\
1024  &  5.10  &  500000  &  10000  & 
    604.335  &  1.023  &  83709.00  &  79.97  & 
    0.00138923  &  0.00000012  &  4.39  &  0.07  &  2.52  &  0.03   \\
1024  &  5.20  &  315000  &  10000  & 
    614.282  &  1.298  &  85148.87  &  97.33  & 
    0.00125153  &  0.00000014  &  4.33  &  0.08  &  2.55  &  0.04   \\
1024  &  5.30  &  300000  &  10000  & 
    619.635  &  1.346  &  86150.77  &  99.78  & 
    0.00112806  &  0.00000014  &  4.46  &  0.09  &  2.44  &  0.04   \\
1024  &  5.40  &  500000  &  10000  & 
    624.864  &  1.029  &  86978.99  &  75.68  & 
    0.00101735  &  0.00000010  &  4.45  &  0.07  &  2.55  &  0.03   \\
1024  &  5.50  &  500000  &  10000  & 
    630.919  &  1.022  &  87719.01  &  72.87  & 
    0.00091759  &  0.00000010  &  4.32  &  0.06  &  2.51  &  0.03   \\
1024  &  5.60  &  500000  &  10000  & 
    634.020  &  1.030  &  88261.17  &  72.93  & 
    0.00082786  &  0.00000009  &  4.40  &  0.07  &  2.53  &  0.03   \\
1024  &  5.70  &  350000  &  10000  & 
    636.439  &  1.243  &  88750.82  &  86.74  & 
    0.00074709  &  0.00000010  &  4.39  &  0.08  &  2.53  &  0.04   \\
1024  &  5.80  &  500000  &  10000  & 
    639.532  &  1.034  &  89200.43  &  71.47  & 
    0.00067443  &  0.00000008  &  4.39  &  0.07  &  2.53  &  0.03   \\
1024  &  5.90  &  400000  &  10000  & 
    640.955  &  1.148  &  89486.27  &  79.44  & 
    0.00060909  &  0.00000009  &  4.36  &  0.07  &  2.54  &  0.03   \\
\hline
1536  &  5.40  &  200000  &  10000  & 
    921.169  &  2.484  &  167548.48  &  247.24  & 
    0.00101730  &  0.00000011  &  4.47  &  0.11  &  2.47  &  0.04   \\
1536  &  5.50  &  300000  &  10000  & 
    929.501  &  2.010  &  169381.15  &  196.44  & 
    0.00091779  &  0.00000008  &  4.45  &  0.09  &  2.56  &  0.04   \\
1536  &  5.60  &  500000  &  10000  & 
    939.605  &  1.537  &  171375.42  &  146.36  & 
    0.00082793  &  0.00000006  &  4.33  &  0.06  &  2.51  &  0.03   \\
1536  &  5.70  &  500000  &  10000  & 
    945.918  &  1.553  &  172479.44  &  147.38  & 
    0.00074711  &  0.00000006  &  4.48  &  0.07  &  2.51  &  0.03   \\
1536  &  6.00  &  200000  &  10000  & 
    963.094  &  2.481  &  175739.34  &  225.37  & 
    0.00055010  &  0.00000008  &  4.39  &  0.11  &  2.46  &  0.04   \\
\hline
\end{tabular}
\vspace{3mm}
\caption{ {\bf [Third page of 3-page table]}}
\end{table}
\clearpage

%
%
%
\begin{table}[h]
\hspace*{-1cm}
\footnotesize
\begin{tabular}{|r|r||r|r||r@{ (}r@{) }|r@{ (}r@{) }|r@{ (}r@{) }||r@{ (}r@{) }|r@{ (}r@{) }|}
\hline\hline  \\[-4mm]
\multicolumn{1}{c|}{$L$} &
\multicolumn{1}{c||}{$\beta$} &
   \multicolumn{1}{c|}{Total} &
   \multicolumn{1}{c||}{Discard} &
      \multicolumn{2}{c|}{$\xi$}  &
      \multicolumn{2}{c|}{$\chi_{\stagg}$}  &
         \multicolumn{2}{c||}{$E$}  &
         \multicolumn{2}{c|}{$\tau_{int,\scrm_{\smstagg}^2}$}  &
         \multicolumn{2}{c|}{$\tau_{int,\scre}$}    \\[1mm]
\hline
32  &  0.50  &  1000000  &  10000  & 
    0.408  &  0.088  &  1.64  &  0.00  & 
    0.33333424  &  0.00004069  &  2.47  &  0.02  &  2.86  &  0.02   \\
32  &  1.00  &  1000000  &  10000  & 
    0.713  &  0.051  &  2.62  &  0.00  & 
    0.20768309  &  0.00003587  &  2.49  &  0.02  &  3.12  &  0.03   \\
32  &  1.50  &  1000000  &  10000  & 
    0.963  &  0.039  &  3.86  &  0.01  & 
    0.12339135  &  0.00002967  &  2.54  &  0.02  &  3.39  &  0.03   \\
32  &  2.00  &  1000000  &  10000  & 
    1.284  &  0.029  &  5.22  &  0.01  & 
    0.07216655  &  0.00002316  &  2.54  &  0.02  &  3.51  &  0.03   \\
32  &  2.50  &  1000000  &  10000  & 
    1.494  &  0.026  &  6.45  &  0.01  & 
    0.04227138  &  0.00001777  &  2.56  &  0.02  &  3.55  &  0.03   \\
32  &  2.60  &  200000  &  10000  & 
    1.597  &  0.055  &  6.72  &  0.03  & 
    0.03809065  &  0.00003844  &  2.52  &  0.05  &  3.57  &  0.08   \\
32  &  2.70  &  200000  &  10000  & 
    1.513  &  0.060  &  6.85  &  0.03  & 
    0.03421029  &  0.00003588  &  2.72  &  0.05  &  3.49  &  0.07   \\
32  &  2.80  &  200000  &  10000  & 
    1.597  &  0.056  &  7.06  &  0.03  & 
    0.03081867  &  0.00003480  &  2.53  &  0.05  &  3.63  &  0.08   \\
32  &  2.90  &  200000  &  10000  & 
    1.707  &  0.053  &  7.32  &  0.03  & 
    0.02778558  &  0.00003256  &  2.53  &  0.05  &  3.51  &  0.08   \\
32  &  3.00  &  1000000  &  10000  & 
    1.694  &  0.023  &  7.48  &  0.01  & 
    0.02498434  &  0.00001345  &  2.59  &  0.02  &  3.53  &  0.03   \\
32  &  3.10  &  200000  &  10000  & 
    1.686  &  0.053  &  7.62  &  0.03  & 
    0.02251044  &  0.00002917  &  2.56  &  0.05  &  3.50  &  0.07   \\
32  &  3.20  &  200000  &  10000  & 
    1.737  &  0.052  &  7.76  &  0.03  & 
    0.02031029  &  0.00002774  &  2.54  &  0.05  &  3.53  &  0.08   \\
32  &  3.30  &  200000  &  10000  & 
    1.727  &  0.052  &  7.90  &  0.03  & 
    0.01829279  &  0.00002590  &  2.60  &  0.05  &  3.45  &  0.07   \\
32  &  3.40  &  200000  &  10000  & 
    1.744  &  0.051  &  8.03  &  0.03  & 
    0.01646475  &  0.00002463  &  2.54  &  0.05  &  3.49  &  0.07   \\
32  &  3.50  &  1000000  &  10000  & 
    1.832  &  0.022  &  8.20  &  0.02  & 
    0.01488912  &  0.00001037  &  2.63  &  0.02  &  3.55  &  0.03   \\
32  &  3.60  &  200000  &  10000  & 
    1.791  &  0.052  &  8.27  &  0.04  & 
    0.01341060  &  0.00002191  &  2.71  &  0.05  &  3.43  &  0.07   \\
32  &  3.70  &  200000  &  10000  & 
    1.855  &  0.050  &  8.44  &  0.04  & 
    0.01207502  &  0.00002158  &  2.61  &  0.05  &  3.64  &  0.08   \\
32  &  3.80  &  200000  &  10000  & 
    1.870  &  0.049  &  8.52  &  0.04  & 
    0.01094776  &  0.00002028  &  2.59  &  0.05  &  3.56  &  0.08   \\
32  &  3.90  &  200000  &  10000  & 
    1.864  &  0.049  &  8.58  &  0.04  & 
    0.00985439  &  0.00001904  &  2.63  &  0.05  &  3.53  &  0.08   \\
32  &  4.00  &  1000000  &  10000  & 
    1.865  &  0.022  &  8.68  &  0.02  & 
    0.00891195  &  0.00000792  &  2.63  &  0.02  &  3.52  &  0.03   \\
32  &  4.10  &  200000  &  10000  & 
    1.837  &  0.050  &  8.69  &  0.04  & 
    0.00805595  &  0.00001717  &  2.58  &  0.05  &  3.53  &  0.08   \\
32  &  4.20  &  200000  &  10000  & 
    1.890  &  0.049  &  8.79  &  0.04  & 
    0.00726329  &  0.00001631  &  2.69  &  0.05  &  3.51  &  0.08   \\
32  &  4.30  &  200000  &  10000  & 
    1.907  &  0.048  &  8.87  &  0.04  & 
    0.00656180  &  0.00001557  &  2.57  &  0.05  &  3.54  &  0.08   \\
32  &  4.40  &  200000  &  10000  & 
    1.917  &  0.048  &  8.96  &  0.04  & 
    0.00591430  &  0.00001478  &  2.61  &  0.05  &  3.54  &  0.08   \\
32  &  4.50  &  1000000  &  10000  & 
    1.907  &  0.021  &  8.99  &  0.02  & 
    0.00536798  &  0.00000617  &  2.62  &  0.02  &  3.54  &  0.03   \\
32  &  4.60  &  200000  &  10000  & 
    1.906  &  0.049  &  9.05  &  0.04  & 
    0.00486859  &  0.00001338  &  2.72  &  0.05  &  3.54  &  0.08   \\
32  &  4.70  &  200000  &  10000  & 
    1.911  &  0.049  &  9.06  &  0.04  & 
    0.00439399  &  0.00001268  &  2.62  &  0.05  &  3.52  &  0.08   \\
32  &  4.80  &  200000  &  10000  & 
    1.905  &  0.049  &  9.11  &  0.04  & 
    0.00395714  &  0.00001187  &  2.61  &  0.05  &  3.45  &  0.07   \\
32  &  4.90  &  200000  &  10000  & 
    1.996  &  0.048  &  9.20  &  0.04  & 
    0.00358136  &  0.00001127  &  2.78  &  0.05  &  3.43  &  0.07   \\
32  &  5.00  &  1000000  &  10000  & 
    1.960  &  0.021  &  9.21  &  0.02  & 
    0.00323220  &  0.00000472  &  2.63  &  0.02  &  3.47  &  0.03   \\
\hline
64  &  2.50  &  1000000  &  10000  & 
    1.559  &  0.093  &  6.47  &  0.01  & 
    0.04228318  &  0.00000895  &  2.52  &  0.02  &  3.60  &  0.03   \\
64  &  3.00  &  1000000  &  10000  & 
    1.765  &  0.083  &  7.47  &  0.01  & 
    0.02497250  &  0.00000678  &  2.51  &  0.02  &  3.56  &  0.03   \\
64  &  3.50  &  1000000  &  10000  & 
    1.729  &  0.085  &  8.17  &  0.01  & 
    0.01488177  &  0.00000517  &  2.51  &  0.02  &  3.53  &  0.03   \\
64  &  4.00  &  1000000  &  10000  & 
    1.754  &  0.084  &  8.65  &  0.02  & 
    0.00890631  &  0.00000397  &  2.49  &  0.02  &  3.53  &  0.03   \\
64  &  4.50  &  1000000  &  10000  & 
    1.890  &  0.078  &  9.00  &  0.02  & 
    0.00536320  &  0.00000308  &  2.52  &  0.02  &  3.56  &  0.03   \\
64  &  5.00  &  1000000  &  10000  & 
    1.894  &  0.078  &  9.18  &  0.02  & 
    0.00323899  &  0.00000236  &  2.52  &  0.02  &  3.48  &  0.03   \\
\hline
\end{tabular}
\vspace{3mm}
\caption{
   Our Monte Carlo data for the 4-state Potts antiferromagnet.
   ``Total'' is the total number of WSK iterations performed;
   ``Discard'' is the number of iterations discarded for equilibration.
   Error bars (one standard deviation) are shown in parentheses.
}
\label{table:q=4}
\end{table}
\clearpage

%
%
%
%
\begin{table}[h]
\hspace*{-1cm}
\small
\begin{tabular}{|c||c|c|c|c|c|c|c|}
\hline
\multicolumn{1}{|c||}{$L_{min}$}  &
  \multicolumn{1}{|c|}{$n=3$}  &
  \multicolumn{1}{c|}{$n=4$}  &
  \multicolumn{1}{c|}{$n=5$}  &
  \multicolumn{1}{c|}{$n=6$}  &
  \multicolumn{1}{c|}{$n=7$}  &
  \multicolumn{1}{c|}{$n=8$}  &
  \multicolumn{1}{c|}{$n=9$}      \\
\hline\hline
 32 &  571.70, 94 &  178.98, 93 &  138.80, 92 & 
       138.62, 91 &  135.95, 90 &  134.00, 89 &  125.94, 88  \\
    &  6.08,  0.0\% &  1.92,  0.0\% &  1.51,  0.1\% & 
       1.52,  0.1\% &  1.51,  0.1\% &  1.51,  0.1\% &  1.43,  0.5\%  \\
\hline
 64 &  263.84, 86 &  113.41, 85 &  100.21, 84 & 
       100.11, 83 &   98.52, 82 &   96.49, 81 &   92.60, 80  \\
    &  3.07,  0.0\% &  1.33,  2.1\% &  1.19,  11.0\% & 
       1.21,  9.7\% &  1.20,  10.3\% &  1.19,  11.5\% &  1.16,  15.9\%  \\
\hline
128 &  206.52, 68 &   87.32, 67 &   75.41, 66 & 
        75.21, 65 &   72.97, 64 &   72.58, 63 &   70.02, 62  \\
    &  3.04,  0.0\% &  1.30,  4.8\% &  1.14,  20.0\% & 
       1.16, 18.1\% &  1.14,  20.7\% &  1.15,  19.1\% &  1.13,  22.6\%  \\
\hline
256 &  119.94, 40 &   63.84, 39 &   58.22, 38 & 
        58.17, 37 &   57.01, 36 &   55.12, 35 &   52.71, 34  \\
    &  3.00,  0.0\% &  1.64,  0.7\% &  1.53,  1.9\% & 
       1.57,  1.5\% &  1.58,  1.4\% &  1.57,  1.7\% &  1.55,  2.1\%  \\
\hline
\end{tabular}
\vspace{3mm}
\caption{
   $\chi^2$ for the fit (\protect\ref{eqfss3}) of
   $\xi(\beta,2L)/\xi(\beta,L)$ versus $\xi(\beta,L)/L$.
   First line is $\chi^2$ followed by DF (number of degrees of freedom).
   Second line is $\chi^2$/DF followed by the confidence level.
   In all cases $\xi_{min} = 10$.
}
\label{table_chisq_xi}
\end{table}

%
%
%
%
\begin{table}[h]
\hspace*{-1cm}
\small
\begin{tabular}{|l|r||r@{ (}r@{) }|r@{ (}r@{ DF, }r@{\%) }||r@{ (}r@{) }|r@{ (}r@{ DF, }r@{\%) }|}
\hline
\multicolumn{1}{|c|}{$\beta$}  &
   \multicolumn{1}{|c||}{$L_{min}$}  &
   \multicolumn{2}{c|}{$\xi_\infty$}  &
   \multicolumn{3}{c||}{${\cal R}$ for $\xi_\infty$}  &
   \multicolumn{2}{c|}{$\chi_{\stagg,\infty}$}  &
   \multicolumn{3}{c|}{${\cal R}$ for $\chi_{\stagg,\infty}$}   \\
\hline\hline
{\rm 2.50} & {\rm  32} & {\rm    13.30} & {\rm     0.03} & 
    {\rm  0.45} & {\rm 3} & {\rm  93.0} & 
    {\rm      200.85} & {\rm        0.31} & 
    {\rm  2.14} & {\rm 3} & {\rm  54.4} \\
{\rm 2.50} & {\rm  64} & {\rm    13.31} & {\rm     0.04} & 
    {\rm  0.27} & {\rm 2} & {\rm  87.4} & 
    {\rm      201.13} & {\rm        0.39} & 
    {\rm  0.84} & {\rm 2} & {\rm  65.9} \\
{\it 2.50} & {\it 128} & {\it    13.32} & {\it     0.13} & 
    {\it  0.27} & {\it 1} & {\it  60.2} & 
    {\it      201.51} & {\it        0.57} & 
    {\it  0.09} & {\it 1} & {\it  76.6} \\
{\sf 2.50} & {\sf 256} & {\sf    13.10} & {\sf     0.42} & 
    {\sf  0.00} & {\sf 0} & {\sf 100.0} & 
    {\sf      201.34} & {\sf        0.80} & 
    {\sf  0.00} & {\sf 0} & {\sf 100.0} \\
\hline
{\rm 2.60} & {\rm  32} & {\rm    15.99} & {\rm     0.08} & 
    {\rm  0.32} & {\rm 2} & {\rm  85.3} & 
    {\rm      272.17} & {\rm        0.69} & 
    {\rm  1.54} & {\rm 2} & {\rm  46.3} \\
{\rm 2.60} & {\rm  64} & {\rm    16.00} & {\rm     0.08} & 
    {\rm  0.31} & {\rm 2} & {\rm  85.8} & 
    {\rm      272.22} & {\rm        0.70} & 
    {\rm  1.38} & {\rm 2} & {\rm  50.3} \\
{\it 2.60} & {\it 128} & {\it    16.00} & {\it     0.12} & 
    {\it  0.31} & {\it 1} & {\it  57.7} & 
    {\it      272.54} & {\it        0.79} & 
    {\it  0.66} & {\it 1} & {\it  41.7} \\
{\sf 2.60} & {\sf 256} & {\sf    16.18} & {\sf     0.36} & 
    {\sf  0.00} & {\sf 0} & {\sf 100.0} & 
    {\sf      273.13} & {\sf        1.10} & 
    {\sf  0.00} & {\sf 0} & {\sf 100.0} \\
\hline
{\rm 2.70} & {\rm  32} & {\rm    19.29} & {\rm     0.05} & 
    {\rm  0.43} & {\rm 3} & {\rm  93.3} & 
    {\rm      369.78} & {\rm        0.69} & 
    {\rm  0.73} & {\rm 3} & {\rm  86.6} \\
{\rm 2.70} & {\rm  64} & {\rm    19.29} & {\rm     0.06} & 
    {\rm  0.27} & {\rm 2} & {\rm  87.6} & 
    {\rm      369.51} & {\rm        0.75} & 
    {\rm  0.13} & {\rm 2} & {\rm  93.7} \\
{\it 2.70} & {\it 128} & {\it    19.27} & {\it     0.12} & 
    {\it  0.22} & {\it 1} & {\it  63.8} & 
    {\it      369.28} & {\it        1.09} & 
    {\it  0.05} & {\it 1} & {\it  83.0} \\
{\sf 2.70} & {\sf 256} & {\sf    19.13} & {\sf     0.33} & 
    {\sf  0.00} & {\sf 0} & {\sf 100.0} & 
    {\sf      369.05} & {\sf        1.51} & 
    {\sf  0.00} & {\sf 0} & {\sf 100.0} \\
\hline
{\rm 2.80} & {\rm  32} & {\rm    23.27} & {\rm     0.09} & 
    {\rm  0.91} & {\rm 3} & {\rm  82.2} & 
    {\rm      504.73} & {\rm        0.92} & 
    {\rm  1.94} & {\rm 3} & {\rm  58.5} \\
{\rm 2.80} & {\rm  64} & {\rm    23.28} & {\rm     0.09} & 
    {\rm  1.00} & {\rm 3} & {\rm  80.0} & 
    {\rm      504.74} & {\rm        0.91} & 
    {\rm  1.97} & {\rm 3} & {\rm  57.8} \\
{\it 2.80} & {\it 128} & {\it    23.29} & {\it     0.12} & 
    {\it  0.86} & {\it 2} & {\it  65.2} & 
    {\it      504.80} & {\it        0.98} & 
    {\it  1.99} & {\it 2} & {\it  37.0} \\
{\sf 2.80} & {\sf 256} & {\sf    23.13} & {\sf     0.27} & 
    {\sf  0.36} & {\sf 1} & {\sf  55.1} & 
    {\sf      504.68} & {\sf        1.07} & 
    {\sf  1.95} & {\sf 1} & {\sf  16.3} \\
\hline
{\rm 2.90} & {\rm  32} & {\rm    28.11} & {\rm     0.11} & 
    {\rm  1.29} & {\rm 3} & {\rm  73.2} & 
    {\rm      690.90} & {\rm        1.31} & 
    {\rm  3.05} & {\rm 3} & {\rm  38.4} \\
{\rm 2.90} & {\rm  64} & {\rm    28.13} & {\rm     0.12} & 
    {\rm  1.34} & {\rm 3} & {\rm  72.0} & 
    {\rm      690.94} & {\rm        1.33} & 
    {\rm  3.32} & {\rm 3} & {\rm  34.4} \\
{\it 2.90} & {\it 128} & {\it    28.14} & {\it     0.14} & 
    {\it  0.94} & {\it 2} & {\it  62.6} & 
    {\it      691.27} & {\it        1.40} & 
    {\it  1.24} & {\it 2} & {\it  53.9} \\
{\sf 2.90} & {\sf 256} & {\sf    28.02} & {\sf     0.24} & 
    {\sf  0.59} & {\sf 1} & {\sf  44.2} & 
    {\sf      691.04} & {\sf        1.51} & 
    {\sf  0.99} & {\sf 1} & {\sf  32.0} \\
\hline
{\rm 2.95} & {\rm  32} & {\rm    30.84} & {\rm     0.18} & 
    {\rm  0.00} & {\rm 0} & {\rm 100.0} & 
    {\rm      804.48} & {\rm        4.06} & 
    {\rm  0.00} & {\rm 0} & {\rm 100.0} \\
{\rm 2.95} & {\rm  64} & {\rm    30.87} & {\rm     0.19} & 
    {\rm  0.00} & {\rm 0} & {\rm 100.0} & 
    {\rm      804.91} & {\rm        4.20} & 
    {\rm  0.00} & {\rm 0} & {\rm 100.0} \\
{\it 2.95} & {\it 128} & {\it    30.84} & {\it     0.20} & 
    {\it  0.00} & {\it 0} & {\it 100.0} & 
    {\it      804.34} & {\it        4.35} & 
    {\it  0.00} & {\it 0} & {\it 100.0} \\
\hline
{\rm 3.00} & {\rm  32} & {\rm    34.19} & {\rm     0.10} & 
    {\rm  2.86} & {\rm 4} & {\rm  58.1} & 
    {\rm      952.35} & {\rm        1.70} & 
    {\rm  3.81} & {\rm 4} & {\rm  43.2} \\
{\rm 3.00} & {\rm  64} & {\rm    34.23} & {\rm     0.11} & 
    {\rm  2.89} & {\rm 3} & {\rm  41.0} & 
    {\rm      952.09} & {\rm        1.80} & 
    {\rm  3.45} & {\rm 3} & {\rm  32.8} \\
{\it 3.00} & {\it 128} & {\it    34.16} & {\it     0.16} & 
    {\it  2.72} & {\it 2} & {\it  25.7} & 
    {\it      951.17} & {\it        1.93} & 
    {\it  2.53} & {\it 2} & {\it  28.3} \\
{\sf 3.00} & {\sf 256} & {\sf    34.11} & {\sf     0.23} & 
    {\sf  2.51} & {\sf 1} & {\sf  11.3} & 
    {\sf      950.81} & {\sf        2.12} & 
    {\sf  2.08} & {\sf 1} & {\sf  14.9} \\
\hline
{\rm 3.10} & {\rm  32} & {\rm    41.56} & {\rm     0.17} & 
    {\rm  1.39} & {\rm 3} & {\rm  70.8} & 
    {\rm     1319.45} & {\rm        3.44} & 
    {\rm  3.11} & {\rm 3} & {\rm  37.5} \\
{\rm 3.10} & {\rm  64} & {\rm    41.59} & {\rm     0.17} & 
    {\rm  1.23} & {\rm 3} & {\rm  74.6} & 
    {\rm     1319.40} & {\rm        3.43} & 
    {\rm  3.36} & {\rm 3} & {\rm  34.0} \\
{\it 3.10} & {\it 128} & {\it    41.57} & {\it     0.18} & 
    {\it  1.46} & {\it 2} & {\it  48.1} & 
    {\it     1318.96} & {\it        3.57} & 
    {\it  4.82} & {\it 2} & {\it   9.0} \\
{\sf 3.10} & {\sf 256} & {\sf    41.77} & {\sf     0.25} & 
    {\sf  0.01} & {\sf 1} & {\sf  92.5} & 
    {\sf     1322.89} & {\sf        4.06} & 
    {\sf  0.48} & {\sf 1} & {\sf  49.0} \\
\hline
{\rm 3.20} & {\rm  32} & {\rm    50.63} & {\rm     0.16} & 
    {\rm  5.25} & {\rm 4} & {\rm  26.2} & 
    {\rm     1831.78} & {\rm        4.49} & 
    {\rm  5.95} & {\rm 4} & {\rm  20.3} \\
{\rm 3.20} & {\rm  64} & {\rm    50.81} & {\rm     0.17} & 
    {\rm  0.60} & {\rm 3} & {\rm  89.7} & 
    {\rm     1834.33} & {\rm        4.75} & 
    {\rm  1.35} & {\rm 3} & {\rm  71.7} \\
{\it 3.20} & {\it 128} & {\it    50.70} & {\it     0.22} & 
    {\it  0.23} & {\it 2} & {\it  89.2} & 
    {\it     1831.27} & {\it        5.25} & 
    {\it  0.16} & {\it 2} & {\it  92.2} \\
{\sf 3.20} & {\sf 256} & {\sf    50.72} & {\sf     0.27} & 
    {\sf  0.17} & {\sf 1} & {\sf  68.3} & 
    {\sf     1831.56} & {\sf        6.16} & 
    {\sf  0.07} & {\sf 1} & {\sf  79.5} \\
\hline
{\rm 3.30} & {\rm  32} & {\rm    61.96} & {\rm     0.22} & 
    {\rm  0.64} & {\rm 3} & {\rm  88.8} & 
    {\rm     2556.28} & {\rm        5.94} & 
    {\rm  0.91} & {\rm 3} & {\rm  82.3} \\
{\rm 3.30} & {\rm  64} & {\rm    62.01} & {\rm     0.22} & 
    {\rm  0.95} & {\rm 3} & {\rm  81.4} & 
    {\rm     2556.71} & {\rm        5.95} & 
    {\rm  1.15} & {\rm 3} & {\rm  76.5} \\
{\it 3.30} & {\it 128} & {\it    61.98} & {\it     0.23} & 
    {\it  0.60} & {\it 2} & {\it  74.1} & 
    {\it     2556.23} & {\it        6.16} & 
    {\it  0.69} & {\it 2} & {\it  70.7} \\
{\sf 3.30} & {\sf 256} & {\sf    61.89} & {\sf     0.27} & 
    {\sf  0.10} & {\sf 1} & {\sf  75.8} & 
    {\sf     2554.73} & {\sf        6.52} & 
    {\sf  0.06} & {\sf 1} & {\sf  79.9} \\
\hline
\end{tabular}
\vspace{3mm}
\caption{
   Estimated infinite-volume correlation lengths $\xi_\infty$ and
   staggered susceptibilities $\chi_{stagg,\infty}$
   as a function of $\beta$, from extrapolations using various $L_{min}$.
   Error bars are one standard deviation (statistical errors only).
   All extrapolations use $n=5$ for $\xi$ and $n=6$ for $\chi_{stagg}$.
   ${\cal R}$ indicates the residual sum-of-squares (\protect\ref{residual})
   for combining estimates from different $L$ at the same $\beta$;
   the number of degrees of freedom (DF) and the confidence level
   are indicated.
   Our preferred fit is shown in {\em italics}\/;
   a more conservative good fit is shown in {\sf sans-serif};
   bad fits are shown in {\rm roman}.
   {\bf [First page of 4-page table]}
}
\label{table:DatiEstrapolatiMediatiConChiQuadro}
\end{table}
\clearpage \addtocounter{table}{-1}
\begin{table}[h]
\hspace*{-1cm}
\small
\begin{tabular}{|l|r||r@{ (}r@{) }|r@{ (}r@{ DF, }r@{\%) }||r@{ (}r@{) }|r@{ (}r@{ DF, }r@{\%) }|}
\hline
\multicolumn{1}{|c|}{$\beta$}  &
   \multicolumn{1}{|c||}{$L_{min}$}  &
   \multicolumn{2}{c|}{$\xi_\infty$}  &
   \multicolumn{3}{c||}{${\cal R}$ for $\xi_\infty$}  &
   \multicolumn{2}{c|}{$\chi_{\stagg,\infty}$}  &
   \multicolumn{3}{c|}{${\cal R}$ for $\chi_{\stagg,\infty}$}   \\
\hline\hline
{\rm 3.40} & {\rm  32} & {\rm    75.63} & {\rm     0.26} & 
    {\rm  0.05} & {\rm 3} & {\rm  99.7} & 
    {\rm     3574.64} & {\rm        8.59} & 
    {\rm  0.29} & {\rm 3} & {\rm  96.2} \\
{\rm 3.40} & {\rm  64} & {\rm    75.68} & {\rm     0.26} & 
    {\rm  0.04} & {\rm 3} & {\rm  99.8} & 
    {\rm     3575.04} & {\rm        8.50} & 
    {\rm  0.24} & {\rm 3} & {\rm  97.0} \\
{\it 3.40} & {\it 128} & {\it    75.64} & {\it     0.27} & 
    {\it  0.01} & {\it 2} & {\it  99.6} & 
    {\it     3573.43} & {\it        8.99} & 
    {\it  0.02} & {\it 2} & {\it  99.1} \\
{\sf 3.40} & {\sf 256} & {\sf    75.60} & {\sf     0.30} & 
    {\sf  0.11} & {\sf 1} & {\sf  73.5} & 
    {\sf     3571.90} & {\sf        9.54} & 
    {\sf  0.04} & {\sf 1} & {\sf  84.6} \\
\hline
{\rm 3.45} & {\rm  32} & {\rm    83.56} & {\rm     0.76} & 
    {\rm  0.00} & {\rm 0} & {\rm 100.0} & 
    {\rm     4220.00} & {\rm       47.16} & 
    {\rm  0.00} & {\rm 0} & {\rm 100.0} \\
{\rm 3.45} & {\rm  64} & {\rm    83.75} & {\rm     0.76} & 
    {\rm  0.00} & {\rm 0} & {\rm 100.0} & 
    {\rm     4223.68} & {\rm       47.21} & 
    {\rm  0.00} & {\rm 0} & {\rm 100.0} \\
{\it 3.45} & {\it 128} & {\it    83.64} & {\it     0.78} & 
    {\it  0.00} & {\it 0} & {\it 100.0} & 
    {\it     4215.80} & {\it       47.28} & 
    {\it  0.00} & {\it 0} & {\it 100.0} \\
\hline
{\rm 3.50} & {\rm  32} & {\rm    92.94} & {\rm     0.25} & 
    {\rm  4.88} & {\rm 5} & {\rm  43.1} & 
    {\rm     5035.19} & {\rm        9.43} & 
    {\rm  6.50} & {\rm 5} & {\rm  26.0} \\
{\rm 3.50} & {\rm  64} & {\rm    93.05} & {\rm     0.26} & 
    {\rm  1.45} & {\rm 4} & {\rm  83.5} & 
    {\rm     5034.94} & {\rm        9.86} & 
    {\rm  1.85} & {\rm 4} & {\rm  76.3} \\
{\it 3.50} & {\it 128} & {\it    93.02} & {\it     0.29} & 
    {\it  0.68} & {\it 3} & {\it  87.9} & 
    {\it     5030.37} & {\it       10.68} & 
    {\it  0.37} & {\it 3} & {\it  94.7} \\
{\sf 3.50} & {\sf 256} & {\sf    92.90} & {\sf     0.34} & 
    {\sf  0.63} & {\sf 2} & {\sf  72.8} & 
    {\sf     5025.27} & {\sf       12.41} & 
    {\sf  0.01} & {\sf 2} & {\sf  99.3} \\
\hline
{\rm 3.60} & {\rm  32} & {\rm   113.85} & {\rm     0.41} & 
    {\rm  1.82} & {\rm 3} & {\rm  61.0} & 
    {\rm     7054.43} & {\rm       22.62} & 
    {\rm  2.11} & {\rm 3} & {\rm  54.9} \\
{\rm 3.60} & {\rm  64} & {\rm   113.96} & {\rm     0.42} & 
    {\rm  1.27} & {\rm 3} & {\rm  73.7} & 
    {\rm     7058.81} & {\rm       23.51} & 
    {\rm  1.34} & {\rm 3} & {\rm  72.0} \\
{\it 3.60} & {\it 128} & {\it   113.81} & {\it     0.46} & 
    {\it  0.08} & {\it 2} & {\it  96.0} & 
    {\it     7053.38} & {\it       25.92} & 
    {\it  0.02} & {\it 2} & {\it  98.9} \\
{\sf 3.60} & {\sf 256} & {\sf   113.70} & {\sf     0.52} & 
    {\sf  0.01} & {\sf 1} & {\sf  91.5} & 
    {\sf     7048.81} & {\sf       30.78} & 
    {\sf  0.00} & {\sf 1} & {\sf  97.9} \\
\hline
{\rm 3.65} & {\rm  32} & {\rm   125.80} & {\rm     0.55} & 
    {\rm  0.00} & {\rm 0} & {\rm 100.0} & 
    {\rm     8350.93} & {\rm       30.21} & 
    {\rm  0.00} & {\rm 0} & {\rm 100.0} \\
{\rm 3.65} & {\rm  64} & {\rm   125.92} & {\rm     0.57} & 
    {\rm  0.00} & {\rm 0} & {\rm 100.0} & 
    {\rm     8354.88} & {\rm       31.24} & 
    {\rm  0.00} & {\rm 0} & {\rm 100.0} \\
{\it 3.65} & {\it 128} & {\it   125.78} & {\it     0.63} & 
    {\it  0.00} & {\it 0} & {\it 100.0} & 
    {\it     8348.35} & {\it       34.51} & 
    {\it  0.00} & {\it 0} & {\it 100.0} \\
{\sf 3.65} & {\sf 256} & {\sf   125.66} & {\sf     0.70} & 
    {\sf  0.00} & {\sf 0} & {\sf 100.0} & 
    {\sf     8341.72} & {\sf       40.35} & 
    {\sf  0.00} & {\sf 0} & {\sf 100.0} \\
\hline
{\rm 3.70} & {\rm  32} & {\rm   139.75} & {\rm     0.58} & 
    {\rm  1.07} & {\rm 3} & {\rm  78.4} & 
    {\rm     9952.36} & {\rm       34.98} & 
    {\rm  1.32} & {\rm 3} & {\rm  72.5} \\
{\rm 3.70} & {\rm  64} & {\rm   139.91} & {\rm     0.59} & 
    {\rm  1.44} & {\rm 3} & {\rm  69.7} & 
    {\rm     9952.21} & {\rm       35.13} & 
    {\rm  1.69} & {\rm 3} & {\rm  63.8} \\
{\it 3.70} & {\it 128} & {\it   139.76} & {\it     0.67} & 
    {\it  0.64} & {\it 2} & {\it  72.5} & 
    {\it     9940.95} & {\it       38.57} & 
    {\it  0.82} & {\it 2} & {\it  66.5} \\
{\sf 3.70} & {\sf 256} & {\sf   139.51} & {\sf     0.77} & 
    {\sf  0.05} & {\sf 1} & {\sf  82.7} & 
    {\sf     9927.15} & {\sf       45.42} & 
    {\sf  0.12} & {\sf 1} & {\sf  72.4} \\
\hline
{\rm 3.80} & {\rm  32} & {\rm   171.68} & {\rm     0.70} & 
    {\rm  4.96} & {\rm 3} & {\rm  17.5} & 
    {\rm    14021.81} & {\rm       49.40} & 
    {\rm  5.11} & {\rm 3} & {\rm  16.4} \\
{\rm 3.80} & {\rm  64} & {\rm   171.85} & {\rm     0.73} & 
    {\rm  3.22} & {\rm 3} & {\rm  35.9} & 
    {\rm    14009.84} & {\rm       51.53} & 
    {\rm  3.37} & {\rm 3} & {\rm  33.8} \\
{\it 3.80} & {\it 128} & {\it   171.59} & {\it     0.79} & 
    {\it  0.18} & {\it 2} & {\it  91.5} & 
    {\it    13982.10} & {\it       54.81} & 
    {\it  0.27} & {\it 2} & {\it  87.4} \\
{\sf 3.80} & {\sf 256} & {\sf   171.23} & {\sf     0.90} & 
    {\sf  0.00} & {\sf 1} & {\sf  97.6} & 
    {\sf    13953.90} & {\sf       61.64} & 
    {\sf  0.03} & {\sf 1} & {\sf  86.8} \\
\hline
{\rm 3.90} & {\rm  32} & {\rm   211.00} & {\rm     0.88} & 
    {\rm  0.98} & {\rm 3} & {\rm  80.6} & 
    {\rm    19803.32} & {\rm       77.69} & 
    {\rm  1.08} & {\rm 3} & {\rm  78.3} \\
{\rm 3.90} & {\rm  64} & {\rm   211.19} & {\rm     0.91} & 
    {\rm  1.09} & {\rm 3} & {\rm  78.0} & 
    {\rm    19790.24} & {\rm       81.81} & 
    {\rm  1.26} & {\rm 3} & {\rm  73.8} \\
{\it 3.90} & {\it 128} & {\it   210.98} & {\it     0.99} & 
    {\it  1.30} & {\it 2} & {\it  52.3} & 
    {\it    19765.50} & {\it       87.33} & 
    {\it  1.48} & {\it 2} & {\it  47.8} \\
{\sf 3.90} & {\sf 256} & {\sf   210.74} & {\sf     1.13} & 
    {\sf  0.34} & {\sf 1} & {\sf  55.9} & 
    {\sf    19737.60} & {\sf       99.20} & 
    {\sf  0.34} & {\sf 1} & {\sf  56.1} \\
\hline
{\rm 3.95} & {\rm  32} & {\rm   237.95} & {\rm     6.17} & 
    {\rm  0.00} & {\rm 0} & {\rm 100.0} & 
    {\rm    24159.76} & {\rm      969.01} & 
    {\rm  0.00} & {\rm 0} & {\rm 100.0} \\
{\rm 3.95} & {\rm  64} & {\rm   236.37} & {\rm     5.97} & 
    {\rm  0.00} & {\rm 0} & {\rm 100.0} & 
    {\rm    23848.23} & {\rm      931.88} & 
    {\rm  0.00} & {\rm 0} & {\rm 100.0} \\
{\it 3.95} & {\it 128} & {\it   235.80} & {\it     6.14} & 
    {\it  0.00} & {\it 0} & {\it 100.0} & 
    {\it    23773.60} & {\it      954.62} & 
    {\it  0.00} & {\it 0} & {\it 100.0} \\
\hline
{\rm 4.00} & {\rm  32} & {\rm   259.76} & {\rm     0.92} & 
    {\rm  4.79} & {\rm 5} & {\rm  44.2} & 
    {\rm    27975.61} & {\rm       95.23} & 
    {\rm  5.05} & {\rm 5} & {\rm  41.0} \\
{\rm 4.00} & {\rm  64} & {\rm   260.24} & {\rm     0.96} & 
    {\rm  1.09} & {\rm 4} & {\rm  89.7} & 
    {\rm    27993.91} & {\rm      103.35} & 
    {\rm  1.27} & {\rm 4} & {\rm  86.6} \\
{\it 4.00} & {\it 128} & {\it   259.98} & {\it     1.07} & 
    {\it  0.92} & {\it 3} & {\it  82.1} & 
    {\it    27970.00} & {\it      116.37} & 
    {\it  1.12} & {\it 3} & {\it  77.2} \\
{\sf 4.00} & {\sf 256} & {\sf   259.73} & {\sf     1.30} & 
    {\sf  0.40} & {\sf 2} & {\sf  81.7} & 
    {\sf    27941.80} & {\sf      143.30} & 
    {\sf  0.58} & {\sf 2} & {\sf  74.8} \\
\hline
{\rm 4.10} & {\rm  32} & {\rm   317.61} & {\rm     1.78} & 
    {\rm  0.61} & {\rm 2} & {\rm  73.6} & 
    {\rm    39274.02} & {\rm      257.27} & 
    {\rm  0.61} & {\rm 2} & {\rm  73.7} \\
{\rm 4.10} & {\rm  64} & {\rm   318.18} & {\rm     1.82} & 
    {\rm  0.28} & {\rm 2} & {\rm  87.0} & 
    {\rm    39291.96} & {\rm      262.01} & 
    {\rm  0.30} & {\rm 2} & {\rm  86.0} \\
{\it 4.10} & {\it 128} & {\it   317.70} & {\it     1.97} & 
    {\it  0.30} & {\it 2} & {\it  86.0} & 
    {\it    39219.30} & {\it      273.74} & 
    {\it  0.34} & {\it 2} & {\it  84.4} \\
{\sf 4.10} & {\sf 256} & {\sf   317.47} & {\sf     2.39} & 
    {\sf  0.16} & {\sf 1} & {\sf  69.3} & 
    {\sf    39222.30} & {\sf      325.32} & 
    {\sf  0.18} & {\sf 1} & {\sf  67.5} \\
\hline
{\rm 4.15} & {\rm  32} & {\rm   355.81} & {\rm     3.47} & 
    {\rm  0.00} & {\rm 0} & {\rm 100.0} & 
    {\rm    47353.74} & {\rm      572.90} & 
    {\rm  0.00} & {\rm 0} & {\rm 100.0} \\
{\rm 4.15} & {\rm  64} & {\rm   356.61} & {\rm     3.49} & 
    {\rm  0.00} & {\rm 0} & {\rm 100.0} & 
    {\rm    47389.86} & {\rm      581.04} & 
    {\rm  0.00} & {\rm 0} & {\rm 100.0} \\
{\it 4.15} & {\it 128} & {\it   356.11} & {\it     3.58} & 
    {\it  0.00} & {\it 0} & {\it 100.0} & 
    {\it    47294.50} & {\it      588.26} & 
    {\it  0.00} & {\it 0} & {\it 100.0} \\
{\sf 4.15} & {\sf 256} & {\sf   355.96} & {\sf     3.93} & 
    {\sf  0.00} & {\sf 0} & {\sf 100.0} & 
    {\sf    47319.30} & {\sf      627.61} & 
    {\sf  0.00} & {\sf 0} & {\sf 100.0} \\
\hline
\end{tabular}
\vspace{3mm}
\caption{
   {\bf [Second page of 4-page table]}
}
\end{table}
\clearpage \addtocounter{table}{-1}
\begin{table}[h]
\hspace*{-1cm}
\small
\begin{tabular}{|l|r||r@{ (}r@{) }|r@{ (}r@{ DF, }r@{\%) }||r@{ (}r@{) }|r@{ (}r@{ DF, }r@{\%) }|}
\hline
\multicolumn{1}{|c|}{$\beta$}  &
   \multicolumn{1}{|c||}{$L_{min}$}  &
   \multicolumn{2}{c|}{$\xi_\infty$}  &
   \multicolumn{3}{c||}{${\cal R}$ for $\xi_\infty$}  &
   \multicolumn{2}{c|}{$\chi_{\stagg,\infty}$}  &
   \multicolumn{3}{c|}{${\cal R}$ for $\chi_{\stagg,\infty}$}   \\
\hline\hline
{\rm 4.20} & {\rm  32} & {\rm   394.98} & {\rm     1.92} & 
    {\rm  1.43} & {\rm 3} & {\rm  69.8} & 
    {\rm    56299.87} & {\rm      272.76} & 
    {\rm  2.48} & {\rm 3} & {\rm  47.9} \\
{\rm 4.20} & {\rm  64} & {\rm   395.38} & {\rm     1.96} & 
    {\rm  1.44} & {\rm 3} & {\rm  69.7} & 
    {\rm    56273.48} & {\rm      284.20} & 
    {\rm  2.48} & {\rm 3} & {\rm  47.9} \\
{\it 4.20} & {\it 128} & {\it   394.85} & {\it     2.12} & 
    {\it  1.23} & {\it 3} & {\it  74.5} & 
    {\it    56161.10} & {\it      300.19} & 
    {\it  2.26} & {\it 3} & {\it  52.1} \\
{\sf 4.20} & {\sf 256} & {\sf   394.00} & {\sf     2.47} & 
    {\sf  1.71} & {\sf 2} & {\sf  42.6} & 
    {\sf    56045.30} & {\sf      332.62} & 
    {\sf  3.02} & {\sf 2} & {\sf  22.1} \\
\hline
{\rm 4.30} & {\rm  32} & {\rm   489.45} & {\rm     3.01} & 
    {\rm  1.82} & {\rm 2} & {\rm  40.2} & 
    {\rm    80567.72} & {\rm      664.23} & 
    {\rm  1.82} & {\rm 2} & {\rm  40.3} \\
{\rm 4.30} & {\rm  64} & {\rm   489.42} & {\rm     3.14} & 
    {\rm  0.99} & {\rm 2} & {\rm  61.1} & 
    {\rm    80425.79} & {\rm      687.24} & 
    {\rm  1.03} & {\rm 2} & {\rm  59.7} \\
{\it 4.30} & {\it 128} & {\it   488.45} & {\it     3.56} & 
    {\it  1.00} & {\it 2} & {\it  60.8} & 
    {\it    80248.30} & {\it      747.13} & 
    {\it  1.05} & {\it 2} & {\it  59.0} \\
{\sf 4.30} & {\sf 256} & {\sf   488.41} & {\sf     4.61} & 
    {\sf  0.94} & {\sf 1} & {\sf  33.2} & 
    {\sf    80379.20} & {\sf      962.86} & 
    {\sf  1.00} & {\sf 1} & {\sf  31.7} \\
\hline
{\rm 4.40} & {\rm  32} & {\rm   608.66} & {\rm     4.77} & 
    {\rm  0.50} & {\rm 2} & {\rm  77.9} & 
    {\rm   116045.88} & {\rm     1258.07} & 
    {\rm  0.50} & {\rm 2} & {\rm  77.9} \\
{\rm 4.40} & {\rm  64} & {\rm   607.55} & {\rm     4.90} & 
    {\rm  0.68} & {\rm 2} & {\rm  71.2} & 
    {\rm   115438.34} & {\rm     1297.88} & 
    {\rm  0.69} & {\rm 2} & {\rm  70.8} \\
{\it 4.40} & {\it 128} & {\it   606.22} & {\it     5.45} & 
    {\it  0.75} & {\it 2} & {\it  68.7} & 
    {\it   115121.00} & {\it     1390.11} & 
    {\it  0.76} & {\it 2} & {\it  68.4} \\
{\sf 4.40} & {\sf 256} & {\sf   607.38} & {\sf     7.01} & 
    {\sf  0.13} & {\sf 1} & {\sf  71.4} & 
    {\sf   115580.00} & {\sf     1724.71} & 
    {\sf  0.13} & {\sf 1} & {\sf  71.5} \\
\hline
{\rm 4.50} & {\rm  32} & {\rm   743.91} & {\rm     4.99} & 
    {\rm 17.63} & {\rm 5} & {\rm   0.3} & 
    {\rm   161644.96} & {\rm     1441.16} & 
    {\rm 18.91} & {\rm 5} & {\rm   0.2} \\
{\rm 4.50} & {\rm  64} & {\rm   741.86} & {\rm     5.25} & 
    {\rm 11.36} & {\rm 4} & {\rm   2.3} & 
    {\rm   160749.98} & {\rm     1524.23} & 
    {\rm 12.20} & {\rm 4} & {\rm   1.6} \\
{\it 4.50} & {\it 128} & {\it   740.18} & {\it     5.90} & 
    {\it  8.57} & {\it 3} & {\it   3.6} & 
    {\it   160184.00} & {\it     1641.99} & 
    {\it  9.19} & {\it 3} & {\it   2.7} \\
{\sf 4.50} & {\sf 256} & {\sf   737.83} & {\sf     7.34} & 
    {\sf  6.81} & {\sf 2} & {\sf   3.3} & 
    {\sf   159681.00} & {\sf     1957.27} & 
    {\sf  7.34} & {\sf 2} & {\sf   2.5} \\
\hline
{\rm 4.60} & {\rm  32} & {\rm   940.15} & {\rm     9.85} & 
    {\rm  2.23} & {\rm 2} & {\rm  32.8} & 
    {\rm   239226.36} & {\rm     3809.46} & 
    {\rm  2.27} & {\rm 2} & {\rm  32.2} \\
{\rm 4.60} & {\rm  64} & {\rm   932.11} & {\rm    10.39} & 
    {\rm  1.00} & {\rm 2} & {\rm  60.6} & 
    {\rm   235469.71} & {\rm     3977.98} & 
    {\rm  1.02} & {\rm 2} & {\rm  60.2} \\
{\it 4.60} & {\it 128} & {\it   929.18} & {\it    11.49} & 
    {\it  1.01} & {\it 2} & {\it  60.4} & 
    {\it   234461.00} & {\it     4288.18} & 
    {\it  1.02} & {\it 2} & {\it  60.1} \\
{\sf 4.60} & {\sf 256} & {\sf   929.73} & {\sf    14.09} & 
    {\sf  0.93} & {\sf 1} & {\sf  33.6} & 
    {\sf   235038.00} & {\sf     5147.64} & 
    {\sf  0.95} & {\sf 1} & {\sf  33.1} \\
\hline
{\rm 4.65} & {\rm  32} & {\rm  1049.65} & {\rm    12.82} & 
    {\rm  2.08} & {\rm 1} & {\rm  14.9} & 
    {\rm   287837.94} & {\rm     5463.27} & 
    {\rm  2.16} & {\rm 1} & {\rm  14.1} \\
{\rm 4.65} & {\rm  64} & {\rm  1036.57} & {\rm    13.29} & 
    {\rm  3.93} & {\rm 1} & {\rm   4.8} & 
    {\rm   281335.40} & {\rm     5552.70} & 
    {\rm  4.10} & {\rm 1} & {\rm   4.3} \\
{\it 4.65} & {\it 128} & {\it  1033.49} & {\it    14.50} & 
    {\it  4.08} & {\it 1} & {\it   4.3} & 
    {\it   280163.00} & {\it     5953.38} & 
    {\it  4.24} & {\it 1} & {\it   3.9} \\
{\sf 4.65} & {\sf 256} & {\sf  1034.53} & {\sf    17.16} & 
    {\sf  4.14} & {\sf 1} & {\sf   4.2} & 
    {\sf   281019.00} & {\sf     6811.95} & 
    {\sf  4.32} & {\sf 1} & {\sf   3.8} \\
\hline
{\rm 4.70} & {\rm  32} & {\rm  1149.03} & {\rm    11.81} & 
    {\rm  0.39} & {\rm 3} & {\rm  94.3} & 
    {\rm   334974.41} & {\rm     5138.61} & 
    {\rm  0.40} & {\rm 3} & {\rm  94.0} \\
{\rm 4.70} & {\rm  64} & {\rm  1140.40} & {\rm    12.25} & 
    {\rm  0.16} & {\rm 3} & {\rm  98.4} & 
    {\rm   330406.55} & {\rm     5250.77} & 
    {\rm  0.17} & {\rm 3} & {\rm  98.3} \\
{\it 4.70} & {\it 128} & {\it  1137.34} & {\it    13.39} & 
    {\it  0.20} & {\it 3} & {\it  97.7} & 
    {\it   329280.00} & {\it     5587.85} & 
    {\it  0.21} & {\it 3} & {\it  97.6} \\
{\sf 4.70} & {\sf 256} & {\sf  1138.49} & {\sf    16.23} & 
    {\sf  0.20} & {\sf 2} & {\sf  90.5} & 
    {\sf   330203.00} & {\sf     6544.22} & 
    {\sf  0.21} & {\sf 2} & {\sf  90.1} \\
\hline
{\rm 4.80} & {\rm  32} & {\rm  1435.78} & {\rm    15.18} & 
    {\rm  3.99} & {\rm 3} & {\rm  26.3} & 
    {\rm   485576.51} & {\rm     7655.78} & 
    {\rm  3.86} & {\rm 3} & {\rm  27.7} \\
{\rm 4.80} & {\rm  64} & {\rm  1423.46} & {\rm    16.09} & 
    {\rm  2.25} & {\rm 3} & {\rm  52.2} & 
    {\rm   478214.99} & {\rm     8078.91} & 
    {\rm  2.09} & {\rm 3} & {\rm  55.3} \\
{\it 4.80} & {\it 128} & {\it  1419.96} & {\it    17.36} & 
    {\it  2.00} & {\it 3} & {\it  57.3} & 
    {\it   476453.00} & {\it     8528.54} & 
    {\it  1.87} & {\it 3} & {\it  60.0} \\
{\sf 4.80} & {\sf 256} & {\sf  1418.50} & {\sf    20.45} & 
    {\sf  0.45} & {\sf 2} & {\sf  79.9} & 
    {\sf   476657.00} & {\sf     9743.68} & 
    {\sf  0.45} & {\sf 2} & {\sf  80.0} \\
\hline
{\rm 4.90} & {\rm  32} & {\rm  1779.47} & {\rm    24.92} & 
    {\rm  3.54} & {\rm 3} & {\rm  31.6} & 
    {\rm   693347.95} & {\rm    15088.08} & 
    {\rm  3.66} & {\rm 3} & {\rm  30.1} \\
{\rm 4.90} & {\rm  64} & {\rm  1759.03} & {\rm    25.37} & 
    {\rm  3.76} & {\rm 3} & {\rm  28.9} & 
    {\rm   679244.94} & {\rm    15148.94} & 
    {\rm  3.90} & {\rm 3} & {\rm  27.2} \\
{\it 4.90} & {\it 128} & {\it  1755.04} & {\it    26.90} & 
    {\it  3.83} & {\it 3} & {\it  28.1} & 
    {\it   677221.00} & {\it    15901.70} & 
    {\it  3.97} & {\it 3} & {\it  26.4} \\
{\sf 4.90} & {\sf 256} & {\sf  1753.30} & {\sf    30.59} & 
    {\sf  3.72} & {\sf 2} & {\sf  15.5} & 
    {\sf   677150.00} & {\sf    17675.40} & 
    {\sf  3.86} & {\sf 2} & {\sf  14.5} \\
\hline
{\rm 5.00} & {\rm  32} & {\rm  2218.69} & {\rm    44.16} & 
    {\rm 10.43} & {\rm 5} & {\rm   6.4} & 
    {\rm  1001133.00} & {\rm    32465.03} & 
    {\rm 10.42} & {\rm 5} & {\rm   6.4} \\
{\rm 5.00} & {\rm  64} & {\rm  2198.02} & {\rm    44.85} & 
    {\rm  4.62} & {\rm 4} & {\rm  32.8} & 
    {\rm   986315.31} & {\rm    32381.73} & 
    {\rm  4.75} & {\rm 4} & {\rm  31.4} \\
{\it 5.00} & {\it 128} & {\it  2189.50} & {\it    47.16} & 
    {\it  5.32} & {\it 3} & {\it  15.0} & 
    {\it   980853.00} & {\it    33603.90} & 
    {\it  5.53} & {\it 3} & {\it  13.7} \\
{\sf 5.00} & {\sf 256} & {\sf  2202.41} & {\sf    53.21} & 
    {\sf  2.84} & {\sf 2} & {\sf  24.2} & 
    {\sf   991980.00} & {\sf    37473.90} & 
    {\sf  2.95} & {\sf 2} & {\sf  22.9} \\
\hline
{\rm 5.10} & {\rm  32} & {\rm  2670.51} & {\rm    51.77} & 
    {\rm  0.02} & {\rm 3} & {\rm  99.9} & 
    {\rm  1371476.00} & {\rm    42469.89} & 
    {\rm  0.02} & {\rm 3} & {\rm  99.9} \\
{\rm 5.10} & {\rm  64} & {\rm  2627.87} & {\rm    51.80} & 
    {\rm  0.23} & {\rm 3} & {\rm  97.3} & 
    {\rm  1332282.00} & {\rm    42007.23} & 
    {\rm  0.21} & {\rm 3} & {\rm  97.6} \\
{\it 5.10} & {\it 128} & {\it  2618.49} & {\it    54.00} & 
    {\it  0.14} & {\it 3} & {\it  98.7} & 
    {\it  1325547.00} & {\it    43166.40} & 
    {\it  0.14} & {\it 3} & {\it  98.7} \\
{\sf 5.10} & {\sf 256} & {\sf  2619.75} & {\sf    60.40} & 
    {\sf  0.10} & {\sf 2} & {\sf  94.9} & 
    {\sf  1328528.00} & {\sf    47282.20} & 
    {\sf  0.10} & {\sf 2} & {\sf  94.9} \\
\hline
\end{tabular}
\vspace{3mm}
\caption{
   {\bf [Third page of 4-page table]}
}
\end{table}
\clearpage \addtocounter{table}{-1}
\begin{table}[h]
\hspace*{-1cm}
\small
\begin{tabular}{|l|r||r@{ (}r@{) }|r@{ (}r@{ DF, }r@{\%) }||r@{ (}r@{) }|r@{ (}r@{ DF, }r@{\%) }|}
\hline
\multicolumn{1}{|c|}{$\beta$}  &
   \multicolumn{1}{|c||}{$L_{min}$}  &
   \multicolumn{2}{c|}{$\xi_\infty$}  &
   \multicolumn{3}{c||}{${\cal R}$ for $\xi_\infty$}  &
   \multicolumn{2}{c|}{$\chi_{\stagg,\infty}$}  &
   \multicolumn{3}{c|}{${\cal R}$ for $\chi_{\stagg,\infty}$}   \\
\hline\hline
{\rm 5.20} & {\rm  32} & {\rm  3488.77} & {\rm    96.81} & 
    {\rm  0.79} & {\rm 3} & {\rm  85.2} & 
    {\rm  2133709.00} & {\rm    96738.46} & 
    {\rm  0.77} & {\rm 3} & {\rm  85.6} \\
{\rm 5.20} & {\rm  64} & {\rm  3431.88} & {\rm   100.37} & 
    {\rm  0.64} & {\rm 3} & {\rm  88.8} & 
    {\rm  2070752.00} & {\rm    98793.74} & 
    {\rm  0.59} & {\rm 3} & {\rm  89.9} \\
{\it 5.20} & {\it 128} & {\it  3416.96} & {\it    99.80} & 
    {\it  0.54} & {\it 3} & {\it  91.0} & 
    {\it  2058294.00} & {\it    97987.50} & 
    {\it  0.52} & {\it 3} & {\it  91.4} \\
{\sf 5.20} & {\sf 256} & {\sf  3419.80} & {\sf   109.80} & 
    {\sf  0.50} & {\sf 2} & {\sf  78.0} & 
    {\sf  2064205.00} & {\sf   106605.00} & 
    {\sf  0.49} & {\sf 2} & {\sf  78.2} \\
\hline
{\rm 5.30} & {\rm  32} & {\rm  4134.51} & {\rm   138.26} & 
    {\rm  0.72} & {\rm 2} & {\rm  69.9} & 
    {\rm  2835978.00} & {\rm   156936.69} & 
    {\rm  0.69} & {\rm 2} & {\rm  70.7} \\
{\rm 5.30} & {\rm  64} & {\rm  4061.19} & {\rm   142.43} & 
    {\rm  0.50} & {\rm 2} & {\rm  77.9} & 
    {\rm  2747493.00} & {\rm   159307.94} & 
    {\rm  0.48} & {\rm 2} & {\rm  78.6} \\
{\it 5.30} & {\it 128} & {\it  4038.28} & {\it   142.59} & 
    {\it  0.47} & {\it 2} & {\it  79.0} & 
    {\it  2725035.00} & {\it   158160.00} & 
    {\it  0.45} & {\it 2} & {\it  79.7} \\
{\sf 5.30} & {\sf 256} & {\sf  4057.82} & {\sf   156.95} & 
    {\sf  0.54} & {\sf 2} & {\sf  76.4} & 
    {\sf  2749120.00} & {\sf   173702.00} & 
    {\sf  0.52} & {\sf 2} & {\sf  77.1} \\
\hline
{\rm 5.40} & {\rm  32} & {\rm  5039.09} & {\rm   148.02} & 
    {\rm  0.28} & {\rm 3} & {\rm  96.4} & 
    {\rm  3955726.00} & {\rm   190762.23} & 
    {\rm  0.27} & {\rm 3} & {\rm  96.6} \\
{\rm 5.40} & {\rm  64} & {\rm  4953.59} & {\rm   148.53} & 
    {\rm  0.71} & {\rm 3} & {\rm  87.1} & 
    {\rm  3831850.00} & {\rm   188716.09} & 
    {\rm  0.66} & {\rm 3} & {\rm  88.2} \\
{\it 5.40} & {\it 128} & {\it  4928.06} & {\it   152.69} & 
    {\it  0.60} & {\it 3} & {\it  89.7} & 
    {\it  3804126.00} & {\it   192397.00} & 
    {\it  0.57} & {\it 3} & {\it  90.4} \\
{\sf 5.40} & {\sf 256} & {\sf  4941.46} & {\sf   163.20} & 
    {\sf  1.01} & {\sf 3} & {\sf  79.8} & 
    {\sf  3823421.00} & {\sf   204251.00} & 
    {\sf  0.92} & {\sf 3} & {\sf  82.1} \\
\hline
{\rm 5.50} & {\rm  32} & {\rm  6377.29} & {\rm   217.63} & 
    {\rm  1.99} & {\rm 2} & {\rm  37.0} & 
    {\rm  5821262.00} & {\rm   327977.26} & 
    {\rm  1.99} & {\rm 2} & {\rm  36.9} \\
{\rm 5.50} & {\rm  64} & {\rm  6223.22} & {\rm   221.23} & 
    {\rm  2.19} & {\rm 2} & {\rm  33.4} & 
    {\rm  5574691.00} & {\rm   326750.54} & 
    {\rm  2.15} & {\rm 2} & {\rm  34.2} \\
{\it 5.50} & {\it 128} & {\it  6203.03} & {\it   224.08} & 
    {\it  1.96} & {\it 2} & {\it  37.5} & 
    {\it  5553237.00} & {\it   330812.00} & 
    {\it  1.92} & {\it 2} & {\it  38.3} \\
{\sf 5.50} & {\sf 256} & {\sf  6197.48} & {\sf   238.44} & 
    {\sf  2.28} & {\sf 2} & {\sf  31.9} & 
    {\sf  5546998.00} & {\sf   347228.00} & 
    {\sf  2.20} & {\sf 2} & {\sf  33.3} \\
\hline
{\rm 5.60} & {\rm  32} & {\rm  7970.66} & {\rm   301.47} & 
    {\rm  0.04} & {\rm 2} & {\rm  97.8} & 
    {\rm  8498373.00} & {\rm   534485.16} & 
    {\rm  0.04} & {\rm 2} & {\rm  98.0} \\
{\rm 5.60} & {\rm  64} & {\rm  7801.45} & {\rm   309.62} & 
    {\rm  0.02} & {\rm 2} & {\rm  99.0} & 
    {\rm  8180645.00} & {\rm   538772.90} & 
    {\rm  0.01} & {\rm 2} & {\rm  99.3} \\
{\it 5.60} & {\it 128} & {\it  7760.02} & {\it   311.34} & 
    {\it  0.01} & {\it 2} & {\it  99.6} & 
    {\it  8116724.00} & {\it   540193.00} & 
    {\it  0.01} & {\it 2} & {\it  99.7} \\
{\sf 5.60} & {\sf 256} & {\sf  7790.68} & {\sf   326.89} & 
    {\sf  0.05} & {\sf 2} & {\sf  97.4} & 
    {\sf  8178031.00} & {\sf   561400.00} & 
    {\sf  0.04} & {\sf 2} & {\sf  97.8} \\
\hline
{\rm 5.70} & {\rm  32} & {\rm  9664.89} & {\rm   456.80} & 
    {\rm  0.23} & {\rm 2} & {\rm  89.2} & 
    {\rm 11718104.00} & {\rm   928104.64} & 
    {\rm  0.22} & {\rm 2} & {\rm  89.8} \\
{\rm 5.70} & {\rm  64} & {\rm  9531.97} & {\rm   469.77} & 
    {\rm  0.02} & {\rm 2} & {\rm  98.9} & 
    {\rm 11427243.00} & {\rm   940367.35} & 
    {\rm  0.02} & {\rm 2} & {\rm  99.0} \\
{\it 5.70} & {\it 128} & {\it  9460.68} & {\it   459.20} & 
    {\it  0.04} & {\it 2} & {\it  97.9} & 
    {\it 11295711.00} & {\it   915422.00} & 
    {\it  0.04} & {\it 2} & {\it  98.1} \\
{\sf 5.70} & {\sf 256} & {\sf  9548.22} & {\sf   514.52} & 
    {\sf  0.04} & {\sf 2} & {\sf  97.8} & 
    {\sf 11476466.00} & {\sf  1029147.00} & 
    {\sf  0.04} & {\sf 2} & {\sf  98.0} \\
\hline
{\rm 5.80} & {\rm  32} & {\rm 12659.85} & {\rm  1409.69} & 
    {\rm  0.00} & {\rm 0} & {\rm 100.0} & 
    {\rm 18383870.00} & {\rm  3478048.00} & 
    {\rm  0.00} & {\rm 0} & {\rm 100.0} \\
{\rm 5.80} & {\rm  64} & {\rm 12557.84} & {\rm  1515.32} & 
    {\rm  0.00} & {\rm 0} & {\rm 100.0} & 
    {\rm 18099541.00} & {\rm  3740026.00} & 
    {\rm  0.00} & {\rm 0} & {\rm 100.0} \\
{\it 5.80} & {\it 128} & {\it 12416.40} & {\it  1475.70} & 
    {\it  0.00} & {\it 0} & {\it 100.0} & 
    {\it 17772288.00} & {\it  3613560.00} & 
    {\it  0.00} & {\it 0} & {\it 100.0} \\
{\sf 5.80} & {\sf 256} & {\sf 12649.50} & {\sf  1593.68} & 
    {\sf  0.00} & {\sf 0} & {\sf 100.0} & 
    {\sf 18354042.00} & {\sf  3979249.00} & 
    {\sf  0.00} & {\sf 0} & {\sf 100.0} \\
\hline
{\rm 5.90} & {\rm  32} & {\rm 14796.02} & {\rm  2144.26} & 
    {\rm  0.00} & {\rm 0} & {\rm 100.0} & 
    {\rm 23832311.00} & {\rm  6025802.00} & 
    {\rm  0.00} & {\rm 0} & {\rm 100.0} \\
{\rm 5.90} & {\rm  64} & {\rm 14844.78} & {\rm  2373.04} & 
    {\rm  0.00} & {\rm 0} & {\rm 100.0} & 
    {\rm 23910024.00} & {\rm  6775206.00} & 
    {\rm  0.00} & {\rm 0} & {\rm 100.0} \\
{\it 5.90} & {\it 128} & {\it 14634.30} & {\it  2366.26} & 
    {\it  0.00} & {\it 0} & {\it 100.0} & 
    {\it 23371893.00} & {\it  6702087.00} & 
    {\it  0.00} & {\it 0} & {\it 100.0} \\
{\sf 5.90} & {\sf 256} & {\sf 15023.40} & {\sf  2645.59} & 
    {\sf  0.00} & {\sf 0} & {\sf 100.0} & 
    {\sf 24447729.00} & {\sf  7839348.00} & 
    {\sf  0.00} & {\sf 0} & {\sf 100.0} \\
\hline
{\rm 6.00} & {\rm  32} & {\rm 25502.30} & {\rm  7489.41} & 
    {\rm  0.00} & {\rm 0} & {\rm 100.0} & 
    {\rm 58916945.00} & {\rm 37012150.00} & 
    {\rm  0.00} & {\rm 0} & {\rm 100.0} \\
{\rm 6.00} & {\rm  64} & {\rm 25937.75} & {\rm  8904.71} & 
    {\rm  0.00} & {\rm 0} & {\rm 100.0} & 
    {\rm 60454213.00} & {\rm 45821760.00} & 
    {\rm  0.00} & {\rm 0} & {\rm 100.0} \\
{\it 6.00} & {\it 128} & {\it 25480.80} & {\it  8389.04} & 
    {\it  0.00} & {\it 0} & {\it 100.0} & 
    {\it 58758630.00} & {\it 41692067.00} & 
    {\it  0.00} & {\it 0} & {\it 100.0} \\
{\sf 6.00} & {\sf 256} & {\sf 26391.60} & $\infty$ & 
    {\sf  0.00} & {\sf 0} & {\sf 100.0} & 
    {\sf 62389831.00} & $\infty$ & 
    {\sf  0.00} & {\sf 0} & {\sf 100.0} \\
\hline
\end{tabular}
\vspace{3mm}
\caption{
   {\bf [Fourth page of 4-page table]}
}
\end{table}
 
\clearpage

%
%
%
%
\begin{table}[h]
\hspace*{-1cm}
\small
\begin{tabular}{|c||c|c|c|c|c|c|c|}
\hline
\multicolumn{1}{|c||}{$L_{min}$}  &
  \multicolumn{1}{|c|}{$n=3$}  &
  \multicolumn{1}{c|}{$n=4$}  &
  \multicolumn{1}{c|}{$n=5$}  &
  \multicolumn{1}{c|}{$n=6$}  &
  \multicolumn{1}{c|}{$n=7$}  &
  \multicolumn{1}{c|}{$n=8$}  &
  \multicolumn{1}{c|}{$n=9$}      \\
\hline\hline
 32 & 1738.06, 94 &  288.25, 93 &  277.13, 92 & 
       244.18, 91 &  243.54, 90 &  236.75, 89 &  228.29, 88  \\
    & 18.49,  0.0\% &  3.10,  0.0\% &  3.01,  0.0\% & 
       2.68,  0.0\% &  2.71,  0.0\% &  2.66,  0.0\% &  2.59,  0.0\%  \\
\hline
 64 &  932.48, 86 &  196.30, 85 &  193.24, 84 & 
       178.47, 83 &  177.49, 82 &  171.16, 81 &  164.95, 80  \\
    & 10.84,  0.0\% &  2.31,  0.0\% &  2.30,  0.0\% & 
       2.15,  0.0\% &  2.16,  0.0\% &  2.11,  0.0\% &  2.06,  0.0\%  \\
\hline
128 &  692.28, 68 &  142.15, 67 &  138.73, 66 & 
       131.45, 65 &  128.69, 64 &  127.09, 63 &  123.49, 62  \\
    & 10.18,  0.0\% &  2.12,  0.0\% &  2.10,  0.0\% & 
       2.02,  0.0\% &  2.01,  0.0\% &  2.02,  0.0\% &  1.99,  0.0\%  \\
\hline
256 &  387.08, 40 &  106.87, 39 &  106.47, 38 & 
       101.90, 37 &  100.23, 36 &   95.69, 35 &   90.82, 34  \\
    &  9.68,  0.0\% &  2.74,  0.0\% &  2.80,  0.0\% & 
       2.75,  0.0\% &  2.78,  0.0\% &  2.73,  0.0\% &  2.67,  0.0\%  \\
\hline
\end{tabular}
\vspace{3mm}
\caption{
   $\chi^2$ for the fit (\protect\ref{eqfss3}) of
   $\chi_{stagg}(\beta,2L)/\chi_{stagg}(\beta,L)$ versus $\xi(\beta,L)/L$.
   First line is $\chi^2$ followed by DF (number of degrees of freedom).
   Second line is $\chi^2$/DF followed by the confidence level.
   In all cases $\xi_{min} = 10$.
}
\label{table_chisq_chi}
\end{table}
 
\clearpage



\begin{figure}[h]
\vspace*{0cm} \hspace*{-0cm}
\begin{center}
\epsfxsize = 1.0\textwidth
\leavevmode\epsffile{estrapola_swaf2d.out_10_128_5.FSSPlotCorrLen.Jan_27_97.ps}
\end{center}
\vspace*{-2cm}
\caption{
   $\xi(\beta,2L)/\xi(\beta,L)$ versus $\xi(\beta,L)/L$.
   Symbols indicate $L=32$ ($+$), 64 ($\times$), 128 ($\Box$),
   256 ($\Diamond$), 512 ($\protect\scriptsize\bigcirc$).
   Error bars are one standard deviation.
   Curve is a fifth-order fit in \protect\reff{eqfss3},
   with $\xi_{min} = 10$ and $L_{min} = 128$.
}
\label{fig:xi_FSSplot}
\end{figure}
\clearpage


\begin{figure}[p]
\vspace*{-0.5cm} \hspace*{-0cm}
\begin{center}
\epsfxsize = 4.3in
\leavevmode
  \epsffile{estrapola_swaf2d.out_10_128_5.DevCurvaCorrLen.Jan_27_97.ps}  \\
\vspace*{-5mm}
\epsfxsize = 4.3in
\leavevmode
  \epsffile{estrapola_swaf2d.out_10_128_5.DevCurvaCorrLen.Jan_27_97.BLOWUP.ps}
\end{center}
\vspace*{-20mm}
\caption{
   Deviation of points from fit to $F_\xi$ with $s=2$, $n=5$,
   $\xi_{min} = 10$ and $L_{min} = 128$.
   Symbols indicate $L=32$ ($+$), 64 ($\times$), 128 ($\Box$),
   256 ($\Diamond$), 512 ($\protect\scriptsize\bigcirc$).
   Error bars are one standard deviation.
   Curves near zero indicate statistical error bars
   ($\pm$ one standard deviation) on the function $F_\xi(x)$.
   Plot (a) shows all points;
   plot (b) is a blow-up showing the $L=32$ points at $x \ge 0.50$.
}
\label{fig:xi_DeviazoniCurva}
\end{figure}
\clearpage


\begin{figure}[p]
\vspace*{-1.8cm}
\begin{center}
\epsfxsize=3.1in
\leavevmode\epsffile{estrapola_swaf2d.out_10_128_6.DevCurvaSus.Jan_27_97.ps}  \\
\vspace*{-3mm}
\epsfxsize=3.1in
\leavevmode
  \epsffile{estrapola_swaf2d.out_10_128_6.DevCurvaSus.Jan_27_97.BLOWUP1.ps}  \\
\vspace*{-3mm}
\epsfxsize=3.1in
\leavevmode
  \epsffile{estrapola_swaf2d.out_10_128_6.DevCurvaSus.Jan_27_97.BLOWUP2.ps}
\end{center}
\vspace{-0.8in}
\caption{
   Deviation of points from fit to $F_{\chi_{stagg}}$ with $s=2$, $n=6$,
   $\xi_{min} = 10$ and $L_{min} = 128$.
   Symbols indicate $L=32$ ($+$), 64 ($\times$), 128 ($\Box$),
   256 ($\Diamond$), 512 ($\protect\scriptsize\bigcirc$).
   Error bars are one standard deviation.
   Curves near zero indicate statistical error bars
   ($\pm$ one standard deviation) on the function $F_{\chi_{stagg}}(x)$.
   Plot (a) shows all points;
   plot (b) is a blow-up showing the $L=32$ points at $x \ge 0.50$;
   plot (c) is a blow-up showing all points with $x \ge 0.58$.
}
\label{fig:chi_DeviazoniCurva}
\end{figure}
%
%

 
\begin{figure}[h]
\vspace*{0cm} \hspace*{-0cm}
\begin{center}
\epsfxsize = 1.0\textwidth
\leavevmode\epsffile{estrapola_swaf2d.out_10_128_6.FSSPlotSus.Jan_27_97.ps}
\end{center}
\vspace*{-2cm}
\caption{
   $\chi_{stagg}(\beta,2L)/\chi_{stagg}(\beta,L)$ versus $\xi(\beta,L)/L$.
   Symbols indicate $L=32$ ($+$), 64 ($\times$), 128 ($\Box$),
   256 ($\Diamond$), 512 ($\protect\scriptsize\bigcirc$).
   Error bars are one standard deviation.
   Curve is a sixth-order fit in \protect\reff{eqfss3},
   with $\xi_{min} = 10$ and $L_{min} = 128$.
}
\label{fig:chi_FSSplot}
\end{figure}
\clearpage

%
%
\begin{figure}[p]
\vspace*{-1.5cm}
\begin{center}
\epsfxsize=3.1in
\leavevmode\epsffile{swaf2d_fullpaper_fig1a.ps}  \\
\epsfxsize=3.1in
\leavevmode\epsffile{swaf2d_fullpaper_fig1b.ps}  \\
\epsfxsize=3.1in
\leavevmode\epsffile{swaf2d_fullpaper_fig1c.ps}   \\
\end{center}
\vspace{-0.55in}
\caption{
    Infinite-volume correlation length $\xi_\infty$ divided by
    $e^{2\beta} \beta^p$ for (a) $p=0$, (b) $p=1/2$, (c) $p=1$.
    Error bars are one standard deviation.
}
\label{fig1}
\end{figure}

\clearpage

%
%
%
\begin{figure}[p] 
\begin{center} 
\epsfxsize=\textwidth
\leavevmode\epsffile{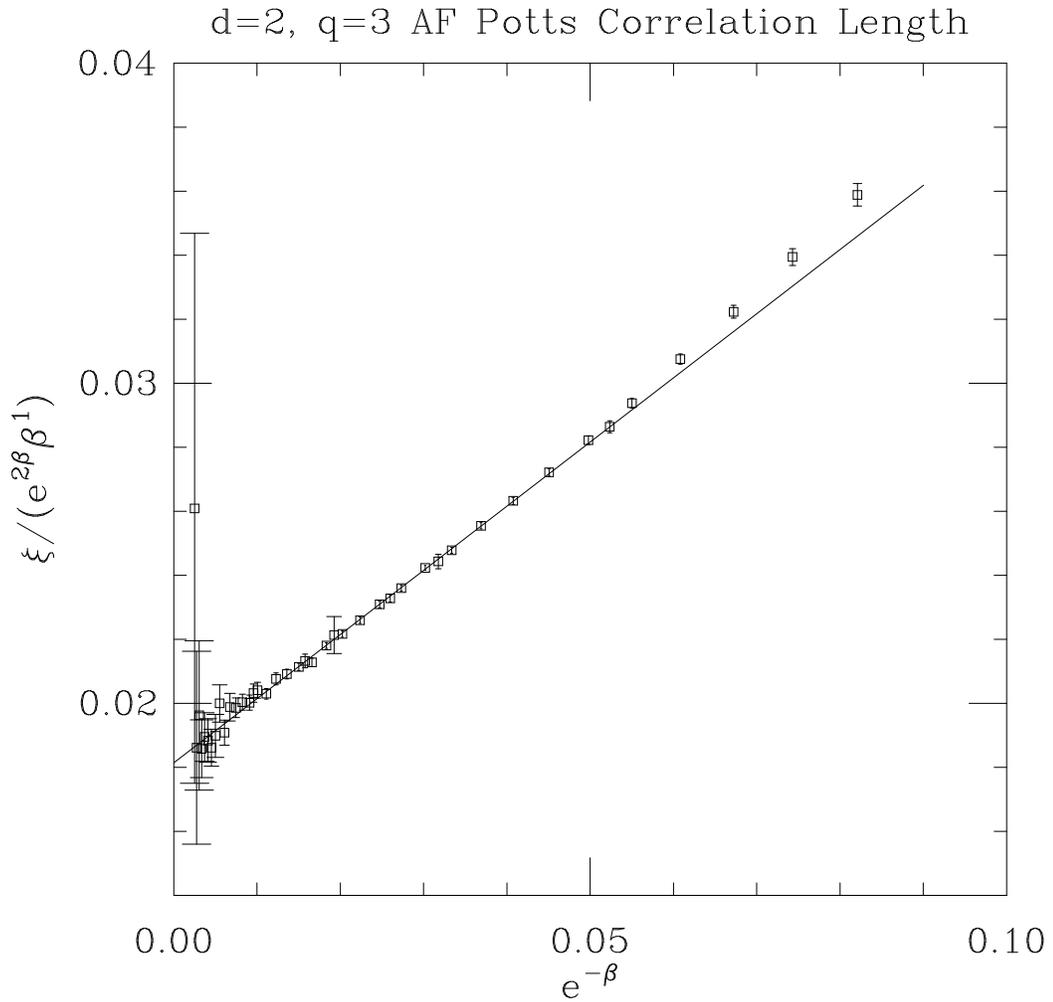}  \\
\end{center} 
\caption{  
    $\xi_\infty/(e^{2\beta} \beta^p)$ with $p=1$,
    plotted versus $e^{-\beta}$.
    Note the nearly linear behavior,
    in good agreement with (\protect\ref{eq5}).
    Straight line is
    $\xi_\infty/(e^{2\beta} \beta) = 0.01814 + 0.20051 e^{-\beta}$,
    which is the least-squares fit to the data with $\beta \ge 2.95$
    ($e^{-\beta} \ltapprox 0.052$).
} 
\label{fig2}
\end{figure}

\clearpage

%
%
%
\begin{figure}[p]
\vspace*{-0.7cm} \hspace*{-0cm}
\begin{center}
\epsfxsize = 0.7\textwidth
\leavevmode\epsffile{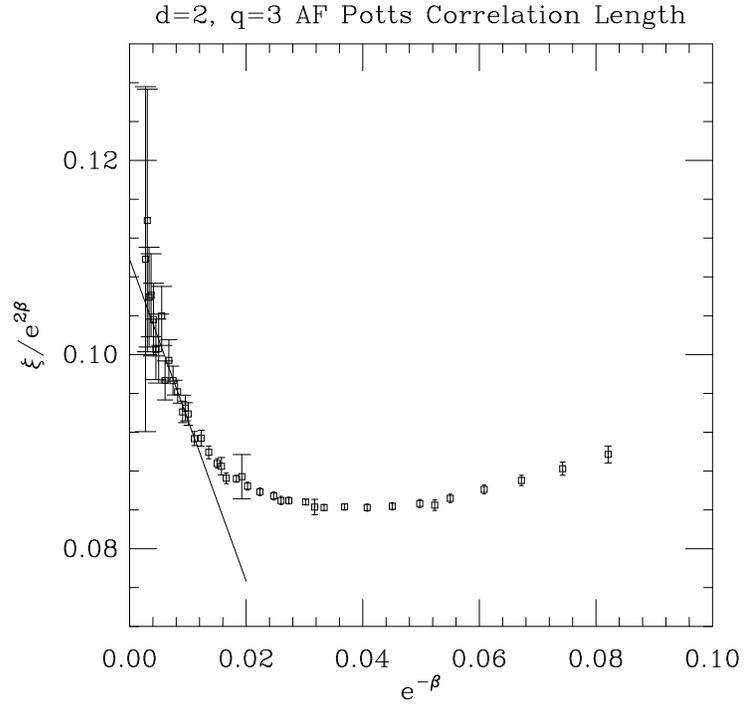} \\
\vspace{0.2cm}
\epsfxsize = 0.7\textwidth
\leavevmode\epsffile{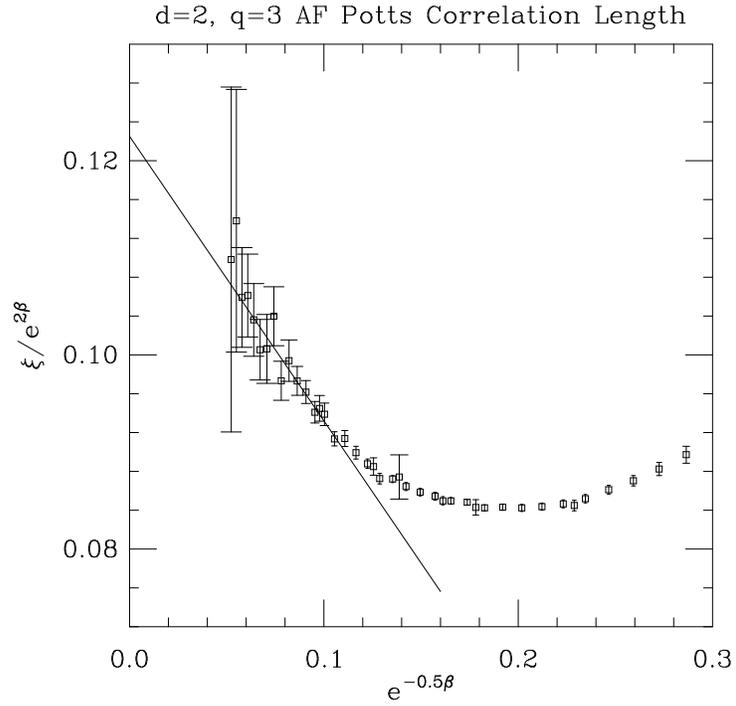}
\end{center}
\vspace*{-1.5cm}
\caption{  
    $\xi_\infty/e^{2\beta}$ plotted versus $e^{-\lambda\beta}$.
    (a) $\lambda=1$.
    Straight line is
    $\xi_\infty/e^{2\beta} = 0.1098 - 1.6574 e^{-\beta}$,
    which is the least-squares fit to the data with $\beta \ge 4.50$
    ($e^{-\beta} \ltapprox 0.011$).
    (b) $\lambda=0.5$.
    Straight line is
    $\xi_\infty/e^{2\beta} = 0.1225 - 0.2929 e^{-0.5\beta}$,
    which is the least-squares fit to the data with $\beta \ge 4.50$
    ($e^{-0.5\beta} \ltapprox 0.105$).
} 
\label{fig_xi_p=0_Delta}
\end{figure}

\clearpage

%
%
\begin{figure}[p]
\vspace*{-1.5cm}
\begin{center}
\epsfxsize=3.1in
\leavevmode\epsffile{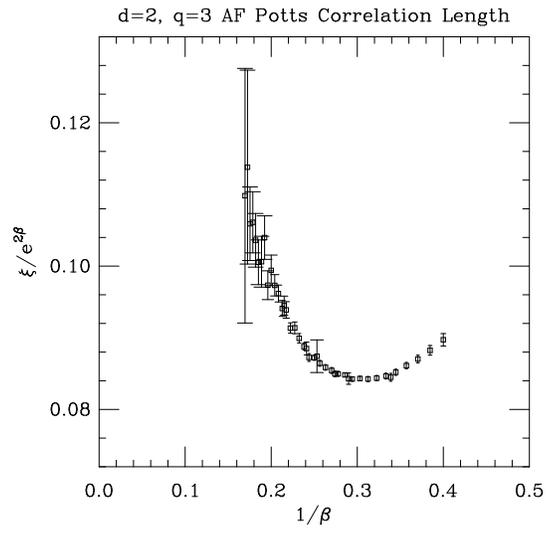}  \\
\epsfxsize=3.1in
\leavevmode\epsffile{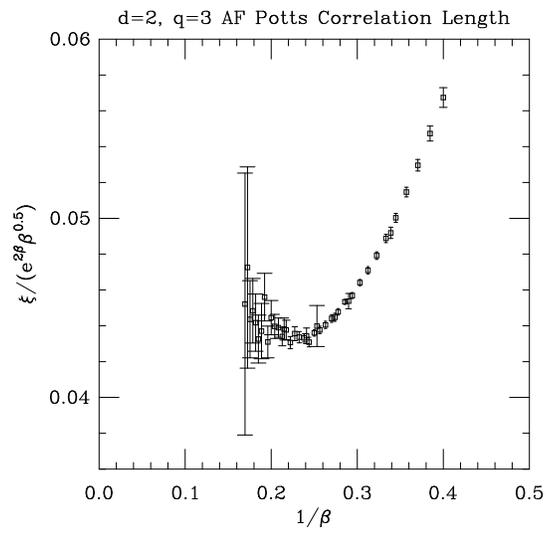}  \\
\epsfxsize=3.1in
\leavevmode\epsffile{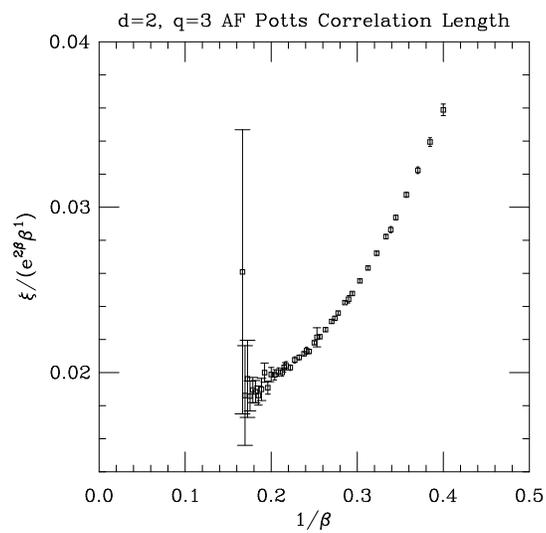}   \\
\end{center}
\vspace{-0.55in}
\caption{
    $\xi_\infty/(e^{2\beta} \beta^p)$ plotted versus $1/\beta$
    for (a) $p=0$, (b) $p=1/2$, (c) $p=1$.
}
\label{fig_referee}
\end{figure}

\clearpage

\begin{figure}[p] 
\vspace*{3cm} 
\begin{center} 
\epsfxsize=\textwidth
\leavevmode\epsffile{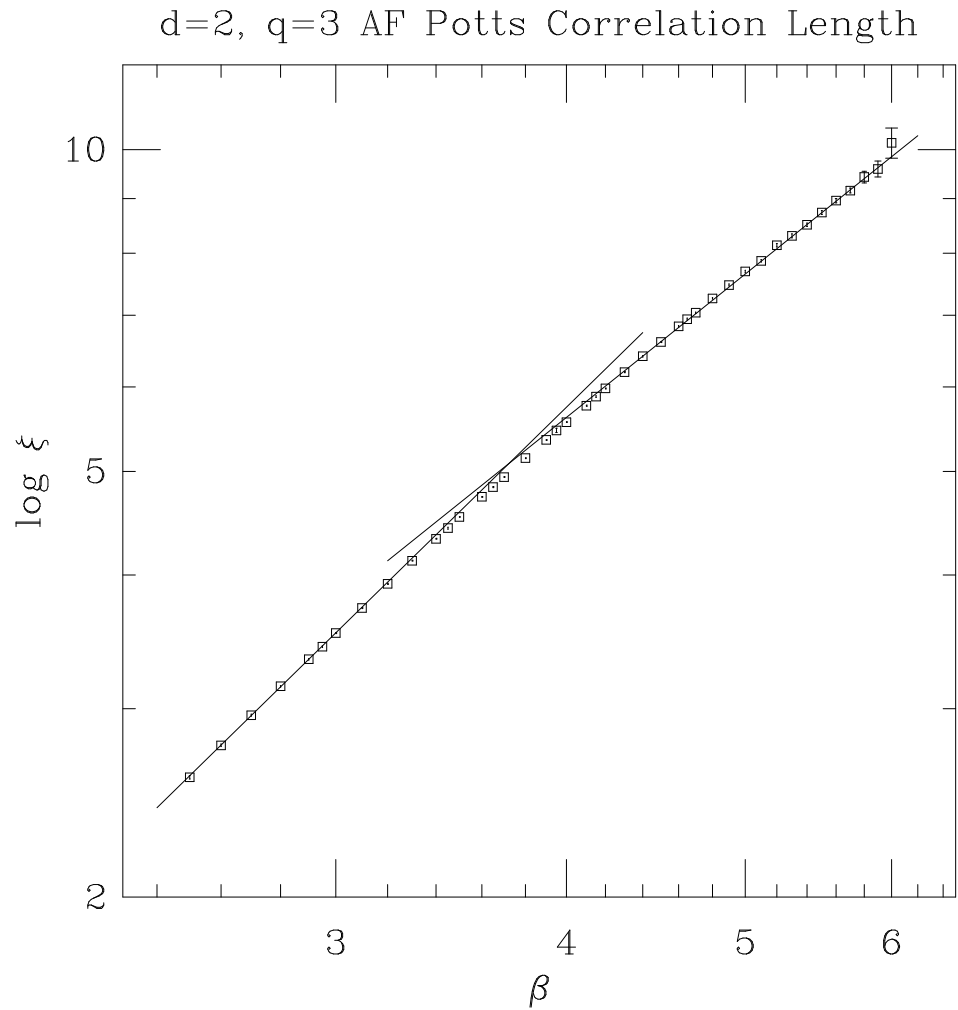}  \\
\end{center} 
\caption{  
    Log-log plot of $\log \xi_\infty$ versus $\beta$.
    The indicated asymptotes are
    $\log\xi_\infty = 0.55300 \beta^{1.68785}$ at small $\beta$, and
    $\log\xi_\infty = 0.82503 \beta^{1.38387}$ at large $\beta$.
} 
 \label{fig3}
\end{figure}

\clearpage

%
%
%
\begin{figure}[p]
\begin{center}
\epsfxsize=\textwidth
\leavevmode\epsffile{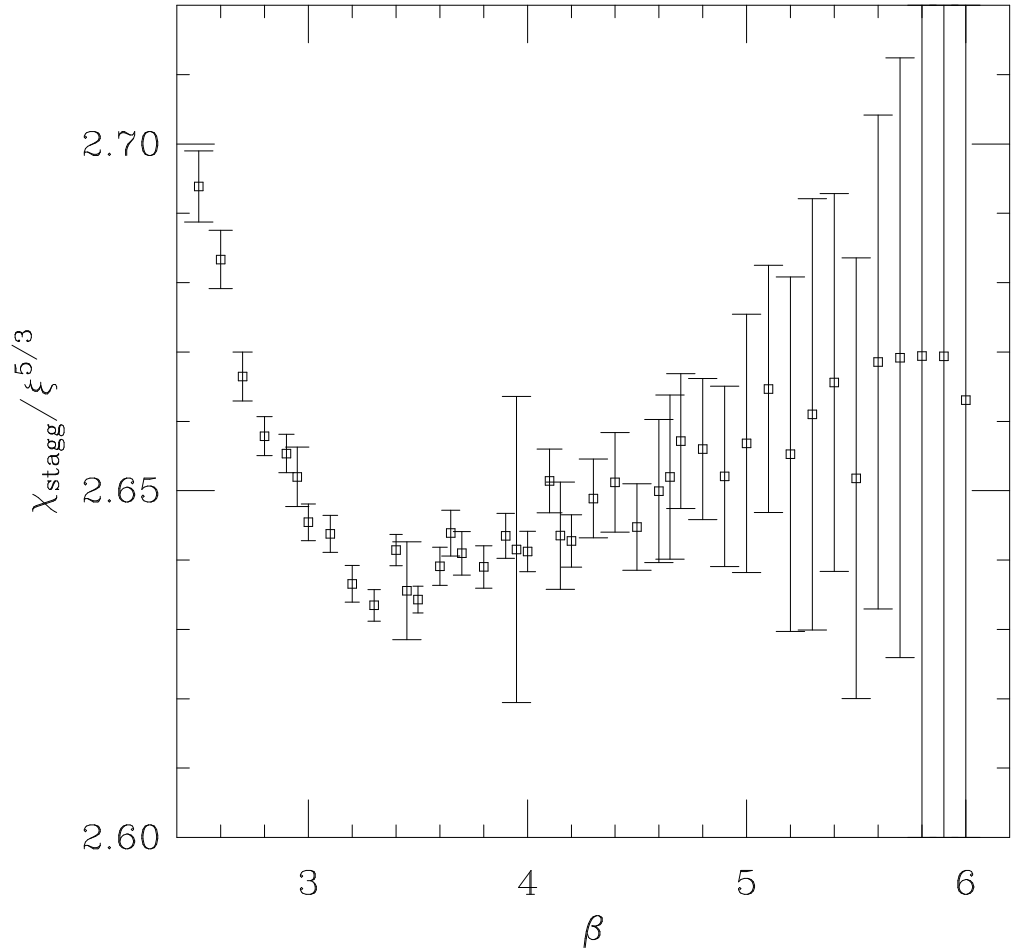}  \\
\end{center}
\caption{
    $\chi_{stagg,\infty}/\xi_\infty^{5/3}$ plotted versus $\beta$.
    Error bars are those given by the triangle inequality,
    {\em reduced by a factor of 10 for visual clarity}\/.
}
\label{fig_chioverxi53}
\end{figure}
 
\clearpage

%
%
\begin{figure}[p]
\vspace*{-1.5cm}
\begin{center}
\epsfxsize=3.1in
\leavevmode\epsffile{swaf2d_fullpaper_fig1a_chi.ps}  \\
\epsfxsize=3.1in
\leavevmode\epsffile{swaf2d_fullpaper_fig1b_chi.ps}  \\
\epsfxsize=3.1in
\leavevmode\epsffile{swaf2d_fullpaper_fig1c_chi.ps}   \\
\end{center}
\vspace{-0.55in}
\caption{
    Infinite-volume staggered susceptibility $\chi_{stagg,\infty}$ divided by
    $e^{(10/3)\beta} \beta^q$ for (a) $q=0$, (b) $q=5/6$, (c) $q=5/3$.
    Error bars are one standard deviation.
}
\label{fig1_chi}
\end{figure}

\clearpage

%
%
%
\begin{figure}[p] 
\begin{center} 
\epsfxsize=\textwidth
\leavevmode\epsffile{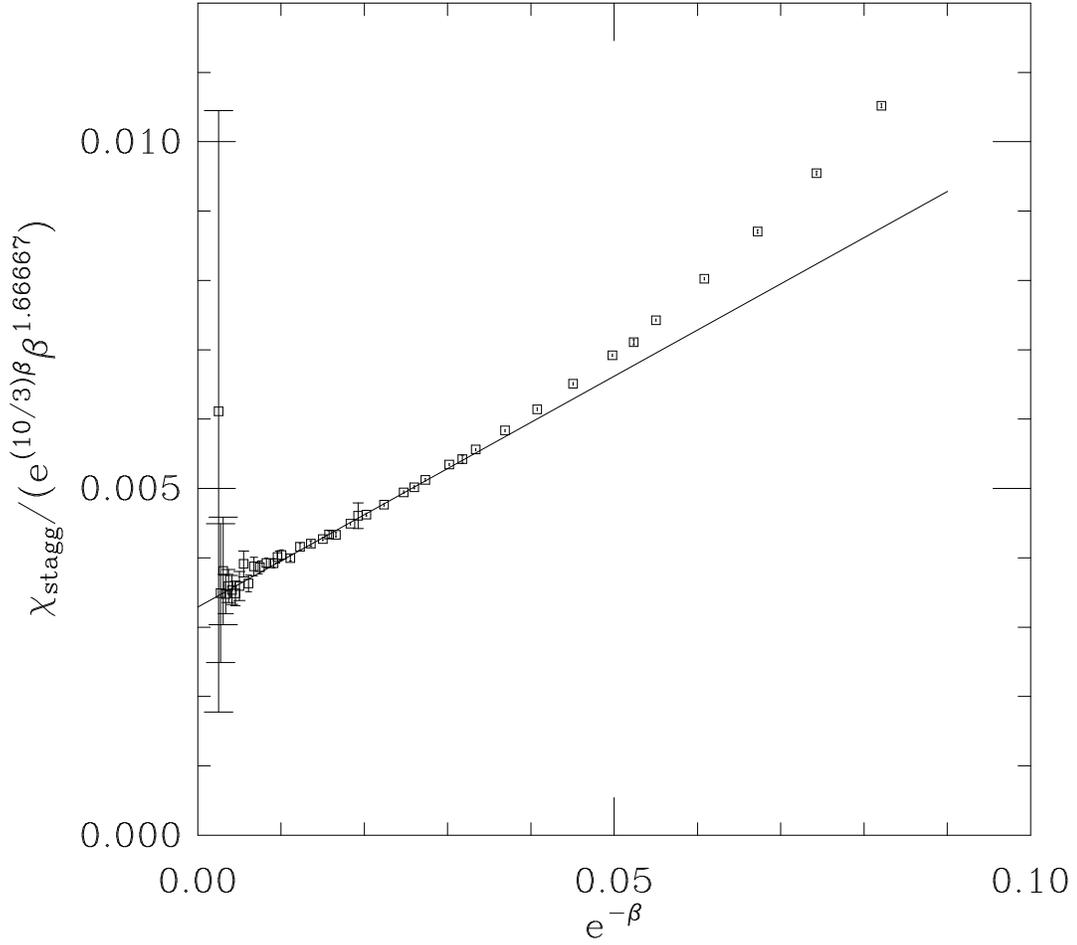}  \\
\end{center} 
\caption{  
    $\chi_{stagg,\infty}/(e^{(10/3)\beta} \beta^q)$ with $q=5/3$,
    plotted versus $e^{-\beta}$.
    The behavior is reasonably linear for $e^{-\beta} \ltapprox 0.03$,
    in good agreement with (\protect\ref{eq5_chi}).
    Straight line is
    $\chi_{stagg,\infty}/(e^{(10/3)\beta} \beta^{5/3})
      = 0.00329 + 0.06661 e^{-\beta}$,
    which is the least-squares fit to the data with $\beta \ge 3.60$.
} 
\label{fig2_chi}
\end{figure}

\clearpage

\begin{figure}[p] 
\begin{center} 
\epsfxsize=\textwidth
\leavevmode\epsffile{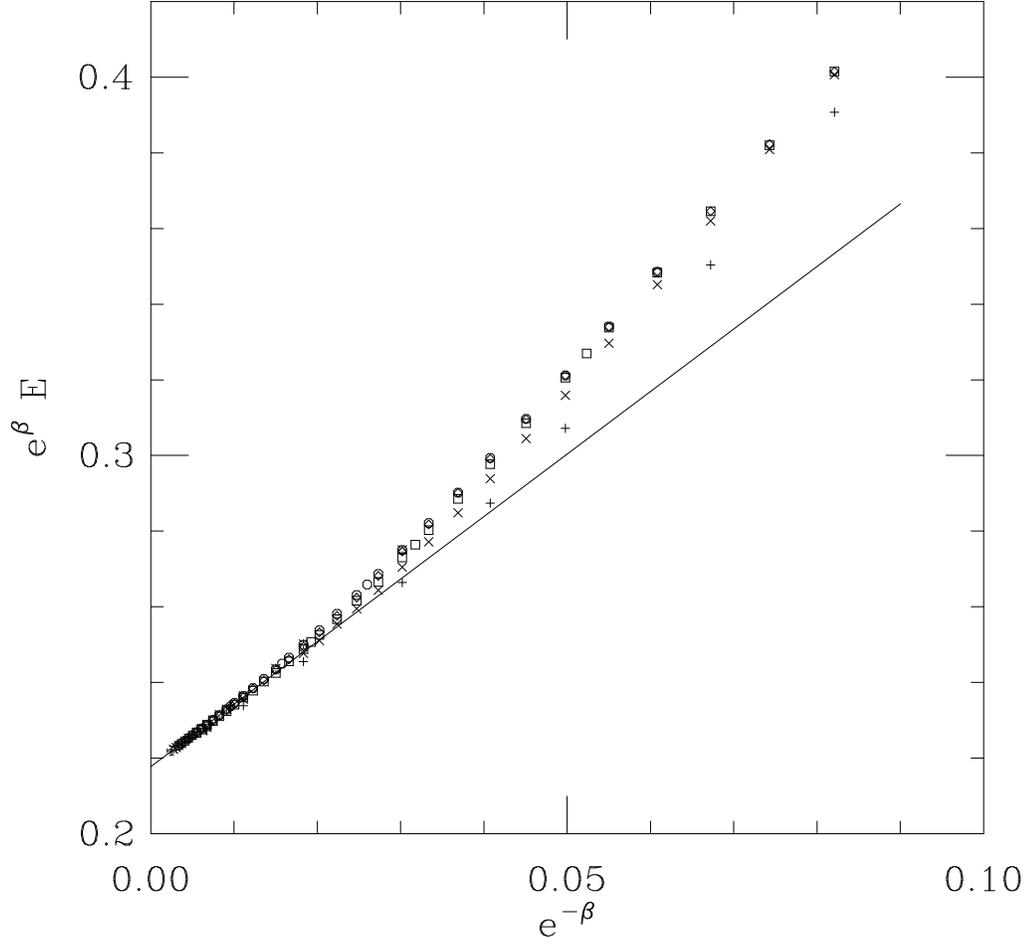}  \\
\end{center} 
\caption{  
    Energy per site $E$ divided by $e^{-\beta}$,
    plotted versus $e^{-\beta}$.
    Symbols indicate $L=32$ ($+$), 64 ($\times$), 128 ($\Box$),
    256 ($\Diamond$), 512 ($\protect\scriptsize\bigcirc$), 1024 ($\ast$),
    1536($\protect\fancyplus$).
    Error bars are invisibly small.
    The uppermost points at each $\beta$ represent the infinite-volume limit.
    Note the nearly linear behavior for small $e^{-\beta}$,
    in good agreement with (\protect\ref{eq6}).
    Straight line is
    $E = 0.21777 e^{-\beta} + 1.65303 e^{-2\beta}$,
    which is the least-squares fit to the data with $L=1024$
    and $\beta \ge 5.00$.
} 
\label{fig4}
\end{figure}

\clearpage

 
\begin{figure}[h]
\vspace*{2cm} \hspace*{-0cm}
\begin{center}
\epsfxsize = 1.0\textwidth
\leavevmode\epsffile{swaf2d_dynFSS_plot_magsq.ps}
\end{center}
\vspace*{-2cm}
\caption{
   Dynamic finite-size-scaling plot of $\tau_{int,\scrm_{\smstagg}^2}$
   versus $\xi(L)/L$,
   assuming dynamic critical exponent $z_{int,\scrm_{\smstagg}^2} = 0$.
   Symbols indicate $L=32$ ($+$), 64 ($\times$), 128 ($\Box$),
   256 ($\Diamond$), 512 ($\protect\scriptsize\bigcirc$),
   1024 ($\ast$), 1536($\protect\fancyplus$).
   Error bars are one standard deviation.
}
\label{fig:magsq_dyn_FSSplot}
\end{figure}

\clearpage

 
\begin{figure}[h]
\vspace*{2cm} \hspace*{-0cm}
\begin{center}
\epsfxsize = 1.0\textwidth
\leavevmode\epsffile{swaf2d_dynFSS_plot_ener.ps}
\end{center}
\vspace*{-2cm}
\caption{
   Dynamic finite-size-scaling plot of $\tau_{int,\scre}$
   versus $\xi(L)/L$,
   assuming dynamic critical exponent $z_{int,\scre} = 0$.
   Symbols indicate $L=32$ ($+$), 64 ($\times$), 128 ($\Box$),
   256 ($\Diamond$), 512 ($\protect\scriptsize\bigcirc$),
   1024 ($\ast$), 1536($\protect\fancyplus$).
   Error bars are one standard deviation.
}
\label{fig:ener_dyn_FSSplot}
\end{figure}

\clearpage

%
%
\begin{figure}[h]
\vspace*{2cm} \hspace*{-0cm}
\begin{center}
\epsfxsize = 1.0\textwidth
\leavevmode\epsffile{scaled_dynamic_plot_allpoints.ps}
\end{center}
\vspace*{-2cm}
\caption{
   Plot of $\rho_{\scrm_{stagg}^2 \scrm_{stagg}^2}(t)$
   versus $t / \tau_{int,\scrm_{stagg}^2}$,
   using {\em all}\/ data points.
   Symbols indicate $L=32$ ($+$), 64 ($\times$), 128 ($\Box$),
   256 ($\Diamond$), 512 ($\protect\scriptsize\bigcirc$),
   1024 ($\ast$), 1536($\protect\fancyplus$).
   Error bars are omitted.
}
\label{scaled_dynamic_plot_allpoints}
\end{figure}
\clearpage

%
%
\begin{figure}[h]
\vspace*{-1.2cm} \hspace*{-0cm}
\begin{center}
\vspace*{0cm} \hspace*{-0cm}
\epsfxsize=0.48\textwidth
\leavevmode\epsffile{scaled_dynamic_plot_00-01.ps}
\epsfxsize=0.48\textwidth
\epsffile{scaled_dynamic_plot_01-02.ps}  \\
%
\epsfxsize=0.48\textwidth
\leavevmode\epsffile{scaled_dynamic_plot_02-05.ps}
\epsfxsize=0.48\textwidth
\epsffile{scaled_dynamic_plot_050-054.ps}  \\
\epsfxsize=0.48\textwidth
\leavevmode\epsffile{scaled_dynamic_plot_054-058.ps}
\epsfxsize=0.48\textwidth
\epsffile{scaled_dynamic_plot_058-063.ps}
\end{center}
\vspace*{-1.2cm}
\caption{
   Plot of $\rho_{\scrm_{stagg}^2 \scrm_{stagg}^2}(t)$
   versus $t / \tau_{int,\scrm_{stagg}^2}$,
   subdivided by ranges of $\xi(L)/L$:
   (a) 0.0--0.1, (b) 0.1--0.2, (c) 0.2--0.5,
   (d) 0.50--0.54, (e) 0.54--0.58, (f) 0.58--0.63.
   Symbols indicate $L=32$ ($+$), 64 ($\times$), 128 ($\Box$),
   256 ($\Diamond$), 512 ($\protect\scriptsize\bigcirc$),
   1024 ($\ast$), 1536($\protect\fancyplus$).
   Error bars are omitted.
   The straight lines correspond to a pure exponential decay
   $\tau_{int,\scrm_{\smstagg}^2} = \tau_{exp,\scrm_{\smstagg}^2}$.
}
\label{scaled_dynamic_plot_subdivided}
\end{figure}
\clearpage

%
%
\begin{figure}
\vspace*{0cm} \hspace*{-0cm}
\begin{center}
\epsfxsize = 0.7\textwidth
\leavevmode\epsffile{swaf2d_RVTP_xi_plot.ps} \\
\vspace{0cm}
\epsfxsize = 0.7\textwidth
\leavevmode\epsffile{swaf2d_RVTP_chi_plot.ps}
\end{center}
\vspace*{-2cm}
\caption{
   Relative variance-time product
   [including errors of types (i) + (ii) + (iii)]
   divided by $\xi_\infty(\beta)^2$,
   plotted versus $\xi_\infty(\beta)/L$, for
   two-dimensional three-state Potts antiferromagnet.
   (a) is for $\scro = \xi$,
   (b) is for $\scro = \chi_{stagg}$.
   Symbols indicate $L=128$ ($\Box$),
   256 ($\Diamond$), 512 ($\protect\scriptsize\bigcirc$), 1024 ($\ast$),
   1536($\protect\fancyplus$).
}
\label{fig_RVTP}
\end{figure}

\clearpage

%
%
\begin{figure}[h]
\vspace*{2cm} \hspace*{-0cm}
\begin{center}
\epsfxsize = 1.0\textwidth
\leavevmode\epsffile{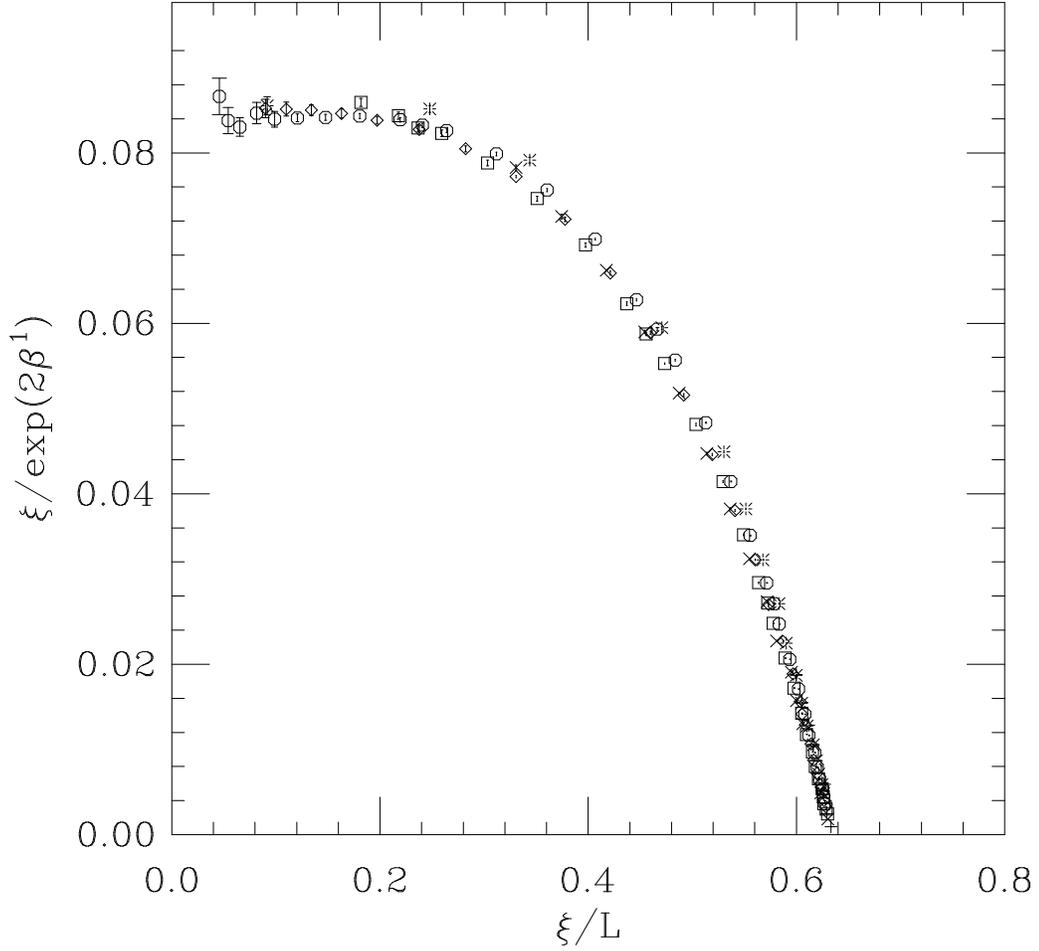}
\end{center}
\vspace*{-2cm}
\caption{
   Traditional finite-size-scaling plot of $\xi(L)/e^{2\beta}$
   versus $\xi(L)/L$, using all data points with $\xi(L) \ge 20$.
   Symbols indicate $L=32$ ($+$), 64 ($\times$), 128 ($\Box$),
   256 ($\Diamond$), 512 ($\protect\scriptsize\bigcirc$),
   1024 ($\ast$), 1536($\protect\fancyplus$).
   Error bars (one standard deviation) are in almost all cases
   smaller than the symbol size.
}
\label{fig_trad_FSS}
\end{figure}

\clearpage

\end{document}